\documentclass[11pt]{article}
\usepackage{amsmath, setspace}
\usepackage[margin=1in]{geometry}

\usepackage{graphicx}
\usepackage[english]{babel}
\usepackage{amsthm}
\usepackage{amssymb}
\usepackage{array}
\usepackage[latin1]{inputenc}
\usepackage{enumerate}
\usepackage{xspace}
\usepackage{authblk}
\usepackage{amsfonts}
\usepackage{amsmath}
\usepackage{hyperref}
\usepackage{caption}
\usepackage{makecell}
\usepackage{multirow}
\usepackage{booktabs}
\usepackage{color}
\usepackage{lmodern}
\usepackage{comment}
\usepackage{enumitem}
\usepackage{pgfplots}
\pgfplotsset{compat=1.18}
\usepackage{lipsum}
\usepackage{cleveref}
\crefname{subsection}{subsection}{subsections}
\usepackage{bbm}
\usepackage{tikz}
\tikzstyle{vertex}=[circle, draw, fill, inner sep=0pt, minimum size=5pt]

\usepackage{svg}
\usepackage{dsfont}

\usetikzlibrary{shapes}
\usetikzlibrary{backgrounds}
\usetikzlibrary{calc}

\theoremstyle{plain}
\newtheorem{theorem}{Theorem}[section]
\newtheorem{lemma}[theorem]{Lemma}

\newtheorem{proposition}[theorem]{Proposition}

\theoremstyle{definition}
\newtheorem{definition}[theorem]{Definition}
\newtheorem{remark}[theorem]{Remark}

\usepackage[english]{babel}
\usepackage[autostyle, english = american]{csquotes}
\MakeOuterQuote{"}

\usepackage[authoryear,round]{natbib}

\usepackage{mathtools}

\usepackage{color}            

\mathcode`l="8000
\begingroup
\makeatletter
\lccode`\~=`\l
\DeclareMathSymbol{\lsb@l}{\mathalpha}{letters}{`l}
\lowercase{\gdef~{\ifnum\the\mathgroup=\m@ne \ell \else \lsb@l \fi}}%
\endgroup

\newcommand{\Ber}{\textup{Ber}}

\usepackage{enumitem}
\makeatletter
\newcommand{\mylabel}[2]{#2\def\@currentlabel{#2}\label{#1}}
\newcommand{\eps}{\varepsilon}

\newcommand{\bE}{\mathbb{E}}
\newcommand{\bP}{\mathbb{P}}

\newcommand{\PLP}{\textsf{LP-POIS}}

\newcommand{\ALG}{\mathsf{ALG}}
\newcommand{\ROCRS}{\mathsf{RE}\text{-}\mathsf{OCRS}}
\newcommand{\SOCRS}{\mathsf{S}\text{-}\mathsf{OCRS}}
\newcommand{\mb}{\mathbb}
\newcommand{\scr}{\mathcal}
\newcommand{\Z}{Z}
\newcommand{\jc}{{j_0}}
\newcommand{\tc}{{t_0}}

\newcommand{\til}{\widetilde}
\newcommand{\ra}{\rightarrow}
\newcommand{\hF}{\hat{F}}
\usepackage{algorithm}
\usepackage[noend]{algpseudocode}

\usepackage{bm}
\newcommand{\Pois}{\mathrm{Pois}}

\allowdisplaybreaks
\begin{document}
\title{\sf{\fontsize{16.0pt}{18pt}\textbf{{Online Contention Resolution Schemes for Network Revenue Management and Combinatorial Auctions}}}}

\author[]{\normalsize Will Ma}
\affil[]{\sf Graduate School of Business and Data Science Institute, Columbia University}

\author[]{\normalsize Calum MacRury}
\affil[]{\sf School of Industrial and Systems Engineering, Georgia Tech}

\author[]{\normalsize Jingwei Zhang}
\affil[]{\sf School of Data Science, The Chinese University of Hong Kong, Shenzhen (CUHK-Shenzhen)}
\affil[ ]{\texttt{wm2428@gsb.columbia.edu, calum.macrury@isye.gatech.edu, zhangjingwei@cuhk.edu.cn}}

\date{}

{\singlespacing \maketitle}
\begin{abstract}%
\singlespacing

We study the Network Revenue Management (NRM) and Online Combinatorial Auctions (OCA) problems, in which products composed of up to $L$ resources are allocated in an online fashion.
We take a randomized rounding approach to these problems, employing the modern tool of Online Contention Resolution Schemes (OCRS) and their random-order analogues (RCRS).

However in NRM and OCA, there can be negative correlation in different products being demanded, making the standard definitions of OCRS/RCRS inapplicable.
This motivates a new definition of \textit{random-element} OCRS/RCRS, in which batches of elements arrive, and at most one element in each batch is demanded.

We first establish a simple random-element OCRS for achieving the well-known guarantee of $1/(1+L)$. We then prove that this guarantee is a fundamental barrier---it is \textit{tight for all prime-power values} of $L$, using a construction based on finite affine planes. Next, we break this barrier in several settings: the standard OCRS setting, on $L$-partite hypergraphs, and in the RCRS setting.  The first setting importantly establishes a separation between standard and random-element OCRS, showing that the negative correlation between elements being demanded in random-element OCRS can actually \textit{worsen} guarantees.
We also establish a separation between standard and random-element RCRS.

Although the general formulations of NRM and OCA require random-element OCRS, we also identify special cases in which standard OCRS or RCRS suffice.  This includes establishing new reductions from NRM with Poisson arrivals to standard OCRS, and NRM with time-homogeneous Poisson arrivals to standard RCRS.

\end{abstract}

\thispagestyle{empty}

\pagebreak

\setcounter{page}{1}

\setstretch{1.46}
\allowdisplaybreaks
\section{Introduction} \label{sec:intro}

Many problems in operations research and computer science can be described as allocating bundles of items, such as resources in a network or hotel occupancy over multiple days, to sequentially arriving agents or customers. Two canonical problems of this form are Network Revenue Management (NRM) and Online Combinatorial Auctions (OCA), which we now define. 

In the (discrete-time) NRM problem, a seller manages a network of items and offers a set of products to customers who arrive over time. Each product is a combination of items and generates a known revenue if sold. We focus on the setting where each product uses at most $L$ items. Time is divided into a finite number of periods. In each period, demand for each product is random, but its probability distribution is known in advance. At the start of a period, the seller must decide for each product whether to make it open for purchase. A product can only be open for purchase if all the associated items have not been sold out. After that, a random product is requested following a known independent probability. If it is open for purchase, then it gets sold; otherwise, nothing happens. The objective is to maximize the expected total revenue from selling products over the finite time horizon.

In the OCA problem, the decision-maker also starts with a finite set of items, and a sequence of agents arrives over time, where each agent has a valuation function over bundles of items. We focus on the case where any agent cares only about bundles of size at most $L$. The valuation function of an agent is independently drawn from a known distribution. When an agent arrives, her realized valuations for bundles are revealed, and the seller must decide immediately which bundle of available items if any to allocate to her. Allocations are irrevocable. The objective is to maximize the expected welfare, i.e.\ the total valuation that agents have for the bundles allocated to them.

Despite differences in application, these problems share a common mathematical structure. In both problems, a central planner must allocate a limited set of items to a sequence of arriving requests. Each request is for a "bundle" of up to $L$ items, and allocation decisions must be made irrevocably and online, knowing the distribution but not the realization of future arrivals. In this paper, we show how both the NRM and OCA problems can be tackled using the Online Contention Resolution Scheme (OCRS) framework.

To elaborate, a standard approach for tackling such online problems is to first formulate a fluid linear programming (LP) relaxation, which replaces random arrivals with their expectations. The solution to this LP provides an optimal fractional allocation, indicating the probability with which each potential product should be sold or each bundle should be allocated. This fluid LP serves as a powerful theoretical benchmark, representing an upper bound on the expected performance of any online policy or even the optimal offline allocation. The central challenge then becomes designing an online algorithm that can round this fractional solution into a feasible set of integral decisions in real-time, while preserving as much of the value as possible.

The OCRS framework is the tool for this online rounding task. In this framework, each potential sale of an element (such as a product in NRM or a bundle in OCA) is considered "active" independently, with a probability prescribed by the LP solution. An OCRS is then an online policy that observes a sequence of active elements
and selects a feasible subset, guaranteeing that every potential sale is ultimately realized with a uniform fraction of its initial probability.

However, a critical assumption of the standard OCRS framework is the independence of the active elements. Many realistic and general versions of NRM and OCA often involve additional subtleties and violate that assumption. For example, in NRM, although the requests are assumed to be independent across time, product $j$ being requested at a particular time $t$ means that a different product $j'$ will not be requested at time $t$, inducing a form of negative correlation. A similar issue arises in OCA if we treat each "agent-bundle" pair as a separate element. For a given arriving agent, there are many possible bundles she might receive, but at most one can actually be allocated. Conceptually, we can think of the agent "activating" at most of these bundles. This structure gives rise to "batched" arrivals: at each time period, a set of elements arrive, at most one of them is active, and the active element can be optionally selected if it preserves feasibility.

The basic negative correlation induced by this batched structure has arisen naturally in previous work, and standard OCRS's have been extended to handle it on a case-by-case basis without worsening the performance guarantees---see \Cref{ssec:further_related}.  In fact, general intuition says that the negative correlation induced by batching is helpful for OCRS guarantees, as discovered by \citet{ezra2022prophet}.

In stark contrast, we show that this batched structure can make the best-possible OCRS guarantees \emph{strictly worse}.
In particular, an OCRS for the standard, independent setting cannot generally be applied to the batched setting without loss in the performance guarantee.
Therefore, we formalize an extended notion of OCRS that handles the basic negative correlation induced by batched arrivals, which we interpret as a "random element" being chosen by each customer. We then establish that within this random-element OCRS framework, a performance guarantee of $1/(1+L)$ is achievable under bundles with size bounded by $L$. Using a construction based on finite affine planes, we also prove that this guarantee is tight, thereby formally quantifying the hardness introduced by negative correlation. Finally, we show that the $1/(1+L)$ benchmark can be strictly beaten in settings where the adversarial impact of this correlation is mitigated. This includes the standard OCRS setting, and also problems with special structure properties such as $L$-partite hypergraphs, and settings where arrivals occur in a uniformly random order. 

\subsection{Overview of Our Results and Contribution}

Our work reduces both the NRM and OCA problems to OCRS, showing for special cases of each problem if standard OCRS is enough or if it needs our extended notion of random-element OCRS.  We also consider RCRS, a variant of OCRS where the active elements are presented in \textit{random order} (instead of adversarial order) and hence leads to better guarantees, and correspondingly distinguish between standard vs.\ random-element RCRS.

\begin{table}[t]
\centering
\begin{tabular}{|rcccl|}
\hline
\multirow{2}{*}{Discrete-time NRM} & \multirow{2}{*}{$\longrightarrow$} & \multirow{2}{*}{Random-element OCRS} & \multirow{2}{*}{$\longleftarrow$} & OCA, agents arrive in \\
& & & & Adversarial order \\
\hline
Discrete-time NRM & \multirow{2}{*}{$\longrightarrow$} & \multirow{2}{*}{Random-element RCRS} & \multirow{2}{*}{$\longleftarrow$} & OCA, agents arrive in \\
w/ time-homogeneous rates 
& & & & Random order\\
\hline
Continuous-time NRM & \multirow{2}{*}{$\longrightarrow$} & \multirow{2}{*}{Standard OCRS} & \multirow{2}{*}{$\longleftarrow$} & OCA, Single-minded agents \\
(or $|\{j:\lambda_{j,t}>0\}|\le1$ $\forall t$ *) & & & & arrive in Adversarial order \\
\hline
Continuous-time NRM & \multirow{2}{*}{$\longrightarrow$} & \multirow{2}{*}{Standard RCRS} & \multirow{2}{*}{$\longleftarrow$} & OCA, Single-minded agents \\
w/ time-homogeneous rates ** & & & & arrive in Random order\\
\hline
\end{tabular}
{*Standard OCRS can also be used for discrete-time arrival models where $\lambda_{j,t}>0$ for at most a single $j$ at each $t$.  In these models, indices $t,j$ typically get merged into one.

**Reduction requires a standard RCRS which is ``inactive-oblivious''. See \Cref{sec:main_body_reductions}.}
\caption{Reductions (indicated by arrows) from variants of NRM and OCA to variants of OCRS.
Special cases of NRM and OCA generally get mapped to Standard (instead of Random-element) and RCRS (instead of OCRS),
in which case performance guarantees get better.
}
\label{tab:mapping}
\end{table}

Our reductions are summarized in \Cref{tab:mapping}. We highlight in particular the surprising distinction between the classical continuous-time formulation of NRM using Poisson arrival processes \citep{gallego1997multiproduct} vs.\ the discrete-time formulation above, where continuous-time reduces to Standard OCRS/RCRS and has better performance guarantees.
Indeed, we explained above how discrete time periods induce negative correlation between two different products $j,j'$ being requested, whereas in the continuous-time formulation these time periods are infinitesimally small causing this negative correlation to "vanish".
Establishing this reduction is non-trivial, and we do so by taking a detour through OCRS in the vanishing regime \citep{nuti2023towards}.
To the best of our knowledge, existing Revenue Management literature has always treated the continuous-time and discrete-time formulations as interchangeable (at least in terms of performance guarantees), whereas we establish a clear distinction between them, stemming from our "surprising separation" between Random-element vs.\ Standard OCRS.

The other distinctions in \Cref{tab:mapping} are more straight-forward, where the special case of NRM with time-homogeneous rates, equivalently i.i.d.\ arrivals, can be captured by RCRS (easier than OCRS); the special case of OCA with single-minded agents can be captured by standard OCRS/RCRS (easier than random-element OCRS/RCRS respectively).  These distinctions are formalized in \Cref{sec:nrm_and_reduction_crs}.

This paper conducts a systematic investigation of the performance guarantee achievable within each of these OCRS models. Throughout the paper, we focus on the case where all products are composed of at most $L$ items, and our performance guarantees are parameterized by this value. These guarantees are expressed as a competitive ratio, where the benchmark of our online algorithm is measured against the optimal value of a fluid LP relaxation. Table \ref{tab:my_label} summarizes our findings, with our new results highlighted in bold, and reveals a surprising separation in performance based on the presence of negative correlation. 

\begin{table}[t]
\centering
\begin{tabular}{c|c|c} \hline
Selection Guarantees & OCRS  & RCRS \\ 
\hline
Random-element & \makecell{$\ge 1/(1+L)$ for $L\ge 1$[\S\ref{ssec:ocrs_warmup}] \\ $\le 1/(1+L)$ \\ for $L$ a prime power  [\S\ref{ssec:tightness}]} & \makecell{$ > 1/(1 + L) $ for $L\ge 2$ [\S\ref{sec:rcrs_random_element}] \\ $\le(1 - 1/(1+L)^{1+L})/L$ \\ for $L$ a prime power [\S\ref{sec:negOffline}]}\\
\hline
Standard & $ > 1/(1+L)$ for $L\ge 2$[\S\ref{ssec:beating_standard}] &$ > (1 - 1/(1+L)^{1+L})/L$ for $L\ge 5$ [\S\ref{ssec:rcrs_standard}]\\ 
\hline
$L$-partite Random-element& $> 1/(1 +L)$ for $L\ge 2$ [\S\ref{sec:Lpartite}]  & inherited from \S\ref{sec:rcrs_random_element} and \S\ref{sec:Lpartite} \\
\hline
\end{tabular}
\caption{Summary of our technical results. "$\ge$" and "$>$" refer to lower bounds on the selection guarantee (algorithmic results), while "$\le$" refers to upper bounds (impossibility results). We establish separation between standard vs.\ random-element OCRS---the tight guarantee for the latter is $1/(1+L)$, whereas the guarantee for the former strictly exceeds $1/(1+L)$.  We also breach the $1/(1+L)$ guarantee in the $L$-partite and RCRS settings.  We establish a similar separation between standard vs.\ random-element RCRS for large $L$.  }
\label{tab:my_label}
\end{table}

\subsection{Discussion of Results}

\paragraph{$1/(1+L)$ benchmark in OCRS} The guarantee of $1/(1+L)$ was already known in both the NRM \citep{ma2020approximation} and OCA \citep{correa2023optimal} problems, with the former result achieved using approximate dynamic programming and the latter result achieved using a particularly simple static item pricing mechanism. Both results have been extended in subsequent works. Therefore, $1/(1+L)$ can be viewed as a benchmark for $L$-bounded products. We begin by designing a simple random-element OCRS that achieves this guarantee using the idea of exact selection from \cite{ezra2022prophet}. Our approach provides a unified proof of the $1/(1+L)$ competitive ratio against the fluid LP for both the NRM formulation introduced above and for general OCA problems where agents desire at most $L$ items. This stands in contrast to prior work, which established this guarantee using disparate, problem-specific techniques.

More importantly, we prove that for general $L$, this benchmark is fundamentally tight under our notion of a random-element OCRS. While the $1/(1+L)$ guarantee was known in NRM, it was an open question whether this was the best possible. Our work closes this gap.  In particular, we construct a worst-case instance by translating the combinatorial properties of a finite affine plane of order $L$ into an NRM problem. Since these planes are known to exist whenever $L$ is a prime power, our result proves that $1/(1+L)$ guarantee is unbeatable for a large class of problems. The $L=2$ result implies that OCRS's for graph matching cannot be better than $1/3$-selectable when edges are batched adversarially, complementing the work of \citet{ezra2022prophet}. Our result more generally shows that $1/(1+L)$ is unbeatable by any analysis that is based on the LP relaxation. We also show that in the offline version of the random-element contention resolution problem, finite affine planes can be similarly used, to construct a weaker upper bound of $(1-\frac1{(1+L)^{1+L}})/L$ for prime power $L$.

Our next finding is that this $1/(1+L)$ barrier can be broken in several settings. To prove this, we develop a new analytical technique that provides a more powerful alternative to simple union bound often used in such analysis. We believe this tool, which allows for a more precise accounting of correlated events, may be of independent interest for beating benchmarks in other online optimization problems. Our argument may be useful whenever feasibility is analyzed using a union bound over the events that individual resources are unavailable. If two resources can become unavailable together, the union bound counts that outcome twice, and lower bound on their overlap gives a sharper estimate. Similar union-bound steps arise in  asymptotic analyses of fluid policies in revenue management when bounding unavailable probabilities of products. Within this paper, our standard RCRS result provides another concrete application of this idea.

Using this new technique, we first establish a guarantee strictly exceeding $1/(1+L)$ for all $L>1$ in the standard OCRS setting where arrivals are independent. While a guarantee strictly exceeding 1/3 was already known in the $L=2$ case, which corresponds to matchings in graphs \citep{ezra2022prophet,macrury2022random}, their "witness" arguments do not easily extend to a general $L>2$. Our proof technique provides a unified proof for all $L$ and offers new insights even for the $L=2$ case. Next, we show that $1/(1+L)$ can be beaten if $L>1$ even for random-element OCRS assuming $L$-partite hypergraphs, where the items come from $L$ groups and each product requires at most one item from each group. This structure arises in applications such as hotel booking \citep{rusmevichientong2023revenue} or selling combo products. Moreover, this captures the prophet inequality problem on the intersection of $L$ partition matroids, whose tight ratio is mentioned as an open problem in \citet{correa2023optimal}. They show that the tight ratio is at least $1/(1+L)$; we now know that the tight ratio is strictly greater than $1/(1+L)$.  Moreover, our work suggests that the tight ratio may depend on whether the elements in the prophet inequality problem are allowed to be random.

\paragraph{Guarantees in RCRS} The standard RCRS setting has been studied in great detail, with many results showing that guarantees strictly improve now that the arrival order is random instead of adversarial (see \citealt{lee2018optimal,adamczyk2018random}).
However, this literature has not considered the notion of random-element RCRS, where adversarially-designed batches arrive in a random order. We show that this setting is strictly easier than its adversarial-order counterpart by designing a random-element RCRS with a guarantee strictly better than $1/(1+L)$ if $L>1$. For example, when $L=2$, our guarantee is  $\approx 0.397$, thus beating the $1/3$ benchmark. In OCA, when the agents arrive in a random order, this is often called a \textit{prophet secretary problem}. Our result for random-element RCRS implies that a competitive guarantee greater than $1/(1+L)$ is attainable for this problem, whereas previously this was only known for $L=1$. Specifically, $1-1/e \approx 0.632$ was proven by \citet{ehsani2018prophet}.

Moving to standard RCRS -- where arrivals are both independent and in a random order -- we can further beat the "baseline" guarantee of $(1-e^{-L})/L$. In the graph matching case of $L=2$, \citet{brubach2021improved} first established the guarantee of $(1 - e^{-2})/2 \approx 0.432$, and this was then beaten by \citet{pollner2022improved}
and \citet{macrury2022random}.
We extend these findings to $L \ge 3$, and show that $(1 - e^{-L})/L$ is beatable. As in the OCRS setting, our result for standard RCRS implies guarantees for special cases of NRM and OCA, including NRM with time-homogeneous Poisson arrivals and OCA with single-minded agent valuations. Finally, we note that \citet{marinkovic2024online} had previously proven a guarantee of $(1 - e^{-L})/L$ for this i.i.d. problem. Our results are incomparable, as their algorithm works for an unknown distribution with minimal sample access,
whereas ours needs access to the distribution but beats $(1 - e^{-L})/L$ and operates under the harder random arrival order. The guarantee for our standard RCRS also exceeds the aforementioned random-element offline CRS upper bound of $(1-\frac1{(1+L)^{1+L}})/L$ (which is greater than $(1-e^{-L})/L$) when $L\ge 5$. This separation implies that for $L\ge 5$, random order is less constraining than random elements: it is easier to design a standard RCRS than it is to design a random-element offline CRS.

\subsection{Paper Outline}
The rest of the paper is organized as follows. Section \ref{sec:nrm_and_reduction_crs} establishes the formal connection between NRM, OCA and the OCRS framework. We begin by defining the discrete-time NRM problem and its fluid relaxation, which motivates the need for our extended random-element OCRS. We then establish the reductions that map various NRM/OCA formulations to these OCRS models. Section \ref{sec:ocrs} presents our results for the adversarial-order (OCRS) setting. We start by presenting a simple policy that achieves the $1/(1+L)$ guarantee in the general random-element OCRS model and then prove that this bound is tight and cannot be beaten. After establishing this benchmark, in Section \ref{ssec:beating_standard}, we discuss how our new analytical technique allows us to strictly exceed the $1/(1+L)$ guarantee for the standard OCRS. Section \ref{sec:rcrs} provides the analogous results for the random-order (RCRS) setting. Finally, Section \ref{sec:conclusion} concludes the paper. 

\subsection{Further Related Work} \label{ssec:further_related}

\paragraph{Random-element CRS.} Our notion of random elements, which imply a basic form of negative correlation, is not new to the vast literature on online Bayesian selection and allocation.
That being said, our work makes the surprising finding that random elements can \textit{worsen} the best-possible guarantee, motivating us to explicitly distinguish between random-element OCRS and the standard OCRS with fixed elements.

Under the simplest online selection constraint where at most $k$ elements can be accepted, there is no difference between fixed vs.\ random elements, because elements are identical.
Under general matroid constraints (in which elements are non-identical),
the prophet inequality of \citet{kleinberg2012matroid} has been extended to handle random elements in the context of combinatorial auctions, with the same guarantee of 1/2 \citep{dutting2020prophet}.
Similarly,
the ex-ante matroid prophet inequality of \citet{lee2018optimal} has been extended to handle random elements in the context of assortment optimization, also with the same guarantee of 1/2 \citep{baek2022bifurcating}.
Under knapsack constraints, \citet{jiang2022tight} establish a tight guarantee of $1/(3+e^{-2})$ for the OCRS problem, which they later extend to elements with random sizes, corresponding to random elements.
In sum, for matroids and knapsacks, guarantees for fixed elements appear to extend to random elements, even though there is no black-box reduction.

Random elements can also be interpreted as a basic form of negative correlation.  In this vein, classical prophet inequalities have been shown to extend to negatively dependent random variables \citep{rinott1987comparisons,samuel1991prophet}.  Meanwhile, \citet{dughmi2019outer} shows that the $(1-1/e)$-selectable offline contention resolution schemes for matroids of \citet{chekuri2011submodular} can be extended under various forms of negative correlation, and even some cases of positive correlation. \citet{qiu2022} expand upon these results, showing that the same $1-1/e$ guarantee is attainable for distributions which satisfy \textit{weak negative regression}, a definition they introduce that generalizes both negative regression and negative association. 

Overall, the findings from the literature suggest that random elements and negative correlation should not worsen guarantees in online Bayesian selection.
In stark contrast, our work finds that they do worsen guarantees for matchings in graphs, and more generally, $L$-bounded products. A related perspective comes from work on approximating the optimal online policy in Bayesian online resource allocation. \cite{anari2019nearly} use LP relaxations that combine ex-ante constraints with online-policy constraints. Their rounding relies on negative dependence and concentration, while still requiring slack to avoid capacity violations. \cite{braverman2025new} use pivotal sampling to induce negative correlations in online Bayesian matching, but the analysis needs delicate bounds. These papers illustrate that negative correlation can help in designing an algorithm while also making its performance more difficult to analyze. 

\paragraph{OCRS with (positive) correlations.}
There is a recent line of work studying how OCRS guarantees worsen under limited positive correlation among the elements. \citet{gupta2024pairwise} study the setting where the elements are pairwise independent and the feasibility constraints are described by various matroid constraints. While they show that constant guarantees are attainable for special classes of matroids, the guarantees are worse than in the independent setting. \citet{dughmi2024limitations} study the same setting, and showed that the situation can become arbitrarily bad for a general matroid. Specifically, they construct a linear matroid of rank $k$ and show that no $\omega(1/k)$-selectable OCRS is possible, where $\omega$ is a function tending to infinity arbitrarily slowly.
Motivated by applications to mechanism design, \citet{bhawalkar2024mechanism} also recently introduced an extension of OCRS called a \textit{two-level OCRS}. Depending on the parametrization of their model, this extension allows for both positive and negative correlation among the active elements. Our \Cref{def:random-ocrs} is in fact a special case of their model, so proving a selection guarantee for their two-level OCRS implies a selection guarantee for our random-element OCRS. However, they study knapsack and a family of "Vertical-Horizontal" constraints, whereas we focus on constraints imposed by $L$-bounded products, so our results are not directly comparable.

\paragraph{Extensions of $1/(1+L)$ results.}
In NRM, the $1/(1+L)$ guarantee of \citet{ma2020approximation} has been extended to both reusable items \citep{baek2022bifurcating}, flexible products \citep{zhu2023performance}
and Markovian-correlated product arrivals \citep{jiang2023constantapproximationnetworkrevenue}.
In OCA, the $1/(1+L)$ guarantee of \citet{correa2023optimal} has been shown to also hold when only a single sample is given about each distribution, if the arrival order is random \citep{marinkovic2024online}.

\paragraph{Integrality gaps for hypergraph matching.}
\citet{chan2012linear} study the randomized rounding problem for fractional matchings satisfying~\eqref{eqn:feasInExpn} on hypergraphs with edge size bounded by $L$.  This represents a relaxation of our problem where elements are always active (i.e., can always be selected if feasible), but the goal is still to accept every element w.p.~$\alpha$ times its "active probability" $x_j$.  The authors show for this relaxed problem that the tight guarantee $\alpha$ is $\frac1{L-1+1/L}$ for general hypergraphs, and $1/(L-1)$ for $L$-partite hypergraphs.  Since $(1-\frac1{(1+L)^{1+L}})/L < \frac1{L-1+1/L}$, our upper bound of $(1-\frac1{(1+L)^{1+L}})/L$ establishes the separation that their guarantee of $\frac1{L-1+1/L}$ for general hypergraphs cannot be attained even by an offline random-element CRS. 

\section{NRM, OCA and Reductions to OCRS Models}\label{sec:nrm_and_reduction_crs}

This section formalizes the models we study and explains how they reduce to variants of OCRS models. We begin with discrete-time and continuous-time formulations of NRM, then describe OCA. Afterward we introduce the theoretical framework of OCRS/RCRS, presenting both the standard model and our key extension, the random-element version, which is designed to handle the negative correlations inherent in "batched" arrivals. Finally, we provide a systematic reduction, mapping the different NRM and OCA problems to the appropriate OCRS models.

\subsection{NRM and OCA Formulations}

\subsubsection{NRM} \label{ssec:NRM}
We begin by formally defining the NRM problem. We start with the discrete-time version, and then define the continuous-time Poisson version.  Orthogonally, the arrival probabilities can be general, or homogeneous over time (equivalently, i.i.d.).

\begin{definition}[Discrete-time NRM Problem]\label{def:discrete_time_nrm} Let $M$ be a set of items, and $N$
be a set of \textit{product types}, where each $j \in N$ has
reward $r_j \ge 0$ and is associated with items $A_j \subseteq M$, where $|A_j| \le L$. For the \textit{discrete-time model}, we assume there are $T$ time periods and items $i\in M$ have positive integer starting inventories $k_i$. At time period $t\in [T]$, the seller must decide for each product whether to make it open for sale. After that, a random product $j\in N$ is requested following a known independent probability vector $(\lambda_{j,t})_{j\in N}$, and no product is requested at time $t$ with probability $1-\sum_{j\in N}\lambda_{j,t}$. If it is open for sale, then it gets sold; otherwise, nothing happens. A product $j$ can only be open for sell if all the items in $A_j$ have not been sold out. The objective is to maximize the expected total revenue from selling products over the finite time horizon.
\end{definition}

A typical approach for NRM solves a fluid linear program (LP) to prescribe how many units of each product should be sold over the horizon in expectation. Letting $x_j$ denote the expected number of units of product $j$ sold across all periods, the fluid LP can be written as
\begin{align}\label{eq:general_fluid_lp}
\textsf{LP-ORIG}: = \max\ &\sum_{j\in N} r_j x_{j} 
\\ \mathrm{s.t. } & \sum_{j:i\in A_j} x_{j} \le k_i   \quad \quad\forall i\in M, \label{eq:expected_selling}
\\ & 0\le x_{j} \le \sum_{t\in [T]}\lambda_{j,t} \quad\quad\forall j\in N. \nonumber
\end{align}
The objective $\sum_{j \in N} r_j x_j$ upper bounds the NRM benchmark, and any solution $(x_{j})_{j \in N}$ to this LP satisfies the constraint \eqref{eq:expected_selling} that the total number of item $i$ sold in expectation cannot exceed its initial inventory $k_i$. Diverging from many NRM papers (e.g., \citealt{gallego1997multiproduct,talluri1998analysis,adelman2007dynamic,reiman2008asymptotically,bumpensanti2020re,vera2021bayesian}), our goal in this paper is to provide a competitive ratio $\alpha$, where our algorithm achieves at least $\alpha$ against the optimal value of the fluid LP \eqref{eq:general_fluid_lp}.

For both our analysis and presentation, it is convenient to make the assumption that all items have only a single unit of inventory and the product indices are disjoint across time. That is, we hereafter assume $k_i=1$ for all $i\in M$, and that $N$ is partitioned as $\dot{\bigcup}_{t=1}^TN_t$ where only products $j\in N_t$ can be requested at time $t$ and we simplify $\lambda_{j,t}$ to $\lambda_j$. This assumption is without loss, as we prove a formal reduction\footnote{The same reduction also applies to the $L$-partite special case studied in \Cref{sec:Lpartite} and the resulting instance preserves the $L$-partite structure. } in \Cref{ssec:reduction_single_unit} by making duplicates of items and products.  Importantly, we can still solve the original LP before making the duplicates, and convert its feasible solution into a feasible solution for the following equivalent LP in which items and products have been duplicated:
\begin{align}\label{eq:fluid_lp}
\textsf{LP-SPLIT}:=\max\ & \sum_{j\in N_1\cup\dots \cup N_T} r_j x_{j} 
\\ \mathrm{s.t. } & \sum_{j:i\in A_j} x_{j} \le 1   \quad \quad\forall i\in M,
\label{eqn:feasInExpn}\\ & 0\le x_{j} \le \lambda_{j} \quad\quad\forall j\in N_1\cup \dots\cup N_T. \nonumber 
\end{align}

We now turn to the continuous-time setting, where we define the \textit{NRM problem with time-varying (respectively, time-homogeneous) Poisson arrivals}. In what follows, $\Pois(\lambda)$ denotes a Poisson distribution of parameter $\lambda \ge 0$
(which by convention is identically $0$ if $\lambda =0$).

\begin{definition}[Poisson-arrival NRM Problem] \label{def:poisson_nrm}
Let $M$ be a set of items and $N$ be a set of \textit{product types}. Each product $j \in N$ yields reward $r_j \ge 0$ and consumes a set of items $A_j \subseteq M$, where $|A_j| \le L$. For \textit{time-varying Poisson arrivals}, we assume that for each $j \in N$, copies of product $j$ arrive according to an independent Poisson point process on $[0,1]$ specified by the intensity function $\lambda_j :[0,1] \rightarrow [0,\infty)$. Concretely, we independently draw $\Lambda_j \sim \Pois(\int_{0}^{1} \lambda_j(y) dy)$ product copies for each $j \in N$. Then, conditional on $\Lambda_j$, we independently draw $\Lambda_j$ \textit{arrival times} from $[0,1]$ using
the probability density function $y \rightarrow \lambda_{j}(y)/ \int_{0}^{1} \lambda_{j}(y) dy$, say
$Y_{1,j}, \ldots , Y_{\Lambda_j,j}$ where $Y_{q,j}$ is the arrival time of
the $q^{th}$ product copy of type $j$.

We assume that the intensity functions are bounded. This implies that there exists some (arbitrarily large) constant $\lambda \ge 0$, such that $\sum_{j \in N} \lambda_{j}(y) \le \lambda$ for all $y \in [0,1]$.
For the special case of \textit{time-homogeneous} Poisson arrivals, we assume
that $\lambda_j$ is constant for all $j \in N$, in which case we can
take $\lambda = \sum_{j \in N} \lambda_j$.

As before, the decision is on which subset of products to keep open over time.  A product $j$ can only be kept open if it is feasible, i.e.\ if the entirety of items in $A_j$ are still available, and if $j$ has an arrival while it is open, then it gets sold---items in $A_j$ get consumed and revenue $r_j$ is collected.  The objective is to maximize the expected total revenue earned over the time horizon $[0,1]$.
\end{definition}

As in the discrete setting, we state the continuous-time model using unit inventories without loss and provide formal details in Appendix \ref{ssec:reduction_single_unit}. We then benchmark performance against the following fluid LP, whose optimal solution we denote by $\textsf{LP-POIS}$:
\begin{align}\label{eqn:poisson_LP}
\textsf{LP-POIS}:=\max\ & \sum_{j\in N} r_j x_j 
\\ \mathrm{s.t. } & \sum_{j:i\in A_j} x_j \le 1   \quad \quad\forall i\in M, \label{eqn:poisson_item_capacity}
\\ & 0\le x_j \le \int_{0}^1 \lambda_j(y) dy \quad\quad\forall j\in N. \label{eqn:poisson_product_arrival}
\end{align}
Here $x_j$ represents the expected number of products of type $j$ sold. Constraint \eqref{eqn:poisson_item_capacity} ensures that no item is sold more than once in expectation. Moreover,
\eqref{eqn:poisson_product_arrival} follows since $\int_{0}^1 \lambda_j(y) dy$ is the expected number of arrivals of product type $j$. 

\subsubsection{OCA}

We next describe the OCA model we study. Like for NRM, there will be four variants: general vs.\ single-minded, and orthogonally, adversarial vs.\ random order arrivals.

\begin{definition}[OCA]\label{def:oca}
    There is a set of items $M$, each with unit capacity, and a set of agents each with a valuation function over bundles of items. Specifically, let $[T]$ denote the set of agents. For each agent $t\in[T]$, we let $N_t$ denote their finite set of possible realized types. An agent $t$ if realizing to type $j\in N_t$ has a given valuation function 
    \[v_j : \{A\subseteq M : |A|\le L\} \rightarrow \mathbb{R}_{\ge 0}, \]
    which assigns a nonnegative value $v_j(A)$ to each bundle $A$ of at most $L$ items. It is assumed that every valuation function satisfies $v_j(\emptyset)=0$ and is monotone, i.e., if $A'\subseteq A$ and $|A|\le L$, then $v_j(A')\le v_j(A)$. Each agent $t$ is of type $j\in N_t$ with probability $\lambda_{j}$, which are known and type realizations are independent across agents. Again, we assume that products have been duplicated so that type sets $N_1,\ldots,N_T$ are mutually disjoint.

    Agents arrive sequentially and the realized valuation function of an agent $t$ is only revealed upon their arrival. At this point, a bundle of up to $L$ available items is allocated to them irrevocably. If agent $t$ of type $j$ receives bundle $A^t$, the contribution to welfare is $v_j(A^t)$. The objective is to maximize the expected welfare. 
\end{definition}
As in NRM, we can formulate a fluid LP benchmark that captures the best possible expected welfare achievable by only requiring capacity constraints to hold in expectation. Letting $x_j(A)$ denote the expected number of times that bundle $A$ is assigned to agent of type $j$, the fluid LP can be written as:
\begin{align}\label{eq:oca_fluid_lp}
\textsf{LP-OCA}:=\max\ & \sum_{j\in N_1\cup\dots \cup N_T}\sum_{A:|A|\le L} v_j(A) x_{j}(A) 
\\ \mathrm{s.t. } & \sum_{j\in N_1\cup\dots \cup N_T}\sum_{A:|A|\le L}\mathds{1}\{i\in A\} x_{j}(A) \le 1   \quad \quad\forall i\in M,
\label{eqn:oca_feasInExpn}\\ & 0\le \sum_{A:|A|\le L}x_{j}(A) \le \lambda_{j} \quad\quad\forall j\in N_1\cup \dots\cup N_T. \label{eqn:oca_arrival} 
\end{align}
Here, constraint \eqref{eqn:oca_feasInExpn} enforces the total expected number of times item $i$ is included in allocated bundles cannot exceed its capacity. The optimal value of \textsf{LP-OCA} upper bounds the expected welfare of any online or offline algorithm.  

In the \textit{single-minded} special case, each agent $t$ is only interested in a particular non-empty bundle $D_t$ with $|D_t|\le L$. The types $j\in N_t$ can then be defined by $r_j$ so that the valuation function $v_j(A)$ always has the structure $v_j(A)=r_j \mathds{1}\{A \supseteq D_t\}$ only defined for $|A|\le L$. In this case, the fluid LP can be simplified because it is always suboptimal to allocate nonempty bundles not equal to $D_t$: 
\begin{align*}
\max\ & \sum_{j\in N_1\cup\dots \cup N_T} r_j x_{j}(D_t) 
\\ \mathrm{s.t. } & \sum_{j\in N_1\cup\dots \cup N_T}\mathds{1}\{i\in D_t\} x_{j}(D_t) \le 1   \quad \quad\forall i\in M,\\ 
& 0\le x_{j}(D_t) \le \lambda_{j} \quad\quad\forall j\in N_1\cup \dots\cup N_T.
\end{align*}

For OCA, in either the general case or the single-minded special case, the agents could arrive in the adversarial\footnote{\label{ftnt:oblivious} Formally, we consider the \textit{oblivious} adversarial order where the worst-case arrival permutation is fixed before seeing the realization of any randomness in the algorithm or environment.} order that is worst for the algorithm, or in uniformly random order.

\subsection{Definition of Random-element OCRS and RCRS}

We now describe the OCRS framework that we use as a common abstraction for NRM and OCA. Taking the discrete-time NRM as an example, in order to develop an algorithm that achieves at least $\alpha$ against \textsf{LP-SPLIT}, a particularly clean target emerges: if we can devise an online policy which accepts each product $j$ with probability exactly $\alpha x_j$, for some constant $\alpha\in [0,1]$, irrespective of the identity of $j$, then by linearity of expectation the resulting policy achieves competitive ratio $\alpha$ against the fluid LP benchmark \textsf{LP-SPLIT}. An OCRS is designed to provide exactly this type of uniform guarantee, and we now describe the abstract setting in which they operate.
\begin{definition}[Standard OCRS and RCRS] \label{def:standard_ocrs}
The standard OCRS framework operates on a set of elements $N$ with a feasibility constraint. Here, elements correspond to products or bundles, and feasibility means that each item $i\in M$ is consumed at most once. Given a fractional solution $(x_j)_{j\in N}$, which satisfies the feasibility constraints in expectation (i.e., \eqref{eqn:feasInExpn} holds), the elements $j$ are presented in sequence, with each one being "active" independently w.p.~$x_j$. The OCRS must immediately and irrecovably decide whether to accept any active element $j$, where $j$ is feasible to accept if and only if $A_j$ does not intersect with $A_{j'}$ for any previously accepted product $j'$ (but the OCRS can optionally not accept $j$ even if it is active and feasible).
We call an \textit{$\alpha$-selectable standard OCRS} one that accepts every element $j$ w.p.~$\alpha x_j$. In the standard OCRS, the sequence of elements is fixed in advance by an oblivious adversary (see \Cref{ftnt:oblivious}). The corresponding standard RCRS is defined identically except that the elements are presented in a uniformly random order, and hence selectability can be higher. We formalize this by assuming that each $j \in N$ draws a uniform arrival time $Y_j \in [0,1]$ independently, and the elements are presented in increasing order of $(Y_j)_{j \in N}$.
\end{definition}

In the special case of discrete-time NRM problems where at most one product can be requested in each period, we can interpret each product as an element in the standard OCRS. An "active" element/product represents that a customer requests product $j$ and the fluid policy sells the product. However, the standard OCRS cannot be directly applied to the general discrete-time NRM problem, because the activeness of products is not independent. In each period, at most one product $j$ is requested, so observing $j$ imples that no other product $j'$ is requested in that period, creating negative correlation within the batch. This motivates an extended notion of OCRS that explicitly models per-period batching, which we interpret as a "random element" being chosen by each customer $t=1,\ldots,T$ (hereafter, we use the terms "product" and "element" interchangeably). We formalize this extended notion of OCRS for feasibility constraints defined by \textit{$L$-bounded products}, but the definition applies more generally to arbitrary feasibility constraints on $N$.

\begin{definition}[Random-element OCRS and RCRS]\label{def:random-ocrs}
A universe $N$ of elements is partitioned into disjoint batches $N_1,\ldots,N_T$. The OCRS is given a fractional solution $(x_j)_{j\in N}$, which is feasible in expectation (i.e., \eqref{eqn:feasInExpn} holds), and further satisfies $\sum_{j\in N_t}x_j\le 1$ for all $t$.  Sequentially over $t\in\{1,\ldots,T\}$, batch $t$ arrives, at which point at most one random element from $N_t$ is drawn to be active following probability vector $(x_j)_{j\in N_t}$, where no element is active w.p.~$1-\sum_{j\in N_t}x_j$.
The OCRS must immediately and irrevocably decide whether to accept any active element $j$, where $j$ is feasible to accept if and only if $A_j$ does not intersect with $A_{j'}$ for any previously accepted element $j'$.
We call an \textit{$\alpha$-selectable random-element OCRS} one that accepts every element $j$ w.p.~$\alpha x_j$, for any choice of element $N$, partitionings $N=N_1\cup\cdots\cup N_T$, and all fractional solutions $(x_j)_{j\in N}$ which are feasible in expectation and satisfy $\sum_{j\in N_t}x_j\le1$ for all $t$. Again we assume an oblivious adversary for OCRS. For random-element RCRS, the same random-order convention applies at the batch level: the batches $N_t$ arrive in a uniformly random order. Once again, we formalize this by assuming that each batch $N_t$ draws a uniform arrival time $Y_t \in [0,1]$ independently, and the batches are presented in increasing order of $(Y_t)_{t \in N}$.
\end{definition}

In the standard notion of OCRS, $\alpha$-selectability only requires the guarantee to hold for the trivial partitioning where $T=|N|$ and $|N_1|=\cdots=|N_T|=1$, and there is typically no index $t$. If $L=2$, then items and products can be interpreted as vertices and edges in a graph respectively, where a set of products is feasible to sell if and only if they form a matching in the graph.
This is a well-studied setting, with~\eqref{eqn:feasInExpn} being the matching polytope.\footnote{Technically our formulation is more general, by allowing for single-vertex products and parallel edges.}
In this setting, \citet{ezra2022prophet} have considered a notion of batched OCRS where the guarantee only has to hold under a specific partitioning ("vertex arrivals"), and shown that the guarantee $\alpha$ can improve. By contrast, our definition of random-element OCRS requires the guarantee to hold under \textit{any adversarial partitioning} of $N$, in which case we show that the guarantee $\alpha$ can actually get worse.

\subsection{Reductions to OCRS and RCRS} \label{sec:main_body_reductions}

We start with the reductions of discrete-time NRM, which are straight-forward given our definitions of random-element OCRS and RCRS.  The time-homogeneous case implies that arrivals occur in random order, and hence can be captured by RCRS.

\begin{theorem}[Reduction of Discrete-time NRM, proved in \Cref{pf:nrm_and_random_OCRS}]\label{thrm:nrm_and_random_OCRS}
Consider any instance of the discrete-time NRM problem and let \textsf{LP-SPLIT} denote the optimal value of its fluid LP relaxation. An $\alpha$-selectable random-element OCRS implies an online algorithm whose total expected reward is at least $\alpha \cdot \textsf{LP-SPLIT}$. If arrivals are time-homogeneous, i.e., $\lambda_{j,1}=\dots=\lambda_{j,T}$ for every product type $j$, an $\alpha$-selectable random-element RCRS implies an online algorithm whose expected rewards is at least $\alpha \cdot \textsf{LP-SPLIT}$.
\end{theorem}

The reductions for continuous-time NRM, and in particular why standard OCRS/RCRS suffice, require a careful argument that to our knowledge is new to this paper.
Intuitively, the negative correlation in different products $j$ being requested at the same time step $t$ disappears in the Poisson regime, where these time steps are infinitessimally small. However, there are some subtleties in the argument which we explain in detail in \Cref{sec:poisson_reduction_time_varying}.
We note that for reducing time-homogeneous arrivals to a standard RCRS, we require the RCRS to be ``inactive-oblivious''. Intuitively, an ``inactive-oblivious'' RCRS executes in a setting where it has less information than in \Cref{def:standard_ocrs}. It is presented only the \textit{active} elements in random-order, and thus does not learn the inactive elements until the end of the process. We defer the formal definition to \Cref{def:inactive-oblivious-rcrs}, and note that the standard RCRS we design in \Cref{ssec:rcrs_standard} (and all the standard RCRS we cite in this paper) are inactive-oblivious.

\begin{theorem}[Reduction of Poisson-arrival NRM Problem, proved in \Cref{sec:poisson_reduction_time_varying}]\label{thm:poisson_reduction}
Consider any instance of the Poisson-arrival NRM problem and let $\PLP$ denote the optimal value of its fluid LP relaxation. An $\alpha$-selectable standard OCRS implies an online algorithm whose total expected reward is at least $\alpha \cdot \PLP$. If arrivals are time-homogeneous, i.e., $\lambda_{j}(y)$ is constant over $y \in [0,1]$ for every product type $j$, an $\alpha$-selectable inactive-oblivious standard RCRS (see \Cref{def:inactive-oblivious-rcrs} in \Cref{sec:random-element_to_standard}) implies an online algorithm whose expected rewards is at least $\alpha \cdot \PLP$.
\end{theorem}

\Cref{thm:poisson_reduction} employs the following auxiliary \namecref{thm:random-element_to_standard}, which says that if the total probability mass in any single random-element batch is small, then the negative correlation between elements in that batch can be essentially ignored, and a standard OCRS suffices. Formally, it shows that for an \textit{arbitrary} random-element OCRS input $(x_j)_{j \in N}$ with $T$ batches $N_1, \ldots ,N_T$ that are  $\eps$-bounded,
i.e., $\max_{t \in [T]} \sum_{j \in N_t} x_j \le \eps$,
a standard OCRS can be used to design a random-element OCRS, with loss in the selection guarantee dependent on $\eps$ and $T$.

\begin{theorem}[proved in \Cref{sec:random-element_to_standard}] \label{thm:random-element_to_standard}
An $\alpha$-selectable standard OCRS (resp. inactive-oblivious standard RCRS) implies a
random-element OCRS (resp. random-element RCRS) with a selection guarantee of
$$(\alpha -(1- e^{-\eps(1 + 2\eps T)}))/(1+\eps),$$ on all inputs with at most $T$ batches
that are $\eps$-bounded, where $\eps \le 1/2$. 
\end{theorem}

\Cref{thm:random-element_to_standard} is used in the reductions for poisson-arrival NRM.
Finally, the reductions of OCA are straight-forward given our delineation of random-element vs.\ standard OCRS/RCRS. For each agent $t$, each possible bundle $A$ can be viewed as an element, which is active w.p. $\sum_{j\in N_t}x_j(A)$. General valuations create a random-element input because at most one bundle can be allocated to an arriving agent; in the single-minded special case, the same reduction collapses to the standard model.

\begin{theorem}[Reduction of OCA, proved in Appendix \ref{pf:oca_reduction}] \label{thm:oca_reduction}
An $\alpha$-selectable random-element OCRS (resp. random-element RCRS) implies an online algorithm for the OCA input with general valuation functions, whose expected reward is at least $\alpha \cdot \textsf{LP-OCA}$. In the single-minded special case, standard OCRS (resp. standard RCRS) suffices.
\end{theorem}

\section{$1/(1+L)$ for Random-element OCRS}\label{sec:ocrs}
In this section, we establish the fundamental limits for the random-element OCRS. We first present a policy that achieves a competitive ratio of $1/(1+L)$. We then prove that this ratio is tight for the general random-element model, specifically when $L$ is a prime power, by constructing an adversarial instance based on finite affine planes. Finally, this benchmark allows us to demonstrate a separation by showing how this $1/(1+L)$ barrier can be strictly beaten under specific models, such as standard OCRS.

\noindent \textbf{General Notation and Terminology.} For a element $j\in N$, let $X_j\in\{0,1\}$ be the indicator random variable for $j$ being active, and $Z_j$ be the event that $j$ is accepted.
Let $F_j$ be the event that $j$ is feasible to accept at the start of time $t$, where $j\in N_t$.
We can write $F_j=\cap_{i\in A_j}F_i(t)$, where $F_i(t)$ is the event that item $i\in M$ is available (i.e., not sold) at the start of time $t$.

\subsection{A Simple Random-element OCRS that is $1/(1+L)$-selectable} \label{ssec:ocrs_warmup}
We now present a simple random-element OCRS $\pi$ that is $\alpha$-selectable for $\alpha = 1/(1+L)$. Our random-element OCRS is based on the idea of exact selection, first used by \citet{ezra2022prophet} for standard OCRS on graphs. To get our $\alpha = 1/(1+L)$-selectable random-element OCRS, we extend the idea of exact selection to arbitrary batches and values of $L$. The idea is to describe the random-element OCRS recursively in terms of the $T$ batches: assuming each element $j' \in N_{t'}$ is selected w.p. $\alpha x_{j'}$ for all $t' < t$, we extend this guarantee to batch $N_t$. Specifically, we wish to design $\pi$ in a way such that for each $t \in [T]$, 
\begin{equation} \label{eqn:induction_hypothesis}
    \mb{P}(Z_j\mid X_j =1) = \alpha, \forall j \in N_t.
\end{equation}
To satisfy this, an active element $j \in N_t$ must be accepted w.p.~$\alpha/\mb{P}(F_j)$, so the crux of the analysis is arguing that this probability is well-defined, i.e., $\alpha \le \mb{P}(F_j)$. Our proof for $1/(1+L)$ guarantee applies a simple union bound over the $L$ items of element $j$, which combined with the feasibility constraint \eqref{eqn:feasInExpn} yields the desired inequality.

We now define $\pi$ recursively in terms of $t \in [T]$. 
\begin{definition} \label{def:ocrs}
For  $t=1$, $\pi$ accepts an active element of $N_1$ (if any) independently w.p. $\alpha$. For
$t > 1$, assume that $\pi$ is defined up until period $t-1$. We extend the definition of $\pi$ to period $t$ in the following way:
If $j \in N_t$ is active and feasible, then $\pi$ accepts $j$ independently w.p. $\min\{1,\alpha/\mb{P}(F_j)\}$.
\end{definition}

We now formalize the result that the policy $\pi$ is well-defined and achieves the desired uniform selection guarantee.
\begin{theorem}\label{thrm:1/1+L_guarantee}
If $\alpha=1/(1+L)$, then $\pi$ is an $\alpha$-selectable random-element OCRS.
\end{theorem}
\begin{proof}[Proof of Theorem \ref{thrm:1/1+L_guarantee}]
It suffices to verify \eqref{eqn:induction_hypothesis} inductively. The base case of $t=1$ clearly
holds, so take $t > 1$, 
and assume that \eqref{eqn:induction_hypothesis} holds for each $t' < t$. We verify \eqref{eqn:induction_hypothesis} holds for $t$.

Fix an arbitrary $j \in N_t$. Observe that due to \Cref{def:ocrs}, conditional on $X_j =1$, $j$ is accepted w.p.
$\mb{P}(F_j) \cdot \min\{1,\frac{\alpha}{\mb{P}(F_j)}\}$.
Thus, in order to complete the inductive step, we must argue that $\alpha \le \mb{P}(F_j)$.
Since $\alpha = 1/(1+L)$, it suffices to show that $\mb{P}(F_j) \ge 1 - \alpha L$.
Now, 
    \begin{equation}\label{eq:union_bound_standard}
        \mb{P}(F_j) =\bP\left(\cap_{i\in A_j} F_{i}(t)\right)=1-\bP\left(\cup_{i\in A_j}\neg{F}_{i}(t) \right)\ge 1-\sum_{i\in A_j}\bP\left(\neg{F}_{i}(t)\right),
    \end{equation}
where $\neg{F}_{i}(t)$ is the complement of $F_{i}(t)$ representing the probability that item $i$ is already consumed before time $t$ and the final inequality uses a union bound. But, $\neg{F}_{i}(t)$ occurs if and only if any element $j'$ using item $i$ was accepted at some previous time $t'$, i.e., there exists some $t' < t$, $j' \in N_{t'}$ with $i \in A_{j'}$ for which $Z_{j'}$ occurs. Yet by the induction hypothesis \eqref{eqn:induction_hypothesis},

    \[\mathbb{P}(\neg{F}_{i}(t)) = \sum_{t' < t} \sum_{j' \in N_{t'}:i\in A_{j'}}\mathbb{P}(\Z_{j'}) = \alpha\sum_{t' < t} \sum_{j' \in N_{t'}:i\in A_{j'}} x_j' \le \alpha,\]
where the inequality follows from constraint \eqref{eqn:feasInExpn}. Thus,
    \[\bP(F_{j})\ge 1-\sum_{i\in A_j}\bP\left(\neg{F}_{i}(t)\right)\ge 1-\alpha\sum_{i\in A_j}\sum_{j':i\in A_{j'}}x_{j'}\ge 1-\alpha |A_j|\ge 1-L\alpha,\] 
and so the proof is complete.
\end{proof}

The union bound \eqref{eq:union_bound_standard} is the key step revisited in the rest of this section. The upper bound in Section \ref{ssec:tightness} shows that, for general random-element OCRS, the adversary can make this union bound essentially tight. In contrast, Section \ref{ssec:beating_standard} shows that under additional structure, such as standard OCRS, one can improve this step by keeping a positive inclusion-exclusion from overlaps among the item-unavailability events. 

Since the event $F_j$ depends on the decisions of $\pi$ strictly before period $t$, $\pi$ is well-defined. Implementing this OCRS requires computing the exact value of $\bP(F_{j})$, which is computationally challenging. However, it can be estimated via Monte Carlo simulation. In \Cref{ssec:implement_ocrs}, we discuss the complexity of implementing the OCRS and provide the number of samples needed in order to achieve a given error tolerance.

\subsection{Upper Bound of $1/(1+L)$ for Random-element OCRS} \label{ssec:tightness}
Having established that a selection guarantee of $1/(1+L)$ is achievable, we now prove that this ratio is fundamentally tight for the general random-element OCRS.\footnote{In \Cref{sec:negOffline} we establish an upper bound of $(1-\frac{1}{(1+L)^{1+L}})/L$ for random-element offline CRS using the same construction.} The proof strategy is to construct an adversarial input that exploits the negative correlation inherent in batched arrivals, forcing the feasibility probability $\bP(F_{j})$ to become arbitrarily close to the lower bound established by the simple union bound \eqref{eq:union_bound_standard}. 

To provide more intuition for the unbeatability of $1/(1+L)$, we first provide an explicit counterexample for $L=2$ to illustrate how the adversarial design forces the union bound inequality to become an equality, thereby proving $1/3$ cannot be surpassed under random-element OCRS. In this illustrative example, there are $3$ periods and $4$ items: $\{1,2,3,4\}$. The figure below represents the possible elements in each period, where each edge denotes one element: 
\begin{center}
\begin{tikzpicture}[scale=1]
    \node[circle, draw, fill=white, inner sep=0pt, minimum size=5mm] (1) at (0,0) {3};
    \node[circle, draw, fill=white, inner sep=0pt, minimum size=5mm] (2) at (1,0) {4};
    \node[circle, draw, fill=white, inner sep=0pt, minimum size=5mm] (3) at (0,1) {1};
    \node[circle, draw, fill=white, inner sep=0pt, minimum size=5mm] (4) at (1,1) {2};
    \draw (1) -- (2);
    \draw (3) -- (4);
    \node at (0.5,-0.5) {period 1};
    \node[circle, draw, fill=white, inner sep=0pt, minimum size=5mm] (1) at (3,0) {3};
    \node[circle, draw, fill=white, inner sep=0pt, minimum size=5mm] (2) at (4,0) {4};
    \node[circle, draw, fill=white, inner sep=0pt, minimum size=5mm] (3) at (3,1) {1};
    \node[circle, draw, fill=white, inner sep=0pt, minimum size=5mm] (4) at (4,1) {2};
    \draw (1) -- (3);
    \draw (2) -- (4);
    \node at (3.5,-0.5) {period 2};
    \node[circle, draw, fill=white, inner sep=0pt, minimum size=5mm] (1) at (6,0) {3};
    \node[circle, draw, fill=white, inner sep=0pt, minimum size=5mm] (2) at (7,0) {4};
    \node[circle, draw, fill=white, inner sep=0pt, minimum size=5mm] (3) at (6,1) {1};
    \node[circle, draw, fill=white, inner sep=0pt, minimum size=5mm] (4) at (7,1) {2};
    \draw (1) -- (4);
    \draw (2) -- (3);
    \node at (6.5,-0.5) {period 3};
\end{tikzpicture}
\end{center}
For example, in the first period, there are two possible elements: $(1,2)$ and $(3,4)$. If elements are labeled by the items contained (i.e.\ the two endpoints of the edge), then this construction amounts to $N_1=\{(1,2),(3,4)\}, N_2=\{(1,3),(2,4)\}$, and $N_3=\{(1,4),(2,3)\}$.

Additionally, the active probability of an element within the first two periods is $(1-\varepsilon)/2$ and the active probability of an element in the final period is $\varepsilon$, ensuring the constraint \eqref{eqn:feasInExpn} is satisfied. Formally, we have $x_{(1,2)}=x_{(3,4)}=x_{(1,3)}=x_{(2,4)}=(1-\varepsilon)/2$ and $x_{(1,4)}=x_{(2,3)}=\varepsilon$. We now explain why $1/3$ is unbeatable in this example. Note that for the element $(1,4)$, the probability that this element is feasible is calculated as follows: 
\begin{align*}
    &\mathbb{P}\left(\text{both items $1$ and $4$ are available}\right)\\
    =&1-\bP(\text{$1$ is used})-\bP(\text{$4$ is used})+\bP\left(\text{both $1$ and $4$ are used}\right)\\
    =&1-\alpha \left( x_{(1,2)}+x_{(1,3)}+x_{(3,4)}+x_{(2,4)}\right)+\bP\left(\text{both $1$ and $4$ are used}\right)\\
    =&1-2\alpha(1-\varepsilon)+\bP\left(\text{both $1$ and $4$ are used}\right),
\end{align*}
where the second equality holds because the probability that an element $j$ is accepted is $\alpha x_j$ under OCRS. Moreover, $\bP\left(\text{both $1$ and $4$ are used}\right)=0$ because it is not possible for two distinct edges to be selected before period 3---the non-conflicting edges are in the same batch and hence cannot both be active. Therefore, for the OCRS to remain valid, it must hold that $1-2\alpha (1-\varepsilon)\ge \alpha$ for any $\varepsilon>0$, which implies $\alpha\le 1/3$.

Expanding this intuition to general $L$, we find that as long as there exists a finite affine plane of order $L$, we can make a similarly adversarial construction where the union bound is tight and $1/(1+L)$ is unbeatable.
The construction here with $L=2$ is a special case of a finite affine plane of order 2, with 3 parallel classes of 2 lines each. 

\begin{figure}
    \centering
    \includegraphics[scale=0.25]{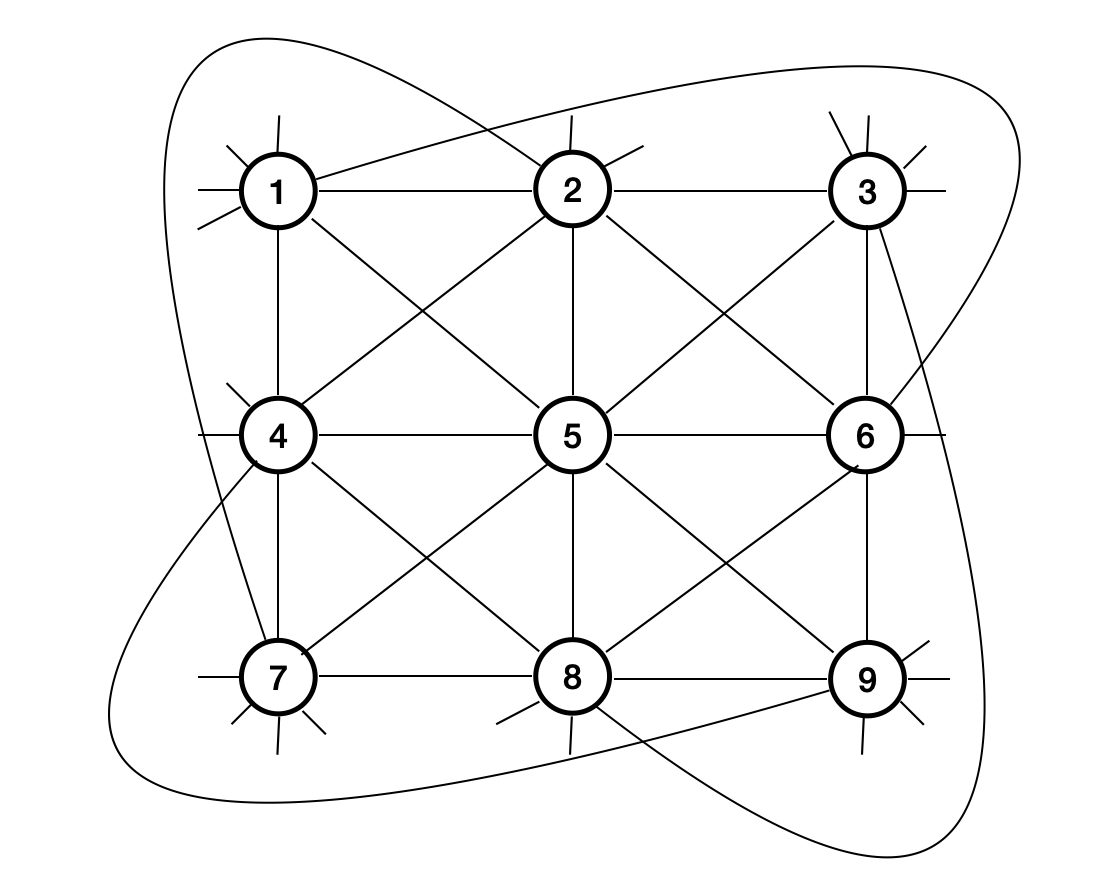}
    \caption{Finite affine plane with order $3$. The nine labeled circles are the points. Each drawn black straight segment or curved arc represents one line, i.e., one subset of three points. Curved arcs are used only to depict the wrap-around lines in the incidence structure and should not be interpreted as Euclidean geometry.}
    \label{fig:finite_affine_plane}
\end{figure}

\begin{definition}[Finite Affine Plane] \label{def:finiteAffPlane}
For our purpose, the definition below is purely combinatorial. A point is an abstract element of a finite ground set, and a line is a subset of points. Two lines intersect if the corresponding subsets have a common point, and lines in the same parallel class are called parallel because they are disjoint subsets of points. In a finite affine plane of order $L$, there are $L^2$ points and $L(1+L)$ distinct lines, each containing exactly $L$ points.
These lines can be grouped into $1+L$ classes of $L$ parallel lines each, where the lines within a class are mutually disjoint and collectively contain all $L^2$ points.
Finally, any two lines from two different classes intersect at exactly one point.
\end{definition}
We display the finite affine plane of order 3 in \Cref{fig:finite_affine_plane}.
Finite affine planes can be constructed from a finite field whenever $L$ is the power of a prime number, and we refer to \citet{moorhouse2007incidence} for further background. We are now ready to state our hardness result.

\begin{theorem}[proved in \Cref{ssec:proof_of_tightness}]\label{thrm:tightness}
No random-element OCRS is better than $1/(1+L)$-selectable when $L$ is a prime power. 
\end{theorem}
As in the standard OCRS setting of \cite{lee2018optimal}, the LP-duality equivalence between selectability and reward maximization against the ex-ante relaxation extends to the random-element model. Thus, Theorem \ref{thrm:tightness} can also be read as the same tight impossibility result for reward maximization against the ex-ante relaxation. To make the connection with revenue management explicit, we prove the upper bound by translating the finite affine plane construction into a discrete-time NRM instance and analyzing its fluid relaxation. Theorem \ref{thrm:tightness} then follows from the reduction in Theorem \ref{thrm:nrm_and_random_OCRS}.
\begin{definition}[NRM Configuration]\label{def:worst_case_problem}
Construct an item for each point in the affine plane, so that $|M|=L^2$.
Construct a product $j$ for each line, where $A_j$ consists of the items corresponding to the $L$ points in that line.
Construct a batch $N_t$ for each class of parallel lines, consisting of the products corresponding to the $L$ lines in that class.
In sum, we have $|N|=L(1+L)$, with $N$ being a disjoint union of the batches $N_t$ for $t=1,\ldots,1+L$.

By the properties in \Cref{def:finiteAffPlane}, this NRM configuration satisfies the following:
\begin{enumerate}
    \item[(i)] For each $t=1,\ldots,1+L$ and $j,j'\in N_t$, if $j \neq j'$, then $A_j\cap A_{j'}=\emptyset$;
    \item[(ii)] For each $1 \le t < t' \le 1+L$ and $j \in N_{t}$, $j' \in N_{t'}$ it holds that $|A_j \cap A_{j'}|=1$.
\end{enumerate} 
\end{definition}

\subsection{Beating $1/(1+L)$ for Standard OCRS}\label{ssec:beating_standard}
While Theorem \ref{thrm:tightness} shows the $1/(1+L)$ guarantee is tight for the general random-element OCRS, this barrier can be broken when the adversarial correlations are mitigated. In this section, we focus our main exposition in the standard OCRS model to illustrate the core analytical framework. In particular, we show that when elements arrive at most one at a time, i.e., $|N_t|=1$ for all $t$, the adversary can no longer construct the tight union-bound instance, and it is possible to achieve a strictly larger competitive ratio.

We develop a new framework aimed at enhancing the guarantee. To do this, we improve the crude union bound used in the proof of Theorem \ref{thrm:1/1+L_guarantee} by a more precise inclusion-exclusion argument. Specifically, we analyze the same recursively defined OCRS $\pi$ of Section \ref{ssec:ocrs_warmup}, yet with $\alpha > 1/(1+L)$. In order to simplify the indices later, let us assume that there are $T+1$ batches.
For each $1 \le t \le T+1$, we again define the induction hypothesis,
\begin{equation} \label{eqn:induction_hypothesis_repeat}
    \mb{P}(Z_j \mid X_j =1) = \alpha, \forall j \in N_t.
\end{equation}
Observe that when verifying \eqref{eqn:induction_hypothesis_repeat}, we can assume
without loss of generality that we are working with an element $\jc$ from the final batch $N_{T+1}$
for which $A_{\jc} = \{1, \ldots ,L\}$. 
Recalling \Cref{def:ocrs} which defines the OCRS, it suffices to argue that $\mb{P}(F_{\jc}) \ge \alpha$, where $F_{\jc}$ denotes the event that $\jc$ is feasible upon arrival. Now, because $F_{i}(T+1)$ is the event that item $i$ is available at period $T+1$, we can write
\begin{equation} \label{eqn:de_morgan_law}
\bP(F_{\jc})=\bP\left(\cap_{i =1}^{L} F_{i}(T+1) \right) = 1 - \mb{P}( \cup_{i=1}^{L} \neg F_{i}(T+1)). 
\end{equation}
In \Cref{thrm:1/1+L_guarantee}, we lower bounded \eqref{eqn:de_morgan_law} by applying a simple union bound
to $\mb{P}( \cup_{i =1}^{L} \neg{F}_{i}(T+1))$, which pessimistically treating the events $\neg F_i(T+1)$ as mutually exclusive. 

In order to improve on this, we first argue that with respect to minimizing \eqref{eqn:de_morgan_law},
or equivalently maximizing $\mb{P}( \cup_{i =1}^{L} \neg{F}_{i}(T+1))$, the worst-case input for $\pi$ occurs
when the constraints \eqref{eqn:feasInExpn} on the items $\{1, \ldots , L\}$ of $\jc$ are tight:
\begin{align} 
    \sum_{j: i\in A_{j}} x_{j}&=1, &\forall i\in \{1, \ldots , L\}.\label{eq:worst_capacity_bound}
\end{align}
To justify this assumption, observe that if \eqref{eq:worst_capacity_bound} does \textit{not} hold for items $M' \subseteq \{1, \ldots , L\}$,
then we can always consider an auxiliary input identical to the original one except with an additional element
for each item of $M'$, all of which arrive before time $T+1$. Due to the definition of $\pi$,
it is clear that adding these elements can only \textit{increase} $\mb{P}( \cup_{i=1}^{L} \neg{F}_{i}(T+1))$, and thus decrease \eqref{eqn:de_morgan_law}.
Finally, by a similar argument, the worst-case input for \eqref{eqn:de_morgan_law} occurs when $\sum_{j' \in N_{T+1}} x_{j'}$ is arbitrarily small. Thus, in the following computations we abuse notation slightly
and write that $\sum_{j' \in N_{T+1}} x_{j'} = 0$, with the understanding that we actually mean
$\sum_{j' \in N_{T+1}} x_{j'} \le \varepsilon$ for some arbitrarily small constant $\varepsilon > 0$.

The remainder of our framework can be summarized in the following three steps:
\begin{enumerate}
    \item[(i)] \textbf{Improve the union bound.} Using the inclusion-exclusion formula, we lower bound \eqref{eqn:de_morgan_law} and improve on the union bound
    by introducing an additional term that accounts for pairs of items not being available.
    This additional term can be further lower bounded by a sum over 
    $\mb{P}(Z_j \cap Z_{j'})$ for certain elements $j, j'$.

    \item[(ii)] \textbf{Lower-bound joint acceptance.} For any pair of elements $j$, $j'$ satisfying certain conditions, we show that $\mb{P}(Z_j \cap Z_{j'}) \ge C(\alpha, L) x_jx_{j'}$,
    where $C(\alpha, L)$ is some absolute constant, dependent only on $\alpha$ and $L$. We prove the inequality by reducing it to a concave optimization problem, in which the coefficient matrix for constraints is totally unimodular and thus the optimal solution can be explicitly characterized.
    \item[(iii)] \textbf{Solve the adversary's problem.} Combining steps (i) and (ii), the problem is reduced to lower bounding a sum over terms of the form $x_j x_{j'}$ (see \eqref{eqn:club_suit_proof}). This can then be reformulated as an optimization problem. For standard OCRS, the optimal value of the optimization problem must be strictly positive,
    which allows us to beat $1/(1+L)$.
\end{enumerate}
We begin with step (i) by establishing a sharpened bound for $\bP\left(\cap_{i=1}^L F_{i}(T+1)\right)$. We claim the following sequence of inequalities (with explanations following afterwards):
\begin{align*}
    &\bP\left(\cap_{i=1}^L F_{i}(T+1)\right)\ge 1- \bP\left(\cup_{i=1}^L \neg{F}_i(T+1)\right)\\
    \ge& 1-\sum_{i=1}^L \bP\left(\neg{F}_{i}(T+1)\right)+\max_{i}\sum_{i'\neq i}\bP\left(\neg{F}_{i}(T+1)\cap \neg{F}_{i'}(T+1)\right)\\
    \ge&1-\sum_{i=1}^L\bP\left(\neg{F}_{i}(T+1)\right)+\frac{1}{L}\sum_{i=1}^L\sum_{i'\neq i}\bP\left(\neg{F}_{i}(T+1)\cap \neg{F}_{i'}(T+1)\right)\\
    \ge&1-\alpha L+\frac{1}{L}\sum_{\substack{i,i' \in [L]: \\ i \neq i'}}\bP\left(\neg{F}_{i}(T+1)\cap \neg{F}_{i'}(T+1)\right)\\    
    \ge&1-\alpha L+\frac{1}{L}\sum_{\substack{i,i': \\ i \neq i'}}\left(\sum_{j:\{i,i'\}\subseteq A_j}\bP(\Z_j)+\sum_{\substack{t,t'\in[T]:\\ t\neq t'}}\sum_{\substack{j\in N_t,j'\in N_{t'}:\\A_j\cap A_{j'}=\emptyset\\ A_j\cap[L]=i,  A_{j'}\cap [L]=i'}}\bP\left(\Z_j \cap \Z_{j'}\right)\right)\\
    =&1-\alpha L+\frac{1}{L} \sum_{\substack{i,i': \\ i \neq i'}}\left(\sum_{j:\{i,i'\}\subseteq A_j}\alpha x_j+\sum_{\substack{t,t':\\ t\neq t'}}\sum_{\substack{j\in N_t,j'\in N_{t'}:\\A_j\cap A_{j'}=\emptyset\\ A_j\cap[L]=i,  A_{j'}\cap [L]=i'}}\bP\left(\Z_j \cap \Z_{j'}\right)\right).
\end{align*}
The second inequality follows by inclusion-exclusion. Specifically, fix any item $i$, every outcome in which item $i$ and another item $i'$ are both unavailable is counted once through $\neg{F}_i(T+1)$ and once again through $\neg{F}_{i'}(T+1)$. Since the overlap event $\neg{F}_i(T+1)\cap\neg{F}_{i'}(T+1)$ is already contained in $\neg{F}_i(T+1)$, this overlap should not be double counted. Subtracting all overlaps involving the fixed item $i$ is safe. Choosing item $i$ to maximize the summation of overlaps gives the second inequality. 
The third follows by an averaging argument, and the fourth by an application of the induction hypothesis \eqref{eqn:induction_hypothesis_repeat} in the same
way as done in the proof of \Cref{thrm:1/1+L_guarantee}. The fifth inequality holds by considering a subset of the events in which $\neg{F}_{i}(T+1)\cap\neg{F}_{i'}(T+1)$ holds, where either $i$ and $i'$ are taken by the same element or by two disjoint elements from different batches. Finally, the last equality applies the induction hypothesis \eqref{eqn:induction_hypothesis_repeat} again.

We now describe step (ii), where our goal is to lower bound 
$\bP\left(\Z_j \cap \Z_{j'}\right)$ for $j \in N_t$ and $j' \in N_{t'}$, with $t \neq t'$. Recall that when $\pi$ is presented a element $j$,
it draws a random bit, say  $B_j$,  which is $1$ independently w.p. $\alpha/\mb{P}(F_j)$ (note that indeed $\alpha/\mb{P}(F_j) \le 1$, due to the induction hypothesis \eqref{eqn:induction_hypothesis_repeat}). 
We say that $j$ \textit{survives} if $B_j X_j =1$. Otherwise, we say that $j$ \textit{dies}. Using this terminology, we describe
a sufficient condition in order for $Z_j \cap Z_j'$ to occur. Specifically, suppose that
each element $j'' \notin N_t \cup N_{t'}$ which shares an item with $j$ or $j''$ dies. Then, $Z_j \cap Z_{j'}$ occurs, provided both $j$ and $j'$ survive. Using independence, the joint
probability of these events is easily computed, and so we get that
\begin{align*}
    \bP\left(\Z_j\cap \Z_{j'}\right)
    \ge&\frac{\alpha x_{j}}{\bP\left(F_{j}\right)}\frac{\alpha x_{j'}}{\bP\left(F_{j'}\right)}\prod_{\tau \notin \{t,t'\}}\left(1-\sum_{j''\in N_\tau:A_{j''}\cap \left(A_{j}\cup A_{j'}\right)\neq\emptyset}\frac{\alpha x_{j''}}{\bP(F_{j''})}\right)\\
    \ge&\alpha^2 x_{j}x_{j'} \prod_{\tau \notin \{t,t'\}}\left(1-\sum_{j''\in N_{\tau}:A_{j''}\cap \left(A_{j}\cup A_{j'}\right)\neq\emptyset}\frac{\alpha x_{ j''}}{\bP(F_{ j''})}\right) \\
    \ge &\alpha^2 x_{j}x_{j'} \prod_{\tau =1}^{T}\left(1-\sum_{j''\in N_{\tau}:A_{j''}\cap \left(A_{j}\cup A_{j'}\right)\neq\emptyset}\frac{\alpha x_{ j''}}{\bP(F_{ j''})}\right)
\end{align*}
where the second inequality uses the trivial upper bound of $1$ on $\mb{P}(F_j)$ and $\mb{P}(F_{j'})$,
and the final inequality uses that each term in the element takes its value in $[0,1]$.
\begin{lemma}[proved in \Cref{pf:lem:pair_prob}]\label{lem:pair_prob}
   Suppose $\alpha \le 1-\alpha L +\alpha/(2L)$. Then, for any elements $j$ and $j'$ with $A_j\cap A_{j'}=\emptyset$, 
    \[\prod_{\tau=1}^T\left(1-\sum_{j''\in N_{\tau} :A_{j''}\cap \left(A_{j}\cup A_{j'}\right)\neq \emptyset}\frac{\alpha x_{j''}}{\bP(F_{j''})}\right) \ge  \left(\frac{1-\alpha(1+L)+\alpha/(2L)}{1-\alpha L+\alpha/(2L)}\right)^{2L}.\]
\end{lemma}
The proof of \Cref{lem:pair_prob} bounds each $\bP(F_{j''})$ using the various $x_{j''}$ and then converts the element term into an expression depending only on the $x_{j''}$. By analyzing an optimization problem which minimizes the element term via the $x_{j''}$, we can then characterize the optimal solution, which leads to the result above. By Lemma \ref{lem:pair_prob}, we can conclude that $\mb{P}(Z_j \cap Z_{j'}) \ge C(\alpha, L) x_jx_{j'}$, where $C(\alpha,L)$ is a constant characterized above. It says that for \textit{every pair of disjoint edges}, the OCRS of \citet{ezra2022prophet} has positive probability of accepting both of them. To the best of our knowledge, this fact was not apparent from \cite{ezra2022prophet} and \cite{macrury2022random}.

We now explain the last step (iii). By \Cref{lem:pair_prob}, in order to lower bound $\bP(\cap_{i=1}^{L} F_{i}(T+1))$, it remains to analyze
\[\sum_{\substack{i,i': \\ i \neq i'}}\left(\sum_{j:\{i,i'\}\subseteq A_j}\alpha x_j+\alpha^2\left(\frac{1-\alpha(1+L)+\alpha/(2L)}{1-\alpha L+\alpha/(2L)}\right)^{2L}\sum_{\substack{t,t':\\ t\neq t'}} \sum_{\substack{j\in N_t,j'\in N_{t'}:\\A_j\cap A_{j'}=\emptyset\\ A_j\cap [L]=i,  A_{j'}\cap [L]=i'}} x_{j}x_{j'}\right).\]
We claim that in the worst case, $x_j=0$ for any $j$ such that $|A_j\cap [L]|\ge 2$. To see this, note that 
\[\left(\frac{1-\alpha(1+L)+\alpha/(2L)}{1-\alpha L+\alpha/(2L)}\right)^{2L} \]
is decreasing in $\alpha$. Thus, since $\alpha\ge 1/(1+L)$, this is upper bounded by $1/(2L+1)^{2L}$. Therefore, in order to minimize the summand for $\{i,i'\}$, it is never optimal to set $x_j>0$ if $\{i,i'\}\subseteq A_j$. We can thus restrict our attention to the case where $|A_j\cap [L]|\le 1$ for every element $j$. That is, we analyze 
\begin{equation} \label{eqn:club_suit_proof}
\sum_{\substack{i,i': \\ i \neq i'}}\sum_{\substack{t,t':\\ t\neq t'}} \sum_{\substack{j\in N_t,j'\in N_{t'}:\\A_j\cap A_{j'}=\emptyset\\ i\in A_j,  i'\in A_{j'}}} x_{j}x_{j'}, 
\end{equation}
subject to the constraints $\sum_{j:i\in A_j}x_j=1$ for any $i\in [L]$ and $|A_j\cap [L]|\le 1$ for any element $j$.

In general, \eqref{eqn:club_suit_proof} can be as small as zero even with these two constraints satisfied (e.g., in our worst case configuration in \Cref{def:worst_case_problem}). However, under certain assumptions, it is possible to show $\eqref{eqn:club_suit_proof}>0$. In the standard OCRS problem, there exists at most one possible element in each time period, i.e., $|N_t|=1$ for all $t$. With such a restriction, it is not possible to choose the elements in such a way that $\eqref{eqn:club_suit_proof}=0$. In fact, we show in the following result that $\eqref{eqn:club_suit_proof}\ge L-1$.
\begin{lemma}[proved in \Cref{pf:lem:prophet_disjoint_mass}]\label{lem:prophet_disjoint_mass}
    Under standard OCRS, it holds that $\eqref{eqn:club_suit_proof}\ge L-1$.
\end{lemma}
We provide a brief sketch of the proof here. Using the fact that constraint \eqref{eq:worst_capacity_bound} is binding, and that every element intersects with $[L]$ in at most one item, we can rephrase \eqref{eqn:club_suit_proof} as an optimization problem maximizing 
\[\sum_{i=1}^L \sum_{i''\in M\backslash [L]} \left(\sum_{j:i,i''\in A_j}x_j\right)\left(\sum_{i'\neq i}\sum_{j':i',i''\in A_{j'}}x_{j'}\right). \]
The problem can be further rewritten as an optimization problem with a bilinear objective and linear constraints. Interestingly, we are able to characterize the optimal solution, which leads to the lemma. By combining \Cref{lem:prophet_disjoint_mass} with the derivation
preceding \eqref{eqn:club_suit_proof}, we get the following result:
\begin{theorem}\label{thrm:prophet_guarantee}
Given $L \ge 2$, suppose that $\pi$ of \Cref{def:ocrs} is passed $\alpha$  which satisfies
    \[\kappa(\alpha):=1-\alpha (1+L)+\alpha^2\frac{L-1}{L}\left(\frac{1-\alpha(1+L)+\alpha/(2L)}{1-\alpha L+\alpha/(2L)}\right)^{2L}\ge 0. \]
Then, $\pi$ is $\alpha$-selectable on standard OCRS inputs.
\end{theorem}
It can be verified that $\kappa(\alpha)\le 1-\alpha(1+L)+\alpha/ (2L)$ and so the assumption in \Cref{lem:pair_prob} is without loss. More, since the derivative of $\kappa$ is negative, the function $\kappa(\alpha)$ is monotonically decreasing in $\alpha$. Since $\kappa(1/(1+L))>0$, this implies that there exists $\alpha^*>1/(1+L)$ such that $\kappa(\alpha^*)=0$. Thus, $\pi$ is $\alpha^*$-selectable,
and so $1/(1+L)$ is beatable. For any given $L$, we can numerically find the value of $\alpha^*$. In particular, when $L=2$, we have $\alpha^*\approx 0.33336$. For large $L$, the explicit improvement certified by this analysis is positive but asymptotically very small. More precisely, the additive gain over $1/(1+L)$ decays roughly as $\exp\left(-\Theta(L\log L)\right)$. Thus, the main significance of the result is the strict separation between standard and random-element OCRS.  

We conclude by noting that this analytical framework extends beyond the standard OCRS model. As detailed in Appendix \ref{sec:Lpartite}, these techniques also allow us to improve upon the $1/(1+L)$ guarantee for random-element OCRS when the underlying constraints form an $L$-partite hypergraph.

\section{Positive Results for RCRS} \label{sec:rcrs}

Having established the limits of adversarial arrivals, we now turn to the RCRS. Before proving our various results for random-order arrivals, we state the arrival model via random \textit{arrival times}. Specifically, let us assume that there are $T+1$ batches, which we index from $t=0, \ldots , T$ (i.e., $N_0, N_1, \ldots ,N_{T}$). More,
assume that each batch $N_t$ has an independent and uniformly at random (u.a.r.) \textit{arrival time} $Y_t \in [0,1]$. The batches are then presented to the RCRS in increasing order of arrival times (we assume that the arrival times are distinct, as this occurs with probability $1$). For the special case of a \textit{standard RCRS}, the batches all each contain a single
element (i.e., $|N_1| = \ldots |N_T|=1$). In this case, there is a one-to-one correspondence between batches and elements, and so we define each $j \in N$ as having an arrival time $Y_j \in [0,1]$ drawn independently and u.a.r. 

Given an arbitrary random-element RCRS, we say that an item $i \in M$ is \textit{available at time} $y \in [0,1]$, provided $i \notin A_{j'}$ for each $j' \in N$ accepted by the random-element RCRS before time $y$.  We denote this event by $F_{i}(y)$, and also use $Z_{j}(y)$ to denote the event the random-element RCRS accepts $j$ by time $y$. Finally, we say that element $j$ is \textit{feasible at time} $y$, provided $\cap_{i \in A_j} F_{i}(y)$ occurs.

\subsection{Beating $1/(1+L)$ for Random-element RCRS} \label{sec:rcrs_random_element}
In this section, we show that a random-element RCRS can attain a guarantee greater
than $1/(1+L)$ for any $L \ge 2$. 

It is easy to see that in the random-order setting, a \textit{greedy} RCRS which accepts each active element whenever possible attains a guarantee of exactly $1/(1+L)$. 
To see this, let us assume that $\jc \in N_0$ and $A_{\jc} = \{1, \ldots ,L\}$. The greedy RCRS ensures that
\begin{equation} \label{eqn:greedy_analysis}
    \mb{P}(\text{$\jc$ is accepted} \mid Y_0 = y, \text{$\jc$ active}) \ge \prod_{t=1}^{T} (1 - \sum_{\substack{j \in N_t: \\ A_{j} \cap \{1, \ldots ,L\} \neq \emptyset}} x_j y ) \ge (1 -y)^{L}.
\end{equation}
To see this, conditional on $Y_0=y$ and on $\jc$ being active, a conflicting element from batch $N_t$ can block $\jc$ only if the batch arrives before time $y$ and the realized active element in that batch intersects $A_{\jc}$. This happens with probability at most $y\sum_{j\in N_t:A_j\cap A_{\jc}\neq\emptyset}x_j$. The batches are independent, giving the product term. For the second inequality, since 
\[\sum_{t=1}^T\sum_{j \in N_t: A_{j} \cap A_{\jc} \neq \emptyset} x_j\le \sum_{j:A_j\cap A_{\jc} \neq \emptyset}x_j\le \sum_{i\in A_{\jc}}\sum_{j:i\in A_j}x_j \le L,\]
we have
\[ \prod_{t=1}^{T} (1 - \sum_{\substack{j \in N_t: \\ A_{j} \cap \{1, \ldots ,L\} \neq \emptyset}} x_j y )\ge (1-y)^{\sum\limits_{t=1}^T\sum\limits_{j \in N_t: A_{j} \cap A_{\jc} \neq \emptyset} x_j }\ge (1-y)^{L}.\] After integrating the inequality \eqref{eqn:greedy_analysis}, this implies a guarantee of $\int_{0}^{1} (1 -y)^{L} dy = 1/(1+L)$, and the analysis is tight.
Thus, to beat $1/(1+L)$, we need to improve upon this simple strategy. For standard RCRS in the graph matching case (i.e.,  $L=2$), previous works \citep{brubach2021improved,pollner2022improved,macrury2022random} apply \textit{edge-based attenuation}. These techniques easily generalize to items and elements when $L \ge 3$, so we discuss them in this context. Ahead of time, one chooses an \textit{attenuation function} $b: [0,1] \rightarrow [0,1]$.
Then, when an element $j$ arrives and is active, a random bit $B_j$ with parameter $b(x_j)$ is drawn independently. If $B_j =1$,
and the items of $j$ are available, then $j$ is accepted. By ensuring $b$ satisfies certain analytic properties, one can characterize the worst-case probability that $j$ is accepted. Concretely, if $b(x):= \frac{(L-x) (1 - e^{-L})}{L ( 1- e^{-(L-x)})}$, then this worst-case probability is attained in the \textit{Poisson regime} (i.e., $\max_{j \in N} x_j \le \varepsilon$ for $\varepsilon \rightarrow 0$), and so this RCRS is easily seen to be $(1 - e^{-L})/L$-selectable (here $(1 - e^{-L})/L > 1/(1+L)$ for all $L \ge 1$).

The challenge with extending this approach to \textit{random-element} RCRS is that due to the negative correlation induced by the batches, it no longer suffices to just consider the $x_j$ value of an element $j$. In order to see this, consider an input with $T=1+L$,
where $\jc$ is the only element of $N_0$, and $x_{\jc} \approx 0$. We can then choose batch $N_t$ such that each $j \in N_t$ includes precisely item $t \in \{1, \ldots ,L\}$, and $\sum_{j \in N_t} x_j \approx 1$.
In this case, \eqref{eqn:greedy_analysis} is now
\begin{equation} \label{eqn:first_try}
    \mb{P}(\text{$\jc$ is accepted} \mid Y_0 = y, \text{$\jc$ active}) \ge  b(x_{\jc}) \prod_{t=1}^{L} (1 - \sum_{j \in N_t} x_j b(x_j) y).
\end{equation}
If we then set $\max_{j \in N_t} x_j \approx 0$ for each $t =1, \ldots ,L$, then no attenuation occurs, and the right-hand side is roughly $b(0) (1-y b(0))^{L}$. Thus, we do not beat $1/(1+L)$, no matter the choice of $b(0)$. 

At the opposite extreme, one could consider dropping an element based on the fractional value of its batch,
i.e., with probability $b(x(N_t))$ where $x(N_t):=\sum_{j \in N_t} x_j$. However, this fails to beat $1/(1+L)$ for similar reasons. If the batch of $\jc$
has $x(N_0) \approx 1$, yet none of its elements intersect with the items of $\jc$, then
$b(x(N_0))$ will be small, and so $\jc$ will be dropped/attenuated too aggressively. 

We propose an attenuation based framework which lies in-between these two extremes.
Specifically, for $t=0, \ldots , T$ and element $j \in N_t$, we define $x_{j,t}$
to be the fractional value of the elements of $N_t$ which  intersect with $A_{j}$. Formally, 
$x_{j,t} := \sum_{j' \in N_t: A_{j'} \cap A_{j} \neq \emptyset} x_{j'}$, where we note that the sum includes $x_j$.
Then, we define $B_j$ to have parameter $b(x_{j,t})$. Observe
that in the construction of \eqref{eqn:first_try}, all the elements $j \in N_t$
include an item of $\jc$, and so $x_{j,t} =1$. Thus, any element is dropped
with probability $b(1)$, and so we avoid the worst-case of \eqref{eqn:first_try}, provided $b(1) < 1$.

The actual analysis of our RCRS involves establishing a slightly different worst-case input via Schur-convexity theory \citep{peajcariaac1992convex}.
While Schur-convexity theory has been used to prove positive results for prophet inequalities in \citet{correa2021prophet}, we are unaware of a prior application to contention resolution schemes.
By again setting $b(x):= \frac{(L-x) (1 - e^{-L})}{L ( 1- e^{-(L-x)})}$, we then show that
the performance of our RCRS on this worst-case input is greater than $1/(1+L)$.

Given an element $j \in N$ and a subset $S \subseteq N$, define
$\partial_{S}(j) := \{j' \in S: A_{j'} \cap A_{j} \neq \emptyset\}$ to be the elements
of $S$ incident to $j$, and $x(S) := \sum_{j' \in S} x_{j'}$ to be the sum of the fractional values of $S$. To simplify the resulting notation,
we define
\begin{equation}\label{eqn:x_star}
x_{j,t} := x( \partial_{N_t}(j)) = \sum_{j' \in \partial_{N_t}(j)} x_{j'},
\end{equation}
where $N_t$ is an arbitrary batch (it need not include $j$).
Let $b: [0,1] \rightarrow [0,1]$ be defined
as 
\begin{equation} \label{eqn:attenuation_function}
b(x):= \frac{(L-x) (1 - e^{-L})}{L ( 1- e^{-(L-x)})}.
\end{equation}
We refer to $b$ as an \textit{attenuation} function, and first observe that $b$ is decreasing on $[0,1]$.
Our RCRS is defined in the following way:

\begin{algorithm}[H] 
\caption{Attenuate Greedy RCRS}
\label{def:random_element_order}
\begin{algorithmic}[1] 
\Require items $M$, elements $N$, batches $(N_t)_{t=1}^T$, and $(x_j)_{j \in N}$ which satisfies \eqref{eqn:feasInExpn}.
\Ensure a subset of active elements which satisfy the feasibility constraints.
\For{arriving batch $N_t$ in increasing order of $Y_t$}
\If{$j \in N_t$ is active}
\State Draw $B_{j}$ from $\Ber \left(b(x_{j,t})\right)$ independently, where $x_{j,t}= \sum_{j'\in N_t: A_{j'}\cap A_j\neq \emptyset}x_{j'}$ as in \eqref{eqn:x_star}.
\If{$j$ is feasible and  $B_{j}=1$}
\State Accept $j$.
\EndIf
\EndIf
\EndFor
\State \Return the accepted elements.
\end{algorithmic}
\end{algorithm}

\begin{remark}
 For each $j \in N$, we say that $j$ \textit{survives},
 provided $B_j X_j =1$. Then, we can reinterpret \Cref{alg:recursive_rcrs} as processing the batches in random order, and
 accepting each surviving element (whenever possible).  
\end{remark}

\begin{theorem}[proved in \Cref{ssec:proof_random_element_pos}] \label{thm:random_element_positive}
Fix $L \ge 2$, and define $\alpha > 1/(1+ L)$ where
\begin{itemize}
    \item $\alpha:= \frac{3 - 6 e^{1/2} + e + 8 e^{3/2} + 21 e^2 + 14 e^{5/2} + 
 7 e^3}{16 e (1 + e^{1/2} + e)^2} \approx 0.397$ if $L=2$.
    
    \item  Otherwise, if $L \ge 3$ $$\alpha := \frac{(e^L - e^{1/L}) \left(1- \left(1 + \frac{(e^L -1) (L^2 -1)}{(e^{1/L} - e^L) L^2} \right)^L  \right)}{(e^L - e)(1+L)}.$$
\end{itemize}
Then, \Cref{def:random_element_order} is an $\alpha$-selectable random-element RCRS.
\end{theorem}

In order to prove \Cref{thm:random_element_positive}, fix element $\jc \in N$. It will be convenient to index
from $0$, and assume that the batches are $N_0, N_1, \ldots, N_T$ for $T \ge L$. We can then assume that $\jc \in N_0$ and $A_{\jc} = \{1, \ldots ,L\}$ w.l.o.g. Let us say that a batch $N_{t}$ with $t \ge 1$ is \textit{dangerous} for $\jc$, provided 
\begin{enumerate}
    \item $Y_{t} < Y_{0}$.
    \item $B_{j} X_{j} = 1$ for some $j \in \partial_{N_t}(\jc)$.
\end{enumerate}
Otherwise, we say that $N_{t}$ is \textit{safe} for $\jc$.
Observe then that if $\jc$ survives, and all the batches $(N_{t})_{t =1}^T$ are safe for $\jc$, then $\jc$ will be accepted by \Cref{def:random_element_order}. Thus,
\begin{align*}
    \mb{P}(Z_{\jc}(y) \mid Y_0 = y, X_{\jc} = 1) & \ge \mb{P}(B_{\jc} =1 \cap \bigcap_{t= 1}^T \text{$N_{t}$ is safe for $\jc$} \mid Y_0 = y) \\
    &= b(x_{0,\jc}) \prod_{t=1}^T \mb{P}(\text{$N_{t}$ is safe for $\jc$} \mid Y_0 =y) \\
    &= b(x_{0,\jc}) \prod_{t=1}^T\left(1- \mb{P}(\text{$N_{t}$ is dangerous for $\jc$} \mid Y_0 =y)\right)
\end{align*}
where the first equality uses that conditional on $Y_0 = y$, the batches $(N_{t})_{t =1}^T$ 
are safe independently, and do not depend on $B_{\jc} =1$. Now, using the definition of
dangerous,
\begin{align}
    \mb{P}(\text{$N_{t}$ is dangerous for $\jc$} \mid Y_0 =y) &= \sum_{j \in \partial_{N_t}(\jc)} \mb{P}(B_{j} X_{j} = 1, Y_{t} < y) \nonumber \\
    &= \sum_{j \in \partial_{N_t}(\jc)} b(x_{j,t}) x_{j} y \label{eqn:dangerous_bound}.
\end{align}
Our goal is now to upper bound \eqref{eqn:dangerous_bound}. We do so by arguing that \eqref{eqn:dangerous_bound} is maximized when each $j \in \partial_{N_t}(\jc)$ includes exactly one item $i \in \{1,\ldots ,L\} = A_{\jc}$. More, for any $i \in \{1, \ldots ,L\}$, the fractional value of those $j \in \partial_{N_t}(\jc)$ which contain $i$ is exactly
$x(\partial_{N_t}(\jc))/L = x_{t,\jc}/L$. The following result formalizes this intuition.
\begin{lemma}[proved in \Cref{ssec:pf_batch_upper_bound}] \label{lem:batch_upper_bound}
For any batch $N_t$ with $t \ge 1$, 
$\sum_{j \in \partial_{N_t}(\jc)} b(x_{j,t}) x_{j} \le x_{t,\jc} b(x_{t,\jc}/L)$.
\end{lemma}

By applying \Cref{lem:batch_upper_bound} to \eqref{eqn:dangerous_bound}, and integrating over $y \in [0,1]$, 
\begin{align} 
    \mb{P}(Z_{\jc}(1) \mid X_{\jc} = 1) \ge b(x_{0,\jc}) \int_{0}^{1} \prod_{t =1}^{T}\left(1-  y x_{t,\jc} b(x_{t,\jc}/L)\right) dy. \label{eqn:accepted_lower_bound}
\end{align}
For convenience, denote $z_{t}:= x_{\jc,t}$ for each $0 \le t \le T$. We first view $z_{0}$ as fixed, and ignore the $b(z_{0})$ term of \eqref{eqn:accepted_lower_bound}. Our goal
is then to identify the minimum of
\begin{align}
     \int_{0}^{1} \prod_{t = 1}^{T}\left(1-  y z_{t} b(z_{t}/L)\right) dy \label{eqn:objective_minimize},
\end{align}
over all such inputs for which $z_{t} \le 1$ for all $t =1, \ldots ,T$, and $\sum_{t=1}^{T} z_{t} \le L - z_0$. In order to do so, we rephrase the problem slightly, and interpret \eqref{eqn:objective_minimize} as a function $\psi: [0,1]^{T} \rightarrow [0,1]$
of the vector $\bm{z} = (z_t)_{t=1}^T$. Our goal is then to minimize $\psi$
over all $\bm{z} \in [0,1]^{T}$ with $\sum_{t=1}^{T} z_t \le L - z_0$.

\begin{lemma} \label{lem:minimum_attained}
For any $z_0 \in [0,1]$ and $T \ge L$, the minimum of $\psi$ occurs at $\bm{z}^* \in [0,1]^{T}$, where $z^*_{t} = 1$ for $1 \le t \le L-1$, $z^{*}_{L} = 1 - z_0$, and $z^*_t = 0$ for $t > L$. Thus, for any $\bm{z}=(z_1, \ldots ,z_T) \in [0,1]^{T}$
with $\sum_{t=1}^{T} z_t \le L - z_0$,
$$
    \psi(\bm{z}) \ge \int_{0}^{1}\left(1- y(1- z_0)b\left(\frac{1- z_0}{L} \right) \right) (1 -yb(1/L))^{L-1}) dy.
$$
\end{lemma}
In order to prove \Cref{lem:minimum_attained}, first observe that w.l.o.g., we can restrict our attention to vectors $\bm{z} \in [0,1]^{T}$ with $\sum_{t=1}^{T} z_t = L - z_0$. It then suffices to argue that
$\psi$ is a \textit{Schur-concave} function (i.e., the negation of $\psi$ is Schur-convex), as the definition of Schur-concavity immediately implies that the minimum of $\psi$ occurs at $\bm{z}^*$ from \Cref{lem:minimum_attained}. (We refer the reader to \citealt{peajcariaac1992convex} for an overview of Schur convexity theory.)

To prove that $\psi$ is Schur-concave, we verify that $\psi$ satisfies the Schur-Ostrowski criterion \citep{peajcariaac1992convex}. The partial derivatives of $\psi$ all exist, and the function $\psi$ is permutation symmetric, so it suffices to check that
\begin{align} 
    (z_2 - z_1)( \partial_2 \psi(\bm{z}) - \partial_1 \psi(\bm{z})) &\le 0  &\forall \bm{z} =(z_1, \ldots ,z_T) \in [0,1]^{T}, \sum_{t=1}^T z_t = L -z_0. \label{eqn:schur_concave}
\end{align}
(Here $\partial_{i}( \bm{z})$ is the $i^{th}$ partial derivation of $\psi$ evaluated
at $\bm{z}$). We establish \eqref{eqn:schur_concave} in \Cref{pf:eqn:schur_concave},
which implies \Cref{lem:minimum_attained} by the above discussion. 

Combined with the previous discussion, \Cref{thm:random_element_positive} follows.

\subsection{An Improved Guarantee for Standard RCRS} \label{ssec:rcrs_standard}
In this section, we focus on the special setting of a standard RCRS, and prove a guarantee greater
than $(1- e^{-L})/L$ for $L \ge 2$, and greater than $(1-\frac1{(1+L)^{1+L}})/L$ for $L \ge 5$. We note that the RCRS we design is inactive-oblivious, as defined in \Cref{def:inactive-oblivious-rcrs}. This means that it can be used in \Cref{thm:poisson_reduction} for the NRM problem with homogeneous Poisson arrivals.

For standard RCRS, if no assumption is placed on the constraints \eqref{eqn:feasInExpn}, simple attenuation based techniques previously discussed
cannot beat $(1 -e^{-L})/L$. This limitation was originally observed for $L=2$ by \citet{pollner2022improved,macrury2022random}, though the same construction generalizes to $L \ge 3$. Instead, these papers beat $(1-e^{-2})/2$ by first
reducing\footnote{This reduction was first proposed by \citet{fu2021random} for contention resolution with random-order vertex arrivals.} to the case when the constraints \eqref{eqn:feasInExpn} are tight. After applying this reduction, they then make use of the graph structure of $L=2$ to get an improvement on $(1-e^{-2})/2$. While one can reduce to the setting when \eqref{eqn:feasInExpn} is tight for $L \ge 3$, this is no longer as helpful for the analysis. Even for $L=3$, one can construct an input which tightly satisfies \eqref{eqn:feasInExpn}, and for which an attenuation based analysis is not easily seen to do better than $(1 -e^{-L})/L$.
Thus, since $(1 -e^{-L})/L < (1-\frac1{(1+L)^{1+L}})/L$, we use a different approach to beat $(1-\frac1{(1+L)^{1+L}})/L$.

Our solution is to again use the idea of exaction selection as done in the OCRS setting. However, the probability an element is accepted will now depend on the arrival order. Specifically, let us assume that each element $j \in N$ has an arrival time $Y_j \in [0,1]$ drawn u.a.r.
and independently. For a carefully engineered \textit{selection function} $c:[0,1] \rightarrow [0,1]$, our goal is to ensure for each $z \in [0,1]$,
\begin{align}
\mb{P}( \text{$j$ is accepted}\mid Y_j =z, \text{$j$ is active}) = c(z)&   &  \forall j \in  N. \label{eqn:induction_hypo_cont}
\end{align}
This would then imply that the RCRS is $(\int_{0}^{1} c(z) dz)$-selectable. This approach was recently proposed by \citet{macrury2024random} for vertex arrivals and graph matchings, and we adapt it to the setting of items and elements for arbitrary $L \ge 2$. 

To implement this approach, we process the time horizon recursively. Fix $K \in \mb{N}$ to be a \textit{discretization constant}. 
For each $q=0, \ldots ,K$, define $y_q=q/K$.
We divide $(0,1]$ into $K$ intervals, $(y_{q-1},y_q]$ for each $1 \le q \le K$. 
Our RCRS is defined in $K$ \textit{phases}. In each phase $q$, it
processes elements $j \in N$ which arrive in the interval $(y_{q-1},y_q]$ (i.e., $Y_j \in (y_{q-1},y_q]$) and attempts to accept them with probability proportional to $c(y)$, adjusted by the estimated probability that the element is currently feasible, denoted by $\hat{F}_j(q)$.\footnote{By using Monte-Carlo sampling, one can approximate the $\hat{F}_{j}(q)$ probabilities, and get a poly-time RCRS whose guarantee is arbitrarily close to that of \Cref{alg:recursive_rcrs}. We omit the details, and refer the reader to \citet{macrury2024random} for how
this can be done.}    Our RCRS is defined
recursively with respect to the indices $q=0,1,\ldots,K-1$.  That is, assuming we've defined an RCRS for elements which arrive up until
the end of time $y_{q}$, we extend its definition to elements which arrive in the interval $(y_{q},y_{q+1}]$, thus defining an RCRS for
elements which arrive up until time $y_{q+1}$.

\begin{algorithm}[H]
\caption{Recursive Standard RCRS}
\label{alg:recursive_rcrs}
\begin{algorithmic}[1] 
\Require items $M$, elements $N$, $(x_j)_{j \in N}$ which satisfies \eqref{eqn:feasInExpn}, $K \in \mb{N}$, and selection function $c$.
\Ensure a subset of active elements which satisfy the feasibility constraints.
\For{$q=0, \ldots ,K-1$}
\If{$q=0$}
    \State Set $\hF_{j}(0):=1$ for each $j \in N$
\ElsIf{$q \ge 1$}
    \State Based on the definition of the RCRS up until time $y_q$, for each $j \in N$ compute $\hF_{j}(q):=\mb{P}(\cap_{i \in A_{j}} F_{i}(y_q) \mid Y_j > y_q )$.
\EndIf
\For{arriving elements $j$ with $Y_j \in (y_{q}, y_{q+1}]$}
\State Set $Y_j = y$ and draw $B_{j}$ from $\Ber \left(\min \left( \frac{c(y)}{\hF_{j}(q)}, 1\right) \right)$ independently.
\If{$j$ is feasible at time $y$, and  $B_{j} X_j=1$}
\State Accept $j$.
\EndIf

\EndFor
\EndFor

\State \Return the accepted elements.

\end{algorithmic}
\end{algorithm}
We now specify the choice of the selection function $c$ so that the standard RCRS is well defined and achieves the $(1-e^{-L})/L$ guarantee. 
\begin{theorem}[implied by Theorem \ref{thm:rcrs_standard}] \label{thm:rcrs_standard_result}
Given $L \ge 2$, suppose that $c:[0,1] \rightarrow [0,1]$ satisfies $c(y)= 1- L \int_{0}^{y} c(z) dz$. Then, if $K \ge 2L/e^{-L}$, \Cref{alg:recursive_rcrs} is $\left(1 - \frac{L}{K e^{-L}} \right) \frac{1-e^{-L}}{L}$-selectable on standard RCRS inputs.
\end{theorem}

For any $L \ge 2$, we can take $K \rightarrow \infty$, and ensure that \Cref{alg:recursive_rcrs} gets arbitrarily close to these guarantees. Although our RCRS and formal analysis are discrete, the intuition for Theorem \ref{thm:rcrs_standard_result} is best understood via ``continuous induction'' over the unit interval $[0,1]$ as in the literature \citep{kalantari2007induction}. Given a fixed $y \in [0,1]$, let us assume that we've defined an RCRS which satisfies \eqref{eqn:induction_hypo_cont} for all $z < y$. This should be thought of as a strong induction hypothesis. Our goal is
then to extend the definition of the RCRS to \textit{exactly} time $y$, and prove that \eqref{eqn:induction_hypo_cont} still holds. Towards this goal, imagine that
when $\jc \in N$ arrives at time $y$ and is active, we accept it with probability
\begin{equation} \label{eqn:accept_prob_cont}
c(y)/\mb{P}(\text{every item of $\jc$ is available at time $y$} \mid Y_{\jc} = y),
\end{equation}
provided its items are available. The crux of the analysis then involves showing that this probability is well-defined. Unlike for OCRS, we first must remove the conditioning on $Y_{\jc} = y$ to apply the induction hypothesis. This is the content of \Cref{prop:rcrs_coupling}, where we argue that if $Y_{\jc} = y$, then
this can only \textit{increase} the chance each item of $\jc$ is available at time $y$. That is,
$$
\mb{P}(\text{every item of $\jc$ is available at time $y$} \mid Y_{\jc} = y) \ge \mb{P}(\text{every item of $\jc$ is available at time $y$}).
$$
Here we crucially make use of the fact that we are designing a standard RCRS (for arbitrary batches, the analogous statement is false). Using a union bound argument combined with \eqref{eqn:induction_hypo_cont}, we then get that
\begin{equation} \label{eqn:sample_prob_cont}
\mb{P}(\text{every item of $\jc$ is available at time $y$}) \ge 1 - L \int_0^{y} c(z) dz.
\end{equation}
Thus, in order to upper bound \eqref{eqn:accept_prob_cont} by $1$, $c$ must satisfy $c(y) \le 1- L \int_{0}^{y} c(z) dz$.
At equality, the solution to this integral equation is $c(z) = e^{-L z}$, which satisfies $\int_{0}^{1} e^{-L z} dz = (1 -e^{-L})/L$.

Since we wish to beat $(1 -e^{-L})/L$, we improve on the union bound using a similar style analysis as done for OCRS (see \Cref{lem:pair_prob_random}). This analysis introduces a positive term corresponding to the joint availability of item pairs, leading to a more permissive constraint. Specifically, we refer to $c: [0,1] \rightarrow [0,1]$ as a \textit{selection function}, provided $c$ is decreasing on $[0,1]$, and $c(1) > 0$, and for each $y \in [0,1]$, we have that
\begin{equation} \label{eqn:selection_function}
c(y) \le 1 - L \int_{0}^{y} c(z) dz + \frac{(L-1)}{L}\left( \int_{0}^{y} c(z) (1-z)^L dz\right)^2.
\end{equation}
The solution to \eqref{eqn:selection_function} at equality strictly dominates $y \rightarrow e^{-L y}$,
allowing us to beat $(1 -e^{-L})/L$. More, for $L \ge 5$, the integral of this solution is greater than
$(1-\frac1{(1+L)^{1+L}})/L$, which implies the desired separation from random-element offline contention resolution.

\begin{proposition} \label{prop:integral_solution}
If $c$ is the solution to \eqref{eqn:selection_function} at equality, then $c$ is a selection
function. Moreover,  $\int_{0}^{1} c(y) dy > (1-e^{-L})/L$ for $L \ge 2$, and $\int_{0}^{1} c(y) dy > (1-\frac1{(1+L)^{1+L}})/L$ for $L \ge 5$.
\end{proposition}
When $c$ is taken to be the solution to \eqref{eqn:selection_function} at equality, there is not a closed-form
expression for $\int_{0}^{1} c(z) dz$. However, for any fixed value of $L \ge 2$, we can compute this integral numerically. For instance, if $L=2 $, $\int_{0}^{1} c(z) dz \ge 0.441$, and if $L =3$,
$\int_{0}^{1} c(z) dz \ge 0.321$. We now state our main result.

\begin{theorem}[proved in \Cref{ssec:pf_rcrs_standard}] \label{thm:rcrs_standard}
Given $L \ge 2$, suppose that $c:[0,1] \rightarrow [0,1]$ is
a selection function.
Then, if $K \ge 2L/c(1)$, \Cref{alg:recursive_rcrs} is $\left(1 - \frac{L}{K c(1)} \right) \left(\int_{0}^{1} c(y) dy\right)$-selectable on standard RCRS inputs.
\end{theorem}
 In order to prove \Cref{thm:rcrs_standard}, we define an induction hypothesis dependent on the phase $q \in \{0, \ldots ,K -1\}$: For each $j \in N$ and $y \in (y_{q-1},y_q]$,
\begin{equation} \label{eqn:rcrs_induction_hyp}
    c(y)\left(1 - \frac{L}{K c(1)} \right) \le \mb{P}(Z_{j}(y) =1 \mid Y_j = y, X_j =1) \le c(y), 
\end{equation}
The critical step is showing that if the hypothesis holds up to phase $k-1$, the probability that an element remains feasible at time $y_k$ is sufficiently high.
\begin{lemma}[proved in \Cref{ssec:pf_continuous_induction}] \label{lem:continuous_induction}
Fix $1 \le k \le K-1$. If \eqref{eqn:rcrs_induction_hyp} holds for all $0 \le q \le k -1$,
then $\mb{P}(\cap_{i \in A_{j}} F_{i}(y_k) \mid Y_{j} > y_k ) \ge c(y_k)$ for all $j \in N$.
\end{lemma}
To prove Lemma \ref{lem:continuous_induction}, we first argue that we can remove the conditioning on $Y_{j_0} > y_k$, and instead focus on lower bounding $\mb{P}(\cap_{i =1}^{L} F_{i}(y_k))$. As discussed in
Section \ref{ssec:rcrs_standard}, this step of the proof only works because we are in the standard RCRS setting. We then improve on the union bound using the same inclusion-exclusion analysis as done
for OCRS. As before, we make the tightness assumption \eqref{eq:worst_capacity_bound} for the $L$ focal items in $A_{\jc}=\{1,\dots,L\}$. In fact, we end up using the same term introduced in \eqref{eqn:club_suit_proof}, leading to
the integral inequality \eqref{eqn:selection_function}. Assuming Lemma \ref{lem:continuous_induction}, it is easy to complete the inductive step. 

\section{Conclusion and Open Questions}\label{sec:conclusion}

We recap the main contributions of this paper.  First, we beat the benchmark of $1/(1+L)$ that has appeared in many papers about Network Revenue Management or Online Combinatorial Auctions.  Also, we demonstrate that the subtlety of whether elements are random can affect the best-possible guarantees in OCRS.  Finally, we define an extended notion of "random-element" OCRS that is necessary to handle the general NRM and OCA problems in a black-box manner.

We end by posing a few open questions. First, we would like to determine the optimal guarantee attainable for random-element RCRS. The limits of our algorithmic techniques suggest that it may be
$(1-e^{-L})/L$, however neither our positive or negative results match this value. Since our upper bound of $(1-\frac1{(1+L)^{1+L}})/L$ applies even to an offline CRS, a natural way to improve on our construction would be to make use of the random-order arrivals.
Second, our analysis does not naturally lend itself to improved guarantees if all items have large initial inventories.  It may be interesting to interpolate between our guarantees and \citet{amil2022multi}, whose guarantees for NRM do improve with large inventories.  Finally, our counterexamples have the curious property of relying on a finite affine plane of order $L$.  Might it be possible to beat $1/7$ for random-element OCRS when $L=6$?

\paragraph{Acknowledgements.} The authors thank Huseyin Topaloglu for insightful early discussions. The authors also thank two anonymous referees, an anonymous Associate Editor, and the Area Editor (Ilan Lobel) for insightful comments and suggestions that helped to improve the paper.

\bibliographystyle{agsm}
\setstretch{0.9}
\bibliography{bib}

\newpage
\appendix

\section{Additions to \Cref{sec:nrm_and_reduction_crs}}
\subsection{Reduction to Unit Inventory}\label{ssec:reduction_single_unit}
\noindent\textbf{Discrete-time NRM.} To start with, we first label the initial products $j=1,\dots,N$ and items $i=1,\dots,M$, and relabel each unit of items, e.g., let $i_k$ denote the $k$-th unit for item $i$. We treat different units of the same item as "different" items so that all items have an initial inventory of $1$. Let $x_{j,t} = x_j \lambda_{j,t}/ \sum_{t'\in[T]}\lambda_{j,t'}$ for $t\in [T]$. Algorithm \ref{alg:transformation_single} describes the transformation in detail.

\begin{algorithm}

\caption{Reduction to Single Unit Case}\label{alg:transformation_single}

\begin{algorithmic}[1] 
\Require $(x_{j,t})_{t\in[T],j\in N}$, $N_t=\emptyset, \forall t$, $\ell(j)=1, \forall j$, $k(i)=1, \forall i$, $c_{i_{k}}=1, \forall i_k$.
    \For{$t=1,\dots,T$}
    \For{$j=1,\dots,N$} 
     \While{ $x_{j,t}> 0$}
     \State $\delta = \min\{x_{j,t}, \min_{i\in A_j}c_{i_{k(i)}}\}$
     \State $z_{j_{\ell(j)}} = \delta$, $x_{j,t}\leftarrow x_{j,t}-\delta$, $A_{j_{\ell(j)}}=\cup_{i\in A_j}\{k(i)\}$, $N_t\leftarrow N_t\cup \{j_{\ell(j)}\}$ and $\ell(j)\leftarrow \ell(j)+1$
    \State $c_{i_{k(i)}}\leftarrow c_{i_{k(i)}}- \delta$, $\forall i\in A_j$, If $c_{i_{k(i)}}=0$, $k(i)\leftarrow k(i)+1, \forall i\in A_j$
     \EndWhile
     \EndFor
     \EndFor
    \Ensure $N_t, \forall t, A_{j_{\ell}}, \forall j_\ell$ and $z_{j_{\ell}},  \forall j_{\ell}$.
\end{algorithmic}
\end{algorithm}

By Algorithm \ref{alg:transformation_single}, given any optimal solution to the fluid LP \eqref{eq:general_fluid_lp}, we consider a reduced NRM problem where all items have an initial inventory of $1$, and the products $N$ are partitioned into $N_1\cup\cdots\cup N_T$ where $\lambda_{j_{\ell}} = z_{j_{\ell}} \lambda_{j,t} / x_{j,t}$. By viewing $(z_{j_{\ell}})_{j_{\ell}}$ as the selling probability of product $j_{\ell}$ in the reduced probability, it follows immediately that $(z_{j_{\ell}})_{j_{\ell}}$ is feasible to the fluid LP of the reduced problem and achieves the same objective value as the fluid LP \eqref{eq:general_fluid_lp}. Therefore, the fluid relaxation of the reduced problem provides an upper bound to the original problem, and for any policy provides a constant approximation to this problem against the corresponding fluid LP, it provides same constant approximation to the original problem. Furthermore, note that a dummy product is created only if an unit is "overflowed". Because each item $i\in M$ can be consumed for at most one unit in one period and thus there can be at most one "overflow" for each item $i$. Hence, there are at most $M$ dummy products created for each period, which implies that the reduced fluid problem is still polynomial sized. \\ 

\noindent\textbf{Continuous-time Poisson arrivals.} Suppose item $i$ has initial inventory $k_i$, and arrivals of product type $j$ follow an independent Poisson process with intensity $\lambda_j(y)$, where $y\in [0,1]$. Then $\int_0^1 \lambda_j(y)dy$ is the expected number of arrivals of type $j$. Let $(x_j)_{j\in N}$ be an optimal fluid solution of \textsf{LP-POIS} with inventory $(k_i)_{i\in M}$. As in the discrete-time reduction, label the units of item $i$ by $i_1,\dots,i_k$ and treat them as different unit-inventory items. 
\begin{algorithm}

\caption{Continuous-time Reduction}
\begin{algorithmic}[1] 
\Require $(x_{j})_{j\in N}$, $\ell(j)=1, \forall j$, $k(i)=1, \forall i$, $c_{i_{k}}=1, \forall i_k$.
    \For{$j=1,\dots,N$} 
     \While{ $x_{j}> 0$}
     \State $\delta = \min\{x_{j}, \min_{i\in A_j}c_{i_{k(i)}}\}$
     \State $z_{j_{\ell(j)}} = \delta$, $x_{j}\leftarrow x_{j}-\delta$, $A_{j_{\ell(j)}}=\cup_{i\in A_j}\{k(i)\}$ and $\ell(j)\leftarrow \ell(j)+1$
    \State $c_{i_{k(i)}}\leftarrow c_{i_{k(i)}}- \delta$, $\forall i\in A_j$, If $c_{i_{k(i)}}=0$, $k(i)\leftarrow k(i)+1, \forall i\in A_j$
     \EndWhile
     \EndFor
    \Ensure $A_{j_{\ell}}, \forall j_\ell$ and $z_{j_{\ell}},  \forall j_{\ell}$.
\end{algorithmic}
\end{algorithm}

We then consider a reduced problem where all items have initial inventory of $1$, and each product $j_{\ell}$ follows an independent Poisson process with intensity $z_{j_{\ell}}\lambda_j(u)/x_j$. The expected number of arrivals is $z_{j_{\ell}}\int_0^1 \lambda_j(y)dy/x_{j}\ge z_{j_{\ell}}$. Therefore, $(z_{j_{\ell}})$ is a feasible solution to \textsf{LP-POIS} of the reduced problem and achieves the same objective value as the original one. Given any policy that provides a constant approximation to the reduced problem, in the original system, when a product type $j$ arrives, we independently sample a product $j_{\ell}$ with probability $z_{j_{\ell}}/x_j$ and then route it to the given one-unit policy. If $j_{\ell}$ is accepted, the one-unit policy consumes the copies in $A_{j_{\ell}}$, while the original system consumes the physical items in $A_{j}$ and rewards are same in the two systems. Therefore, any approximation guarantee for the reduced problem transfers to the original problem.  

\subsection{Proof of \Cref{thrm:nrm_and_random_OCRS}} \label{pf:nrm_and_random_OCRS}

Any feasible solution $(x_j)_{j\in N}$ to the reduced problem \textsf{LP-SPLIT}  satisfies the conditions of a random-element OCRS for $L$-bounded elements.  The OCRS, if $\alpha$-selectable, is able to accept every $j$ w.p.~$\alpha x_j$, while only accepting active elements and satisfying the item feasibility constraints.  This can be re-interpreted as follows.
For each $t$ and $j\in N_t$, let $X_j$ indicate whether $j$ is active, i.e.~$\bE[X_j]=x_j$ and $\sum_{j\in N_t}X_j\le 1$ w.p.~1.
For each $t$, based on its present state, the OCRS can \textit{pre-decide} whether to accept each element $j\in N_t$ if it were to be active, indicated by $B_j\in\{0,1\}$. Element $j$ is then accepted if and only if $B_jX_j=1$.
The OCRS guarantees that $\bE[B_jX_j]=\alpha x_j$, which equals $\bE[B_j]\bE[X_j]$ because $X_j$ is independent from everything else. We use these random bits $(B_j)_{j\in N_t}$ in the online algorithm. 

In addition, we must show that the state evolution in the actual problem is consistent with the state evolution expected by the OCRS.
We can define the following coupling: in the actual problem, for each $t$, product $j\in N_t$ is sold if and only if $X_jB_j=1$.  Conditional on any realization $(B_j)_{j\in N_t}$, we will indeed see that product $j$ is sold w.p.~0 if $B_j=0$, and w.p.~$x_j$ if $B_j=1$, correlated across $j$ so that at most one product is sold.  Moreover, the realization of which product (if any) is sold is independent from everything else, which is consistent with the desired state evolution in the actual problem.  Therefore, the state in the actual problem (where we cannot see whether elements are "active" before deciding accept/reject) can be coupled with the state in the OCRS, and hence the OCRS guarantee which implies $\bE[B_j]=\alpha$ for all $j$ can be applied.  Moreover, the OCRS guarantees that $B_j=0$ whenever $j$ is infeasible, leading to a valid algorithm in the actual problem that respects the inventory constraints.  This completes the proof.

For the time-homogeneous case, before the time horizon begins, we can independently draw a uniformly random permutation $\sigma$ of $[T]$. For each $j\in N_{\sigma(t)}$, let $X_j$ denote whether $j$ is active. Because the requests are independent and identically distributed, conditional on $\sigma$, these activations are independent and satisfy $\mathbb{E}[X_j]=x_j$ and $\sum_{j\in N_t}X_j\le 1$ w.p. $1$. Moreover, the batches $N_{\sigma(1)}, \dots, N_{\sigma(T)}$ are presented in the uniformly random order. Thus, this is a valid random-element RCRS input.    

\subsection{Proof of \Cref{thm:poisson_reduction}} \label{sec:poisson_reduction_time_varying}
We begin by explaining the technical challenges that motivated our approach for proving
\Cref{thm:poisson_reduction}. In this overview, we focus on time-varying Poisson arrivals and standard OCRS.

When there is a single copy of each item, a natural approach for reducing to standard OCRS is to first observe that
for each $j \in N$,
constraint \eqref{eqn:poisson_product_arrival} 
can be tightened by replacing its right-hand side with $1 - e^{-\int_{0}^1 \lambda_j(y) dy}$,
as this is the probability that a product of type $j$ arrives. After solving for an optimal solution $(x_j)_{j \in N}$ to this tightened LP,
one can then draw independent $\{0,1\}$-valued random variables $(B_j)_{j \in N}$
where $\mb{P}[B_j =1]:= x_j/(1 - e^{-\int_{0}^1 \lambda_j(y) dy})$. By defining
each $j \in N$ to be \textit{active} provided $B_{j} =1$ and a copy of $j$ arrives,
$$\mb{P}[\text{$j$ is active}] = \mb{P}[B_j =1 ] \mb{P}[\text{a copy of $j$ arrives}] =  \frac{x_j}{1 - e^{-\int_{0}^1 \lambda_j(y) dy}} (1 - e^{-\int_{0}^1 \lambda_j(y) dy}) = x_j.$$ 
Since $(x_j)_{j \in N}$ is feasible in expectation (i.e., \eqref{eqn:feasInExpn} holds) due to \eqref{eqn:poisson_item_capacity}, and each $j \in N$ is active independently,
this defines a valid standard OCRS input. Ideally, we would use an $\alpha$-selectable standard OCRS to design
an online algorithm with $\mb{P}[\text{a copy of $j$ is sold}] \ge \alpha x_j$ for each $j \in N$, which would imply the
online algorithm gets  an $\alpha$-fraction of the LP's value.
The reason this does not work is that our definition of $\alpha$-selectable is with respect 
to an arrival order chosen by an \textit{oblivious adversary}, who decides upon the ordering prior to learning which elements are
active. In contrast, by defining the active elements in the aforementioned way, the arrival order is correlated with
which elements are active, and so our $\alpha$-selectable guarantee need not hold against this arrival order.
The other weakness of this approach is that it does not easily generalize when there are multiple copies of each item.
We thus take an alternative approach which resolves all of these issues. 

Due to \Cref{ssec:reduction_single_unit}, we hereby assume we have a single copy of each item.
The first step is to solve LP \eqref{eqn:poisson_LP} to get an optimal solution $(x_j)_{j \in N}$ of value $\PLP$. We then prove a
weaker version of \Cref{thm:poisson_reduction}. We partition the time horizon $(0,1]$ into $T$ equal sized intervals, $(y_0, y_1], (y_1,y_2], \ldots , (y_{T-1},y_T]$, where $T$ is an integer chosen by the online algorithm. We make $T$ copies of $N$, and assign exactly one copy to each interval, inducing $T$ batches $N_1, \ldots, N_T$, where $N_t = \{ (j,t) : j \in N\}$. By defining $(j,t)$ to be active provided a certain random bit $B_{j}(t) =1$,
and $j$ is the \textit{first} product type to arrive in the interval $(y_{t-1},y_t]$,
we get a random-element OCRS input, whose activeness probabilities $x_{j,t}$ satisfy $$ e^{-\lambda/T} \cdot \frac{x_j  \int_{y_{t-1}}^{y_t} \lambda_{j}(y) dy }{\int_{0}^1 \lambda_j(y) dy} \le x_{j,t} \le \frac{x_j  \int_{y_{t-1}}^{y_t} \lambda_{j}(y) dy }{\int_{0}^1 \lambda_j(y) dy}$$ We then use our $\alpha$-selectable random-element OCRS to get an online algorithm which for any product type $j$
and interval $(y_{t-1},y_t]$, sells a copy of $j$ w.p. at least $\alpha x_{j,t}$. By summing over
the intervals and product types, the expected reward is at least $e^{-\lambda/T} \cdot \alpha \cdot \PLP$. Since $\lambda$ is a fixed constant, and we can take $T$ as large as we please, this tends to $\alpha \cdot \PLP$ as $T \rightarrow \infty$, as desired. 

The second step in the reduction is to replace the use of the $\alpha$-selectable random-element OCRS with an $(\alpha - o_{T}(1))$-selectable standard OCRS via \Cref{thm:random-element_to_standard}, where $o_{T}(1)$ is a function which tends to $0$ as $T \ra \infty$.
We accomplish this by recognizing that our fractional solution $(x_{j,t})_{j \in N, t \in [T]}$
satisfies $\sum_{j \in N} x_{j,t} \le \lambda/T$ for each batch $t \in [T]$, where we recall that $\lambda$ is a constant independent of $T$. Due to \Cref{thm:random-element_to_standard}, this allows us to get a $(\alpha - o_{T}(1))$-selectable random-element OCRS. By once again taking $T$ sufficiently large, we can combine with the first step of the reduction to complete the proof of the time-varying case of \Cref{thm:poisson_reduction}.

\subsubsection{Proof of \Cref{thm:poisson_reduction} assuming \Cref{thm:random-element_to_standard}}

Recall that $\lambda \ge 0$ is any constant which satisfies $\sum_{j \in N} \lambda_{j}(y) \le \lambda$ for all $y \in [0,1]$.
Fix $T \in \mb{N}$ with $T \ge \lambda$ to be a \textit{discretization constant}, and define $y_t=t/T$ for each $t=0, \ldots ,T$.
We divide $(0,1]$ into $T$ intervals, $(y_{t-1},y_t]$ for each $1 \le t \le T$. 
Let $Q_{j}(t)$ be the indicator random variable for the event that $j$
was the \textit{first} product type to have an arrival during the interval $(y_{t-1}, y_t]$.

\begin{lemma} \label{lem:first_arrival}
The random variables $(Q_{j}(t))_{j \in N, t \in [T]}$ satisfy the following properties:
\begin{enumerate}
    \item For each $t \in [T]$, $\sum_{j \in N} Q_{j}(t) \le 1$.

       \item For each $t \in [T]$ and $j \in N$,
$$
    \mb{P}[Q_{j}(t) =1 ] = \int_{y_{t-1}}^{y_t}   \lambda_j(y)  e^{-  \int_{y_{t-1}}^y \sum_{j' \in N}\lambda_{j'}(y')  dy'}  dy,
$$
and 
    $e^{-\lambda/T}  \int_{y_{t-1}}^{y_t} \lambda_{j}(y) dy \le \mb{P}[Q_{j}(t) =1 ] \le \int_{y_{t-1}}^{y_t} \lambda_{j}(y) dy.$
    
        \item If $\bm{Q}(t) := (Q_{j}(t))_{j \in N}$,
    then the random vectors $(\bm{Q}(t))_{t \in [T]}$ are independent.
\end{enumerate}
\end{lemma}

\begin{proof}[Proof of \Cref{lem:first_arrival}]
The first property is immediate by definition.
In order to see the remaining properties, for each $j \in N$, let $(W_{j}(y))_{y \in [0,1]}$
be the Poisson point process for product type $j$, where $W_{j}(y)$
is a random function which specifies the number of arrivals of product type $j$ in the interval $[0,y]$. We use the following basic fact of Poisson point processes:
For any $j \in N$ and $y, y' \in [0,1]$ with $y < y'$
    \begin{equation} \label{eqn:poisson_arrival_distribution} 
        W_{j}(y') - W_{j}(y) \sim \Pois( \int_{y}^{y'} \lambda_{j}(s) ds),
    \end{equation}
Let us now define
\begin{equation}
    Y_{j,t} := \inf\{ y \in (y_{t-1},y_t]: W_{j}(y) > W_{j}(y_{t-1})\},
\end{equation}
where $Y_{j,t} := \infty$ if $W_{j}(y) = W_{j}(y_{t-1})$ for all $y \in [y_{t-1},y_t)$.
That is, $Y_{j,t}$ is the arrival time of the first product copy of type $j$ to come after
time $y_{t-1}$.
Notice that for each $y \in (y_{t-1},y_t]$, $Y_{j,t} \le y$ if and only if $W_{j}(y) >  W_{j}(y_{t-1})$. Thus, since $W_j$ is non-decreasing,
\begin{equation} \label{eqn:cdf_arrival}
    \mb{P}[Y_{j,t} \le y] = 1 - \mb{P}[W_{j}(y_{t-1}) = W_{j}(y)] = 1 - e^{-\int_{y_{t-1}}^y \lambda_j(y') dy'},
\end{equation}
where the final equality uses \eqref{eqn:poisson_arrival_distribution} for $y_{t-1} < y$. By differentiating \eqref{eqn:cdf_arrival} with respect to $y$, the probability density function of $Y_{j,t}$
is 
\begin{equation} \label{eqn:pdf_arrival}
    y \rightarrow \lambda_{j}(y) \cdot e^{-\int_{y_{t-1}}^y \lambda_j(y') dy'}
\end{equation}
Now, let us condition on $Y_{j,t} = y$ for some $y \in (y_{t-1}, y_t]$.
Observe then that $Q_{j}(t) =1$ if and only if $W_{j'}(y) =W_{j'}(y_{t-1})$ for all $j' \neq j$.
Thus, using the independence of the Poisson processes over $j' \neq j$, 
\begin{align}
    \mb{P}[Q_{j}(t) =1 \mid Y_{j,t} = y] &= \prod_{j' \neq j} \mb{P}[W_{j'}(y) = W_{j'}(y_{t-1})] \notag \\
                                          &= \prod_{j' \neq j} e^{-  \int_{y_{t-1}}^y \lambda_{j'}(y')  dy'} 
                                          = e^{-  \int_{y_{t-1}}^y \sum_{j' \neq j}\lambda_{j'}(y')  dy'} \label{eqn:conditional_first_arrival},
\end{align}
where the first equality applies \eqref{eqn:poisson_arrival_distribution} for $y_{t-1} < y$
for each $j' \neq j$. By combining \eqref{eqn:pdf_arrival} with \eqref{eqn:conditional_first_arrival}, and using that $Q_{j}(t) = 0$ if $Y_{j,t} = \infty$,
\begin{align*}
    \mb{P}[Q_{j}(t) = 1] &= \int_{y_{t-1}}^{y_t} e^{-  \int_{y_{t-1}}^y \sum_{j' \neq j}\lambda_{j'}(y')  dy'} \cdot \lambda_{j}(y) e^{-\int_{y_{t-1}}^y \lambda_j(y') dy'} dy \\
    &= \int_{y_{t-1}}^{y_t}   \lambda_j(y)  e^{-  \int_{y_{t-1}}^y \sum_{j' \in N}\lambda_{j'}(y')  dy'}  dy.
\end{align*}
Finally, the lower and upper bounds on $\mb{P}[Q_{j}(t) = 1]$ follow since
$0 \le \sum_{j' \in N}\lambda_{j'}(y') dy' \le \lambda$ 
for all $y' \in [y_{t-1},y_t]$.

In order to prove the final property, we require one more additional fact
of Poisson point processes:
For any $j \in N$, $y, y' \in [0,1]$ with $y < y'$, and $\ell \in \mb{N}_0$,
    \begin{equation} \label{eqn:independent_increments}
            \mb{P}[W_{j}(y') - W_{j}(y) = \ell \mid (W_j(s))_{s \in [0, y]]}] = \mb{P}[W_{j}(y') - W_{j}(y) = \ell].
    \end{equation}
In other words, the distribution of $W_{j}(y') - W_{j}(y)$ is independent of $(W_j(s))_{s \in [0, y]]}$, provided $y' > y$. This can be proven by using the basic fact 
that the Poisson process $(W_{j}(y))_{y \in [0,1]}$ has independent increments. In particular,
for any $K \in \mb{N}$,
$W_{j}(y') - W_{j}(y)$ is independent of $(W_{j}(s_q) - W_{j}(s_{q-1}))_{q=1}^K$,
where $s_k := y \cdot q/K$ for $q=0, \ldots , K$. On the other hand, since
\begin{equation*}
    \lim_{K \rightarrow \infty} \mb{P}[W_{j}(y') - W_{j}(y) = \ell \mid (W_{j}(s_q) - W_{j}(s_{q-1}))_{q=0}^K] =  \mb{P}[W_{j}(y') - W_{j}(y) = \ell \mid (W_j(s))_{s \in [0, y]]}],
\end{equation*}
this is enough to derive \eqref{eqn:independent_increments}.

We now argue that $(\bm{Q}(t))_{t \in [T]}$ are independent random vectors, where we recall that $\bm{Q}(t) = (Q_j(t))_{j \in N}$. Our goal is to show that for any choice of $\bm{b}_1, \ldots , \bm{b}_T \in \{0,1\}^{|J|}$, 
\begin{equation} \label{eqn:independence_goal}
    \mb{P}[\cap_{t \in  [T]} \{ \bm{Q}(t) = \bm{b}_{t}\}] = \prod_{t \in [T]} \mb{P}[\bm{Q}(t) =\bm{b}_t]. 
\end{equation}
For each $t \in [T]$, let us define the history $\scr{H}_{t-1}$ of the processes
up until time $y_{t-1}$ to be the sigma-algebra generated by $(W_{j'}(y))_{y \in [0,y_{t-1}], j' \in N}$. 
Now, because of \eqref{eqn:independent_increments} and the assumption
that the Poisson point processes are independent, we can apply
the same computations as previously done for $\mb{P}[Q_{j}(t) =1]$
to get that for each $\bm{b}_t \in \{0,1\}^{|J|}$,
\begin{equation} \label{eqn:iterate_independence}
\mb{P}[\bm{Q}(t) = \bm{b}_t \mid \scr{H}_{t-1}] = \mb{P}[\bm{Q}(t) =\bm{b}_t].
\end{equation}
On the other hand, notice that for any choice of $\bm{b}_1, \ldots , \bm{b}_{t-1} \in \{0,1\}^{|J|}$,
the event $\cap_{t' < t} \{\bm{Q}(t') = \bm{b}_{t'} \}$ is determined
by the history up until time $y_{t-1}$ (i.e.,  $\cap_{t' < t} \{\bm{Q}(t') = \bm{b}_{t'} \}$ is
$\scr{H}_{t-1}$-measurable). Thus, we can apply \eqref{eqn:iterate_independence}
to get that
\begin{equation}
    \mb{P}[\cap_{t' \le t} \{ \bm{Q}(t') = \bm{b}_{t'}\}] = \mb{P}[\bm{Q}(t) =\bm{b}_t] \cdot \mb{P}[\cap_{t' < t} \{ \bm{Q}(t') = \bm{b}_{t'}\}].
\end{equation}
By iterating this argument starting from $t = T$, \eqref{eqn:independence_goal} follows.
The lemma is thus proven.
\end{proof}

We next define the random-element OCRS input which we use in our reduction.
We assume that we have solved \eqref{eqn:poisson_LP} to get an optimal solution $(x_j)_{j \in N}$ of value $\textsf{LP-POIS}$.

Our construction has the same item set $M$ as used to define the product types $N$. It
makes $T$ copies of $N$, and defines each batch $N_t$ to be one such copy.
\begin{definition} \label{def:random-element_construction}
For each $t \in [T]$, define the batch $N_{t}:= \{ (j,t): j \in N\}$, and $N' = \cup_{t=1}^T N_t$ to be the set of all possible elements. Then, for each $(j,t) \in N'$, we define its associated 
items $A_{j,t} : = A_j$, where $A_j \subseteq M$ is the items of associated with product type $j$. Finally, we set
\begin{equation} \label{eqn:poisson_active}
x_{j,t} := \frac{x_j}{\int_{0}^{1} \lambda_{j}(y) dy} \cdot \mb{P}[Q_{j}(t) =1].
\end{equation}
\end{definition}

We now verify that \Cref{def:random-element_construction} describes a valid random-element OCRS input,
and that its batches are $\lambda/T$-bounded. 
\begin{lemma} \label{lem:valid_random_element_input}
Suppose that $T \ge \lambda$. Then,   $(x_{j,t})_{(j,t) \in N'}$ is feasible in expectation (i.e., \eqref{eqn:feasInExpn} holds), and $\sum_{j \in N} x_{j,t} \le \lambda/T$ for all $t \in [T]$.
\end{lemma}
\begin{proof}
First observe that for any $j \in N$ and $t \in [T]$,
we can apply the bound on $\mb{P}[Q_{j}(t) =1]$ from \Cref{lem:first_arrival},
as well as \eqref{eqn:poisson_product_arrival} to get that
\begin{equation} \label{eqn:poisson_active_upper_bound}
   \text{$x_{j,t} \le \frac{x_j \int_{y_{t-1}}^{y_t} \lambda_{j}(y) dy}{\int_{0}^1 \lambda_j(y) dy}$ and $\frac{x_j}{\int_{0}^{1} \lambda_{j}(y) dy} \le 1$.}
\end{equation}
More, observe that for any $i \in M$, we can use the first inequality of \eqref{eqn:poisson_active_upper_bound}
to get that
$$\sum_{(j,t) \in N': i \in A_{j,t}} x_{j,t} = \sum_{j \in N} \sum_{t=1}^{T}  x_{j,t} \le  \sum_{j \in N:i\in A_j} \frac{x_j \sum_{t=1}^T \int_{y_{t-1}}^{y_t} \lambda_{j}(y) dy}{\int_{0}^1 \lambda_j(y) dy} \le 1,$$
where the final inequality follows since $(x_j)_{j \in N}$ satisfies \eqref{eqn:poisson_item_capacity}.
Thus, the input is feasible in expectation.

Now, if we apply both inequalities from \eqref{eqn:poisson_active_upper_bound},
\begin{equation} \label{eqn:bounded_batches}
    \sum_{j \in N} x_{j,t} \le  \frac{\sum_{j \in N} x_j \int_{y_{t-1}}^{y_t}  \lambda_{j}(y) dy}{\int_{0}^1 \lambda_j(y) dy} \le
  \int_{y_{t-1}}^{y_t}  \sum_{j \in N} \lambda_{j}(y) dy \le \lambda/T,
\end{equation}
where the last inequality follows from definition of $\lambda$. Thus, $ \sum_{j \in N} x_{j,t} \le \lambda/T$ for each $t \in [T]$ as claimed, and so the proof is complete.

\end{proof}
We next show how to use an $\alpha$-selectable random-element OCRS for the input $(x_{(j,t)})_{(j,t) \in N'}$ to design
an online algorithm whose expected reward at least $\alpha  \cdot e^{-\lambda/T} \cdot \textsf{LP-POIS}$, where $\textsf{LP-POIS}$ is the benchmark from
LP \eqref{eqn:poisson_LP}.

\begin{lemma} \label{lem:poisson_arrival_to_random_element}
Suppose that $T \ge \lambda$.
Then, an $\alpha$-selectable random-element OCRS (respectively, RCRS) for $(x_{(j,t)})_{(j,t) \in N'}$ implies
    an online algorithm whose expected reward on the NRM input with time-varying arrivals (respectively, time-homogeneous) is at least $\alpha  \cdot e^{-\lambda/T} \cdot \textsf{LP-POIS}.$

\end{lemma}
\begin{proof}[Proof of \Cref{lem:poisson_arrival_to_random_element}]

Given the input $(x_{(j,t)})_{(j,t) \in N'}$, we need to draw its active elements
so that they are coupled with the arrivals of the Poisson point processes,
while still ensuring that they are distributed as specified in \Cref{def:random-ocrs}.
For each $(j,t) \in N'$, draw
$B_{j}(t) \sim \Ber(x_j/ \int_{0}^{1} \lambda_j(y) dy)$ independently, which we note is well-defined
due to \eqref{eqn:poisson_product_arrival}. We define
$(j,t) \in N'$ to be \textit{active}
provided $B_{j}(t) \cdot Q_{j}(t) =1$. Using the independence of $B_{j}(t)$
and $Q_{j}(t)$ together with the definition of $x_{j,t}$ in  \eqref{eqn:poisson_active}, first note that
\begin{equation} \label{eqn:poisson_active_restatement}
   \mb{P}[B_{j}(t) \cdot Q_{j}(t) =1] =  \mb{P}[B_{j}(t)=1] \cdot \mb{P}[Q_{j}(t)=1]= \frac{x_j}{\int_{0}^{1} \lambda_j(y) dy} \mb{P}[Q_{j}(t)=1]=x_{j,t}.
\end{equation}
More, there is at most one active element in each $N_{t}$, as $\sum_{(j,t) \in N_t} Q_{j}(t) \le 1$ for each $t \in [T]$, due to the first property of \Cref{lem:first_arrival}.  Finally, due to the third property of \Cref{lem:first_arrival}, it is clear that the active elements between distinct batches are drawn independently.  Thus, defining the active elements in this way is in accordance with
the requirements of \Cref{def:random-ocrs}.

We now prove the lemma for OCRS, and afterwards explain how to adapt it to RCRS and time-homogeneous Poisson arrivals.  Assume that we have access to an $\alpha$-selectable random-element OCRS for $(x_{(j,t)})_{(j,t) \in N'}$. 
The online algorithm $\ALG$ is then easy to describe:
For each $t \in [T]$:
\begin{enumerate}
    \item If $Q_{j}(t) \cdot B_{j}(t) =1$ for some $j \in N$ (i.e., $(j,t)$ is active), then query the OCRS for the current element $(j,t)$. Sell the copy of $j$ if and only if the OCRS accepts $(j,t)$.
\end{enumerate}
Clearly, $\ALG$ is well-defined, as for each $t \in [T]$,  the random variables $(B_{j}(t))_{j \in N}$ are drawn in advance, and it may determine $(Q_{j}(t))_{j \in N}$ in an online manner as the product copies arrive in the interval $(y_{t-1},y_t]$. Now, denote $Z_{j,t}$ as the event $\ALG$ sells a copy of $j$ during the interval $(y_{t-1},y_t]$.  Then, since 
$Z_{j,t}$ occurs if and only if the $\alpha$-selectable OCRS accepts the element $(j,t)$,
\begin{align} \label{eqn:relaxed_acceptance}
\mb{P}[Z_{j,t}] \ge \alpha \cdot x_{j,t}
= \alpha \cdot \mb{P}[B_{j}(t)=1] \cdot \mb{P}[Q_{j}(t)=1] \ge \alpha \cdot e^{-\lambda/T} \cdot\frac{x_j \int_{y_{t-1}}^{y_t} \lambda_{j}(y) dy}{\int_{0}^1 \lambda_j(y) dy},
\end{align}
where  the inequality applies \eqref{eqn:poisson_active_restatement} and the lower
bound on $\mb{P}[Q_{j}(t)=1]$ from \Cref{lem:first_arrival}. 
Since at most one copy of $j \in N$ can be sold by $\ALG$, we can apply \eqref{eqn:real_acceptance} over $t \in [T]$ to get that
\begin{align}
\mb{P}[\text{a copy of $j$ is sold by $\ALG$}] = \sum_{t=1}^T \mb{P}[Z_{j,t}]
&\ge  \frac{\alpha \cdot x_j e^{-\lambda/T}  \sum_{t=1}^T \int_{y_{t-1}}^{y_t} \lambda_{j}(y) dy}{\int_{0}^1 \lambda_j(y) dy} \notag \\
&=   \alpha x_j e^{-\lambda/T}. \label{eqn:real_acceptance}
\end{align}
By summing over all $j \in N$ and applying \eqref{eqn:real_acceptance}, the expected revenue of $\ALG$ is at least
\begin{align*}
\alpha e^{-\lambda/T} \sum_{j \in N} r_j x_j &=    \alpha e^{-\lambda/T} \textsf{LP-POIS}
\end{align*}
where the equality follows since $(x_j)_{j \in N}$ is an optimal solution to LP \eqref{eqn:poisson_LP}. The statement of the lemma for OCRS is thus proven.

For the case of time-homogeneous Poisson arrivals, recall that $\lambda_j$ is now
a constant for each $j \in N$. In this case, for any $t,t' \in [T]$, $\mb{P}[Q_{j}(t)] = \mb{P}[Q_{j}(t')]$, which implies $x_{j,t} = x_{j,t'}$ by \eqref{eqn:poisson_active_restatement}. Thus, the batches of active elements,
$( (B_{j}(1) Q_{j}(1))_{j \in N}, \ldots , (B_{j}(T) Q_{j}(T))_{j \in N})$ are identically distributed. As a result, by a standard simulation argument, we can assume without loss
that the batches are revealed in a uniform at random (u.a.r.) order, and thus use a random-element RCRS
in the definition of $\ALG$ instead of an OCRS. 
\end{proof}

We now complete the proof of the time-varying case of \Cref{thm:poisson_reduction}.
This relies on \Cref{thm:random-element_to_standard}, which we prove in the following section.

\begin{proof}[Proof of \Cref{thm:poisson_reduction} (assuming \Cref{thm:random-element_to_standard})]
 We first prove the statement for OCRS, and then explain the slight modification for RCRS. 
 
    Assume that we are given an $\alpha$-selectable standard OCRS. 
    For any discretization constant $T \in \mb{N}$ with $T \ge \lambda$, let us construct the random-element OCRS input $(x_{(j,t)})_{(j,t) \in N'}$ from \Cref{def:random-element_construction}.
    Due to \Cref{lem:valid_random_element_input}, we know
    that $(x_{(j,t)})_{(j,t) \in N'}$ has $\lambda/T$-bounded batches.
    Thus, \Cref{thm:random-element_to_standard} implies that we have an $((\alpha -(1- e^{-\frac{\lambda}{T}(1 + 2 \lambda)}))/(1 +\lambda/T))$-selectable random-element OCRS for $(x_{(j,t)})_{(j,t) \in N'}$. As such, we can apply \Cref{lem:poisson_arrival_to_random_element} to get an online algorithm $\ALG$ whose expected reward is at least
\begin{equation} \label{eqn:final_performance}
    \frac{\alpha -  (1- e^{-\frac{\lambda}{T}(1 + 2 \lambda)})}{1 + \lambda/T} e^{-\lambda/T} \cdot \textsf{LP-POIS}.
\end{equation}
Now,  since $\lambda$ is a fixed constant, it is clear that $\lambda/T \rightarrow 0$,
and so $e^{-\lambda/T} \rightarrow 1, e^{-\frac{\lambda}{T}(1 + 2 \lambda)} \rightarrow 1$ as $T \rightarrow \infty$. Since $T$ is a parameter chosen by $\ALG$, we can always
take $T$ sufficiently large, so as to ensure that \eqref{eqn:final_performance} is arbitrarily close to $\alpha \textsf{LP-POIS}$. 
The proof is thus complete for OCRS.

For RCRS, the only distinction in the proof is that we start with an $\alpha$-selectable inactive-oblivious standard RCRS. However, this is only relevant when applying \Cref{thm:random-element_to_standard}, in which case we get a random-element RCRS for which the exact same computations apply as done for OCRS, allowing us to complete the proof. 
\end{proof}

\subsection{Proof of \Cref{thm:random-element_to_standard}} \label{sec:random-element_to_standard}
    Suppose that the elements $N$ are partitioned into batches $N_1, \ldots , N_T$ for $T \ge 1$.
    We assume that $(x_j)_{j \in N}$ is a fractional solution which is feasible in expectation and satisfies $\sum_{j \in N_t} x_j \le \eps$ for all $t \in [T]$, where $\eps \in (0,1/2]$. We first prove the statement of \Cref{thm:random-element_to_standard} for OCRS, and then explain how to modify our approach to work an RCRS that is ``inactive-oblivious''. 
    
    There are three main steps used in the proof for standard OCRS. Recal that in this model,
    each $j \in N$ is active independently with probability $x_{j}$, and so with probability at least  $f(\epsilon,T) := e^{-\eps(1 + 2 \eps T)}$, each
$N_t$ has at most one active element. We refer to this as the ``good event'' $G_T$, and focus on
the conditional distribution of the active elements, given $G_T$. This conditional probability space
has active elements which satisfy our random-element OCRS definition (i.e., at most one active element per batch). More, since $G_T$ occurs with probability at least $f(\eps,T)$,
the standard OCRS attains an $(\alpha - (1 - f(\eps,T))$-selection guarantee even in this conditional probability space (see \Cref{lem:conditional_standard}). This implies a $(\alpha - (1 - f(\eps,T))$-selectable random-element OCRS on a related input $(\til{x}_j)_{j \in N}$ with $x_j/(1+\eps) \le \til{x}_j \le x_j$ for all $j \in N$ (see \Cref{lem:conditional_distribution} and \Cref{lem:random_element_standard_relation}). We then recover a $(\alpha - (1 - f(\eps,T))/(1 + \eps)$-selectable random-element OCRS for the original input $(x_j)_{j \in N}$ via a standard down-sampling trick (see \Cref{lem:approximate_random_element}), which completes the proof of \Cref{thm:random-element_to_standard}.

We now provide the various details. Recall the setting of an $\alpha$-selectable standard OCRS, which we denote by $\SOCRS$. On the input $(x_j)_{j \in N}$, each $X_j \sim \Ber(x_j)$ is drawn independently for $j \in N$, and the elements are presented in
some worst-case order $\pi$ chosen by an oblivious adversary. The selection guarantee $\alpha$ of $\SOCRS$ holds on $(x_j)_{j \in N}$ for \textit{every} such ordering. In our reduction, it suffices to consider when $\pi$ is an ordering which respects the chronological arrival order of the batches. That is, for any $j \in N_t$ and $j' \in N_{t'}$, if $j$ arrives before $j'$ with respect to $\pi$, then $t \le t'$.

    Define $S_t = \sum_{j \in N_t} X_j$ for each $t \in [T]$, as well as 
    the \textit{good event} $G_T := \cap_{t \in [T]} \{ S_t \le 1 \}$. Clearly, $\mb{P}[G_T] > 0$. 
    For each $j \in N$, let $Z_j$ denote the event that $j$ is accepted by $\SOCRS$ when the elements arrive in the order $\pi$.
    We first lower bound the probability of this event, conditional on the event $G_T$.
    Afterwards, we explain why this is a useful property in the context of designing a random-element OCRS.
    
    \begin{lemma} \label{lem:conditional_standard}
 For each $j \in N$, $\mb{P}[Z_{j} \mid G_T] \ge (\alpha - (1 - e^{-\eps(1 + 2\eps T)})) x_{j}$.
    \end{lemma}
    \begin{proof}[Proof of \Cref{lem:conditional_standard}]
We prove the statement for an arbitrary $\jc \in N$ where $\jc \in N_{\tc}$ for $\tc \in [T]$,
as this will simplify the indices throughout the proof.
First observe that since $\SOCRS$ is $\alpha$-selectable and $\mb{P}[G_T] > 0$, we know that
\begin{equation} \label{eqn:lower_bound_cond}
    \mb{P}[Z_{\jc} \mid G_T] \ge \frac{\mb{P}[Z_\jc] - \mb{P}[Z_{\jc} \cap \neg G_T]}{\mb{P}[G_T]} \ge \alpha x_{\jc} -  \mb{P}[Z_\jc \cap \neg G_T].
\end{equation}
Moreover, since $X_\jc =1$ is a necessary condition for $Z_\jc =1$, and  $(X_j)_{j \in N}$ are independent random variables,
\begin{align}
    \mb{P}[Z_\jc \cap \neg G_T] &\le \mb{P}[\{X_\jc =1\} \cap \neg G_T] \notag \\
                              &= x_\jc (1 - \mb{P}[\cap_{t \in [T]} \{S_{t} \le 1\} \mid X_\jc =1]) \notag \\
                              &= x_\jc (1 - \mb{P}[S_\tc \le 1 \mid X_\jc =1] \cdot \prod_{t \neq \tc} \mb{P}[S_{t} \le 1]) \label{eqn:final_conditional_acceptance}
\end{align} 
However, $
\mb{P}[S_\tc \le 1 \mid X_\jc =1] \ge \prod_{j \in N_\tc} (1 - x_j),
$
and for each $t \in [T] \setminus \{\tc\}$,
\begin{align*}
\mb{P}[S_{t} \le 1] &= \prod_{j \in N_t}(1 - x_j)  + \sum_{j \in N_t} x_j  \prod_{j' \in N_t \setminus \{j\}} (1- x_{j'}) \\
                     &\ge (1 + \sum_{j \in N_t} x_j) \prod_{j' \in N_t}(1 - x_{j'}). 
\end{align*}
After applying these lower bounds, the elementary inequality $1+ z \ge e^{z - z^2}$ for $|z| \le 1/2$, and that $\sum_{j \in N_t} x_j \le \eps \le 1/2$ for each $t \in [T]$:
\begin{align*}
     \mb{P}[S_\tc \le 1 \mid X_\jc =1] \cdot \prod_{t \neq \tc} \mb{P}[S_{t} \le 1] &
     \ge \prod_{j \in N_\tc}(1 - x_j) \cdot \prod_{t \neq \tc} \left((1 + \sum_{j \in N_t} x_j) \prod_{j' \in N_t}(1 - x_{j'}) \right) \\
     &= \prod_{j \in N} (1 - x_j) \cdot \prod_{t \neq \tc}\left(1 + \sum_{j \in N_t}x_j \right) \\
     &\ge \exp\left(-\sum_{j \in N}x_j - \sum_{j \in N} x^2_j\right) \exp\left(\sum_{t \neq \tc} \left( \sum_{j \in N_t} x_j - (\sum_{j \in N_t} x_j)^2 \right) \right) \\
     &\ge \exp\left(- \sum_{j \in N_{\tc}} x_j -  2 \eps T \right) \ge \exp(-\eps(1 + 2\eps \cdot T))
\end{align*}
By combining this final inequality with \eqref{eqn:final_conditional_acceptance}, $
\mb{P}[Z_\jc \mid G_T] \ge (\alpha - (1- e^{-\eps(1 + 2\eps T)})) x_{\jc},
$  
and so the lemma is proven.
    \end{proof}
We next observe that conditional on $G_T$, the active elements are distributed
as if they were drawn from a certain random-element input.
In order to see this, let $(\til{X}_j)_{j \in N}$ be drawn from the conditional distribution 
of $(X_j)_{j \in N}$ given $G_T$:
    \begin{lemma} \label{lem:conditional_distribution}
    The  random variables $(\til{X}_j)_{j \in N}$ satisfy the following properties:
        \begin{enumerate}
            \item For each $t \in [T]$, $\sum_{j \in N_t} \til{X}_j \le 1$.
            \item If $\bm{\til{X}}(t) := (\til{X}_j(t))_{j \in N}$ for each $t \in [T]$,
            then the random vectors $(\bm{\til{X}}(t))_{t \in [T]}$ are independent. 
                    \item For each $j \in N_t$, \begin{equation} \label{eqn:approximate_random_element}
                    \frac{x_j}{1+ \eps} \le \mb{P}[\til{X}_j =1] = \frac{x_j}{(1- x_j)(1 + \sum_{j' \in N_t} x_{j'}/(1-x_{j'}))} \le x_j
                    \end{equation}
        \end{enumerate}
    \end{lemma}
    \begin{proof}[Proof of \Cref{lem:conditional_distribution}]
        The first two properties follow immediately from the definition of $(\til{X}_j)_{j \in N}$
        and the independence of $(X_j)_{j \in N}$. We focus on the remaining property.
Notice that if $j \in N_t$, then using the independence of $(X_{j'})_{j' \in N}$,
        \begin{align}
            \mb{P}[X_j =1 \mid G_T] &= \frac{\mb{P}[ \{ X_j =1\} \cap \{ S_t \le 1 \}]}{\mb{P}[S_t \le 1]} \notag \\
            &=\frac{x_j \cdot \prod_{j' \in N_t: j' \neq j}(1 - x_{j'})}{\prod_{j' \in N_t}(1- x_{j'}) + \sum_{j' \in N_t} x_{j'} \cdot \prod_{j'' \in N_t: j'' \neq j'}(1 - x_{j''})} \notag \\
            &= \frac{x_j}{(1- x_j)(1 + \sum_{j' \in N_t} x_{j'}/(1-x_{j'}))} \label{eqn:conditional_on_good_event}
        \end{align}
Now, $(1-x_j)(1 + \sum_{j' \in N_t} x_{j'}/(1-x_{j'})) \ge (1- x_j)(1 + x_j/(1-x_j)) =1$,
so \eqref{eqn:conditional_on_good_event} is upper bounded by $x_j$. On the other hand,
$(1-x_j)(1 + \sum_{j' \in N_t} x_{j'}/(1-x_{j'})) \le 1 + \eps$ since $\sum_{j' \in N_t} x_{j'} \le \eps$, so \eqref{eqn:conditional_on_good_event} is lower bounded by $x_j/(1+ \eps)$.
The lemma is thus proven.    
    \end{proof}
Let us now set $\til{x}_j := \mb{P}[\til{X}_j =1]$ for each $j \in N$. Observe
first that the upper bound of \eqref{eqn:approximate_random_element} from \Cref{lem:conditional_distribution} implies $(\til{x}_j)_{j \in N}$
is indeed a valid random-element OCRS input, as $(x_j)_{j \in N}$ is feasible in expectation
and satisfies $\sum_{j \in N_t} x_j \le \eps \le 1$  for each $t \in [T]$.

We next explain how to use the execution of $\SOCRS$ on $(x_j)_{j \in N}$ to get a random-element OCRS for $(\til{x}_j)_{j \in N}$. Recall that $\pi$ is an arbitrary arrival order of $N$ which respects the ordering of the batches. We now describe our random-element OCRS, which
we refer to as $\ROCRS$. For each $t \in [T]$:
\begin{enumerate}

    \item Pass $(\til{X}_j)_{j \in N_t}$ in the place of $(X_j)_{j \in N_t}$ to the standard OCRS. Suppose that $\til{X}_j =1$ for some $j \in N_t$. In this case, using the arrival order $\pi$, accept $j$ if and only if the standard OCRS accepts $j$.
\end{enumerate}

Define $\til{Z}_j$ to be the event that $\ROCRS$ accepts $j \in N$ when executing on $(\til{x}_j)_{j \in N}$, and recall that $Z_j$ is the event that $\SOCRS$ accepts $j$ when executing on $(x_j)_{j \in N}$. We observe the following:
\begin{lemma} \label{lem:random_element_standard_relation}
 For each $j \in N$, $\mb{P}[\til{Z}_{j}] = \mb{P}[Z_j \mid G_T]$.
\end{lemma}
\begin{proof}[Proof of \Cref{lem:random_element_standard_relation}]
Fix some arbitrary $j \in N$. Let us first assume that $\SOCRS$ is deterministic. In this case,
there exists some function $\phi$ such that $\bm{1}_{Z_j} = \phi( \bm{X})$, where $\bm{X} = (X_{j'})_{j' \in N}$.
On the other hand, since $\ROCRS$ is defined using $\SOCRS$, we also know
that $\bm{1}_{\til{Z}_j} = \phi( \bm{\til{X}})$, where $\bm{\til{X}} = (\til{X}_{j'})_{j' \in N}$.
As such, since $\bm{\til{X}}$ is distributed as $\bm{X}$ conditional on $G_T$, 
\begin{equation} \label{eqn:random_element_standard_relation} 
   \mb{P}[\til{Z}_j] =  \mb{E}[ \phi(\bm{\til{X}}) ] = \mb{E}[\phi( \bm{X}) \mid G_T] = \mb{P}[Z_j \mid G_T].
\end{equation}
Thus, when $\SOCRS$ is deterministic, the lemma holds. If $\SOCRS$ is randomized, then we can first prove the analogous version of \eqref{eqn:random_element_standard_relation}, conditional on a particular instantiation of
the internal random bits used by $\SOCRS$ (and in turn $\ROCRS$). After taking expectations, the same claim holds.  
\end{proof}
By combining Lemmas \ref{lem:conditional_standard} and \ref{lem:random_element_standard_relation},
we know that $\ROCRS$ is $(\alpha - (1 - e^{-\eps(1 + 2 \eps T)}))$-selectable on $(\til{x}_j)_{j \in N}$. The final step is to translate this into a random-element OCRS for $(x_j)_{j \in N}$ with the selection guarantee claimed in \Cref{thm:random-element_to_standard}. We achieve this via the following lemma:
    \begin{lemma} \label{lem:approximate_random_element}
    Fix $\delta > 0$, and suppose $(x_j)_{j \in N}$ and $(\til{x}_{j})_{j \in N}$ are random-element OCRS inputs
    with respect to $N_1, \ldots , N_T$
    which satisfy
    \begin{equation} \label{eqn:x^*_bounds}
            \text{$\delta x_j \le \til{x}_j \le x_j$ for all $j \in N$.}      
    \end{equation} 
    Then, an $\alpha$-selectable random-element OCRS  for $(\til{x}_{j})_{j \in N}$ implies an $\alpha \cdot \delta$-selectable random-element OCRS 
    for $(x_j)_{j \in N}$.
    \end{lemma}
    \begin{proof}[Proof of \Cref{lem:approximate_random_element}]
    We provide just a sketch of the proof, as the argument is standard.
Notice that if we draw $B_j \sim \Ber(\til{x}_j/x_j)$ for each $j \in N$ independently, then
we can couple the active elements of the inputs $(x_j)_{j \in N}$
and $(\til{x}_j)_{j \in N}$, such that $j$ is active for $(x_j)_{j \in N}$ if and only if
$j$ is active for $(\til{x}_j)_{j \in N}$ and $B_j =1$. By executing an $\alpha$-selectable
OCRS on the active elements of $(\til{x}_j)_{j \in N}$, this coupling ensures each
active element of $(x_j)_{j \in N}$ is accepted with probability at least $\alpha \til{x}_j \ge \alpha \delta x_j$ as desired. 
    \end{proof}

We now complete the proof of \Cref{thm:random-element_to_standard} for the OCRS case.
\begin{proof}[Proof of \Cref{thm:random-element_to_standard} for the OCRS case]
First observe that Lemmas \ref{lem:conditional_standard} and \ref{lem:random_element_standard_relation} imply that \\ $\ROCRS$ is 
$(\alpha - e^{-\eps(1 + 2\eps T)})$-selectable on $(\til{x}_j)_{j \in N}$. On the other
hand \eqref{eqn:approximate_random_element} of \Cref{lem:conditional_distribution} implies that $(x_j)_{j \in N}$ and $(\til{x}_j)_{j \in N}$ are random-element inputs which satisfy
\eqref{eqn:x^*_bounds} of \Cref{lem:approximate_random_element} for $\delta = 1/(1+\eps)$.
Thus, \Cref{lem:approximate_random_element} implies that we have a $\left(\frac{\alpha - (1- e^{-\eps(1 + 2 \eps T)})}{1+ \eps}\right)$-selectable random-element OCRS
as claimed in \Cref{thm:random-element_to_standard}.
\end{proof}

To complete the proof of \Cref{thm:random-element_to_standard} for RCRS, we must formally define an \textit{inactive-oblivious standard RCRS}. As suggested in \Cref{sec:main_body_reductions}, an inactive-oblivious RCRS operates in the same setting as a standard RCRS, yet with less information available to it as it executes.
It starts with the fractional solution $(x_j)_{j \in N}$ of the RCRS input, and each element is active independently with probability $x_j$. More, each $j \in N$ has a uniform arrival time $Y_j \in [0,1]$ drawn independently. The difference
is that only the \textit{active} elements are presented in increasing order of $(Y_j)_{j \in N}$. It has the same requirement that when an active $j$ arrives at time $Y_j$, it makes an irrevocable decision on whether to accept or reject $j$. We provide a formal definition below:
\begin{definition}[Inactive-oblivious standard RCRS] \label{def:inactive-oblivious-rcrs} Consider a standard RCRS input with fractional solution $\bm{x}=(x_j)_{j\in N}$. Independently draw $X_j\sim\Ber(x_j)$ and  $Y_j$ uniformly on $[0,1]$ for each $j\in N$, where $X_j$ indicates whether $j$ is active and $Y_j$ is its arrival time. 

For each $j\in N$, define the \textit{active history} of $j$ by $ \mathcal H_j(\bm X,\bm Y) := \bigl((j_1,Y_{j_1}),\ldots,(j_k,Y_{j_k})\bigr)$,  where $\{j_1,\ldots,j_k\} = \{j'\in N:X_{j'}=1,\ Y_{j'}\le Y_j\}$,  and the elements are indexed so that $Y_{j_1}<\cdots<Y_{j_k}$. A deterministic standard RCRS is said to be \emph{inactive-oblivious}, if for every $j\in N$, the event ``$j$ is accepted'' is a function of $ \mathcal H_j(\bm X,\bm Y)$. Thus, conditional on $j$ being active, the decision whether to accept $j$ may depend on the identities and arrival times of the active elements that arrive no later than $j$, but not on any information concerning inactive elements. A randomized standard RCRS is said to be \emph{inactive-oblivious} if it is a distribution over deterministic inactive-oblivious standard RCRS's.

\end{definition}

\begin{proof}[Proof of \Cref{thm:random-element_to_standard} for the RCRS case]
For the case of RCRS, the only step we must carefully check is \Cref{lem:random_element_standard_relation}.
All of the other lemmas also apply to any standard RCRS. In particular, the good event $G_T$ and its properties from \Cref{lem:conditional_distribution} do not depend on the arrival order of $N$, the statement of \Cref{lem:conditional_standard} extends to standard RCRS's, and \Cref{lem:approximate_random_element} extends to random-element RCRS's. 

Recall that $(\til{X}_j)_{j \in N}$ is drawn from the conditional distribution of $(X_j)_{j \in N}$ given $G_T$ from (see \Cref{lem:approximate_random_element}) and $\til{x}_j := \mb{P}[\til{X}_j =1]$ for each $j \in N$. Our goal is to define a random-element RCRS for $(\til{x}_j)_{j \in N}$ via a standard RCRS for $(x_j)_{j \in N}$ that is inactive-oblivious. We do this in the same way as for OCRS, yet we have to carefully decide which arrive times to pass it.

For a random-element RCRS, we first generate independent arrival times $(\til{Y}_t)_{t \in [T]}$ for the batches,
where $\til{Y}_t \in [0,1]$ is uniform.
The batches are presented in increasing order of $(\til{Y}_t)_{t \in [T]}$.
When $(\til{X}_{j'})_{j' \in N_{t}}$ and $\til{Y}_t$ are revealed, we pass $(\til{X}_{j'})_{j' \in N_{t}}$ to the standard RCRS in the place of $(X_{j'})_{j' \in N_{t}}$, and set the arrival time of each $j' \in N_t$
to be $Y_{j'} := \til{Y}_{t}$. Suppose that $\til{X}_j =1$
for some $j \in N_{t}$. We then process the elements of $N_t$ via the standard RCRS
in an arbitrary order, and accept $j$ if and only if
the standard RCRS accepts it.

The main technicality is that by applying this approach, we end up executing our standard RCRS with arrival times of $N$ which are \textit{not} independently drawn. However, conditional on $G_T$, there is at most one active element in each batch (i.e., $j \in N_t$ with $X_j =1$), which implies that the \textit{active elements} receive arrival times which are independently drawn. Since the inactive elements do not affect the decisions of a standard RCRS that is inactive-oblivious, this suffices to derive the analogous form of \Cref{lem:random_element_standard_relation} for RCRS.
\end{proof}

\subsection{Proof of \Cref{thm:oca_reduction}} \label{pf:oca_reduction}
Let $(x_j(A))$ denote an optimal solution to \textsf{LP-OCA}. For every agent $t$ and every nonempty bundle $A\subseteq M$ with $|A|\le L$, create a dummy element $(t,A)$ that consumes precisely the items in $A$ and define $z_{t,A}= \sum_{j\in N_t}x_j(A)$. These elements form one batch for each agent $t$. The arrival constraints in \textsf{LP-OCA} imply
\[\sum_{A\subseteq M:1\le |A|\le L}z_{t,A} = \sum_{j\in N_t}\sum_{A\subseteq M:1\le |A|\le L}x_j(A)\le \sum_{j\in N_t}\lambda_j \le 1, \]
and the capacity constraints imply
\[\sum_{t\in [T]}\sum_{A:i\in A}z_{t,A} = \sum_{t\in [T]}\sum_{j\in N_t}\sum_{A:i\in A}x_j(A)\le 1, \  \forall i\in M.\]
Thus $(z_{t,A})$ is feasible in expectation and satisfies the mass constraint in every batch. 

We next explain how the OCA input generates this random-element input online. When agent $t$ realizes type $j\in N_t$ with $\lambda_j >0$, we independently draw at most one candidate bundle $A$ with probability $x_j(A)/\lambda_j$. Thus, $\mathbb{P}((t,A) \text{ is active})=\sum_{j\in N_t} \lambda_j x_j(A)/ \lambda_j=z_{t,A}$. At most one element is active in each agent batch, and the batches are independent. We then pass the element $(t,A)$ to the random-element OCRS. If the scheme accepts $(t,A)$, we allocate $A$ to the agent; otherwise allocate nothing. For every dummy element $(t,A)$ with $z_{t,A}>0$, $\alpha$-selectable guarantee gives acceptance probability $\alpha$ conditional on that element being active, which implies $\mathbb{P}(\text{type }j \text{ receives bundle } A)= \alpha x_j(A)$. The corresponding expected welfare is then $\alpha \cdot \textsf{LP-OCA}$. 

If the agents arrive in an adversarially fixed order, their batches form a valid random-element OCRS input. The same construction extends to the uniformly random order. Finally, in the single-minded case, we may choose an optimal LP solution supported only on the prescribed bundle. Each agent batch then contains a single relevant element, so the random-element model reduces to the standard model.

\section{Additions to \Cref{sec:ocrs}}
\subsection{Implementing the $1/(1+L)$-selectable Random-element OCRS} \label{ssec:implement_ocrs}
The policy $\pi$ defined in Definition \ref{def:ocrs} cannot be implemented directly because it requires the knowledge of the probability $\bP(F_j)$ for every element $j$. In what follows, we provide a policy with the aid of simulation so that it can be implemented.  
\begin{definition}\label{def:simulate_ocrs}
    For each time $t$, run a Monte Carlo simulation with $K$ trails: starting from time $\tau=1$ to $\tau=t-1$, implement the policy $\pi$ in Definition \ref{def:ocrs} with $\hat{\bP}(F_{j})$ for $j\in N_1\cup \dots\cup N_{t-1}$ and set $\alpha=(1-\varepsilon)/(1+L)$. Let $\hat{\bP}(F_{j})$ denote the empirical estimation of the probability that the element $j\in N_t$ is feasible, that is,
    \[\hat{\bP}(F_{j})=\frac{1}{K}\sum\limits_{k\in [K]}\mathds{1}\{\textnormal{element {$j$} is feasible in {$k$}-th trial}\}. \]
\end{definition}

Let $\hat{\pi}$ denote the simulation algorithm and $\hat{\bP}(F_{j})$ denote the output of the simulation algorithm. Moreover, let $\bP^{\hat{\pi}}(F_{j})$ denote the true probability that element $j$ is feasible under policy $\hat{\pi}$, which is a random variable depending on the previous sample paths. Note that by construction, $\hat{\bP}(F_{j})$ is an unbiased estimate of $\bP^{\hat{\pi}}(F_{j})$. Let $V^{\hat{\pi}}$ denote the expected rewards of the simulation based policy $\hat{\pi}$.

\begin{lemma}\label{lem:simu_ocrs_lem}
    For any time $t$, given that $\alpha=(1-\varepsilon)/(1+L)$ and $\bP^{\hat{\pi}}(F_{j})/\hat{\bP}\left(F_{j}\right)\le 1/(1-\varepsilon)$ for all $\tau<t$ and $j\in N_\tau$, it holds that $\bP^{\hat{\pi}}(F_{j})\ge 1/(1+L)$ for any $j\in N_t$.
\end{lemma}
\begin{proof}[Proof of \Cref{lem:simu_ocrs_lem}]
    Note that for any $j\in N_t$, 
\begin{align*}
    \bP^{\hat{\pi}}\left(F_{j}\right)=&\bP^{\hat{\pi}}\left(\cap_{i\in A_j}F_{i}(t)\right)=1-\bP^{\hat{\pi}}\left(\cup_{i\in A_j}\neg{F}_{i}(t)\right)\ge 1-\sum_{i\in A_j}\bP^{\hat{\pi}}\left(\neg{F}_{i}(t)\right)\\
    =&1-\sum_{i\in A_j}\frac{1-\varepsilon}{1+L}\sum_{\tau=1}^{t-1}\sum_{j'\in N_\tau :i\in A_{j'}}x_{j'}\cdot \frac{1}{\hat{\bP}(F_{ j'})}\cdot \bP^{\hat{\pi}}\left(F_{ j'}\right)\\
    \ge&1-\sum_{i\in A_j}\frac{1-\varepsilon}{1+L}\sum_{\tau=1}^{t-1}\frac{1}{1-\varepsilon}\sum_{j'\in N_\tau :i\in A_{j'}}x_{ j'}\\
    \ge&1-\frac{1}{1+L}\left|A_j\right|\ge \frac{1}{1+L}.
\end{align*}
where the first inequality holds due to the assumption. 

\end{proof}

\begin{theorem}\label{thrm:simulate_ocrs}
    For any $\varepsilon\in (0,1)$, by taking $K=\frac{3(1+L)}{\varepsilon^2}\log\left(\frac{2TM}{\varepsilon}\right),$ it holds that $V^{\hat{\pi}}\ge\frac{(1-\varepsilon)^2}{1+\varepsilon}\frac{1}{1+L}V^*$.
\end{theorem}
\begin{proof}[Proof of Theorem \ref{thrm:simulate_ocrs}]
    
 By the union bound and Bayes rule, we have
 \begin{align*}
     &\bP\left(\frac{1}{1+\varepsilon}\le \frac{\bP^{\hat{\pi}}(F_{j})}{\hat{\bP}(F_{j})}\le \frac{1}{1-\varepsilon}, \forall j\right)\\
     =&\bP\left(\left|\bP^{\hat{\pi}}(F_{j})-\hat{\bP}(F_{j})\right|\le \varepsilon \bP^{\hat{\pi}}(F_{j}), \forall j \right)\\
     =&\prod_{t=1}^T\bP\left(\left|\bP^{\hat{\pi}}(F_{j})-\hat{\bP}(F_{j})\right|\le \varepsilon \bP^{\hat{\pi}}(F_{j}), \forall j\in N_t \middle|\left|\bP^{\hat{\pi}}(F_{j})-\hat{\bP}(F_{j})\right|\le \varepsilon \bP^{\hat{\pi}}(F_{j}), \forall (\tau<t,j\in N_\tau)\right)\\
     \ge&1-\sum_{t=1}^T \bP\left(\exists j\in N_t, \left|\bP^{\hat{\pi}}(F_{j})-\hat{\bP}(F_{j})\right|>\varepsilon \bP^{\hat{\pi}}(F_{j}) \middle|\left|\bP^{\hat{\pi}}(F_{j})-\hat{\bP}(F_{j})\right|\le \varepsilon \bP^{\hat{\pi}}(F_{ j}), \forall (\tau<t,j\in N_\tau)\right)\\
     \ge&1-\sum_{t=1}^T\sum_{j\in N_t}\bP\left(\left|\bP^{\hat{\pi}}(F_{j})-\hat{\bP}(F_{j})\right|>\varepsilon \bP^{\hat{\pi}}(F_{j})\middle|\left|\bP^{\hat{\pi}}(F_{j})-\hat{\bP}(F_{j})\right|\le \varepsilon \bP^{\hat{\pi}}(F_{j}), \forall (\tau<t,j\in N_\tau) \right)\\
     \stackrel{(a)}{\ge}&1-\sum_{t=1}^T\sum_{j\in N_t}2\bE^{\hat{\pi}}\left[\exp\left(-\frac{K}{3}\varepsilon^2 \bP^{\hat{\pi}}(F_{j})\right)\middle|\left|\bP^{\hat{\pi}}(F_{j})-\hat{\bP}(F_{ j})\right|\le \varepsilon \bP^{\hat{\pi}}(F_{ j}), \forall (\tau<t,j\in N_\tau)\right] \\
     \stackrel{(b)}{\ge}&1-2TM\exp\left(-\frac{\varepsilon^2K}{3(1+L)}\right),
 \end{align*}
 where inequality $(a)$ follows from Chernoff bound and inequality $(b)$ follows from Lemma \ref{lem:simu_ocrs_lem}. Therefore, by taking 
 \[K=\frac{3(1+L)}{\varepsilon^2}\log\left(\frac{2TM}{\varepsilon}\right), \]
 we have
 \[\bP\left(\frac{1}{1+\varepsilon}\le \frac{\bP^{\hat{\pi}}(F_{j})}{\hat{\bP}(F_{j})}\le \frac{1}{1-\varepsilon}, \forall j\right) \ge 1-\varepsilon. \]
 Thus, we have
\begin{align*}
    &V^{\hat{\pi}}=\bE^{\hat{\pi}}\left[\sum_{t=1}^T\sum_{j=1}^M r_j \Z_{j}\right]=\sum_{t=1}^T\sum_{j\in N_t}r_j\bE^{\hat{\pi}}\left[\Z_j\right]\\
    \ge &\sum_{t=1}^T\sum_{j\in N_t} r_j\bP\left(\frac{1}{1+\varepsilon}\le \frac{\bP^{\hat{\pi}}(F_{j})}{\hat{\bP}(F_{j})}\le \frac{1}{1-\varepsilon}\right)\bE^{\hat{\pi}}\left[\Z_{j}\middle|\frac{1}{1+\varepsilon}\le \frac{\bP^{\hat{\pi}}(F_{j})}{\hat{\bP}(F_{j})}\le \frac{1}{1-\varepsilon}\right]\\
    =&\alpha\sum_{t=1}^T\sum_{j\in N_t} r_jx_{j}\bP\left(\frac{1}{1+\varepsilon}\le \frac{\bP^{\hat{\pi}}(F_{j})}{\hat{\bP}(F_{j})}\le \frac{1}{1-\varepsilon}\right) \bE^{\hat{\pi}}\left[\frac{\bP^{\hat{\pi}}(F_{j})}{\hat{\bP}(F_{j})} \middle|\frac{1}{1+\varepsilon}\le \frac{\bP^{\hat{\pi}}(F_{j})}{\hat{\bP}(F_{j})}\le \frac{1}{1-\varepsilon}\right]\\
    \ge&\frac{1-\varepsilon}{(1+\varepsilon)(1+L)}\sum_{t=1}^T\sum_{j\in N_t} r_j x_{j}\bP\left(\frac{1}{1+\varepsilon}\le \frac{\bP^{\hat{\pi}}(F_{j})}{\hat{\bP}(F_{j})}\le \frac{1}{1-\varepsilon}\right)\\
    \ge&\frac{(1-\varepsilon)^2}{(1+\varepsilon)}\frac{V^*}{1+L}.
\end{align*}
\end{proof}

\subsection{Proof of \Cref{thrm:tightness}}\label{ssec:proof_of_tightness}
Assuming $L$ is a prime power, the NRM configuration in \Cref{def:worst_case_problem} exists.
In this case, take $\varepsilon < 1/L$. We first set the remaining parameters
necessary to describe an input to the accept-reject NRM problem. For each $t=1,\ldots,1+L$ and $j\in N_t$, if $t \le L$ then
set $\lambda_j=(1-\varepsilon)/L$ and $r_j=1$, else set $\lambda_j=\varepsilon$ and $r_j=1/(\varepsilon L)$.

We argue that no online algorithm can attain a competitive ratio
better than $1/(1+L)$ against the fluid LP on this input. First observe it is always possible to accept all products in the fluid relaxation. That is,
if we set $x_j=\lambda_j$ for each product $j$, then for each $i \in M$, 
\[\sum_{j:i\in A_j}x_j = \sum_{j:i\in A_j}\lambda_j=\sum_{t=1}^{1+L}\sum_{j\in N_t:i\in A_j} \lambda_j =L \cdot \frac{1-\varepsilon}{L}+\varepsilon=1, \]
where the penultimate equality holds because by (i) in \Cref{def:worst_case_problem}.
The optimal value of the fluid LP is thus equal to
\[\sum_{j}r_jx_j=\sum_{j}r_j\lambda_j=L^2\cdot \frac{1-\varepsilon}{L}+L\cdot \varepsilon \cdot \frac{1}{\varepsilon L}=1+L(1-\varepsilon). \]
Now, because of condition (ii) of \Cref{def:worst_case_problem}, it is impossible to accept more than one product.
This is because any two products of \textit{distinct} batches share an item, and there is only one copy of each item. On the other hand, any online algorithm which accepts
at most one product has an expected reward of at most $1$. To see this, observe
that if it accepts a product $j$ in one of the first $L$ batches, then $r_j =1$, so this holds.
Otherwise, it waits until the final batch, leading to an expected reward of $L \cdot \varepsilon \cdot \frac{1}{\varepsilon L} =1$. In either case, the claim holds.  By taking $\varepsilon \rightarrow 0$, this implies that no online algorithm can attain a competitive ratio better than $1/(1+L)$.  If a random-element OCRS with a larger selection guarantee existed, Theorem \ref{thrm:nrm_and_random_OCRS} would convert it into an online algorithm with the same competitive guarantee for this NRM instance, giving a contradiction. This proves the theorem.

\subsection{Proof of Lemma \ref{lem:pair_prob}} \label{pf:lem:pair_prob}

Since each element consists of at most $L$ items and $A_j\cap A_{j'}=\emptyset$, we have $\left|A_j\cup A_{j'}\right|\le 2L$. For simplicity, let $\mathcal{L}=\left|A_j\cup A_{j'}\right|\ge 2$ and assume the $\mathcal{L}$ items are indexed by $\{1,\dots,\mathcal{L}\}$ without loss of generality. We can then partition the set $\{j'':A_{j''}\cap (A_j\cup A_{j'})\neq \emptyset\}$ into $\mathcal{L}$ disjoint sets $(J_i)_{i=1}^{\mathcal{L}}$ such that
\[J_i\subseteq \{j: i\in A_j\}, \ \sum_{j\in J_i}x_{j}\le 1, \ \forall i=1,\dots,\mathcal{L}.\]
Moreover, for any element $j\in J_i$, let $t$ be the time such that $j\in N_t$, recall that $\Z_{j}$ denote the event that element $j$ is accepted and we have $\bP(\Z_j)=\alpha x_{j}$. Therefore, it holds that
\begin{align*}
    &\bP\left(F_{j}\right)=\mathbb{\bP}\left(\cap_{i'\in A_j}F_{i'}(t)\right)\ge 1-\sum_{i'\in A_j}\bP(\neg{F}_{i'}(t))\\
    \ge&1-\sum_{i'\in A_j}\sum_{\tau< t}\sum_{j'\in N_\tau:i'\in A_{j'}}\bP(\Z_{j'})\ge 1-\sum_{i'\in A_j}\sum_{t=1}^{T}\sum_{j'\in N_t:i'\in A_{j'}}\bP(\Z_{j'})+\sum_{j'\in N_\tau:i\in A_{j'}}\bP(\Z_{j'})\\\
    \ge& 1-\alpha L+\alpha\sum_{j'\in N_t:i\in A_{j'}}x_{ j'}\ge 1-\alpha L+\alpha \sum_{j'\in N_t\cap J_i}x_{j'},
\end{align*}
where the second inequality holds because the item $i'$ is available at time $t$ only if no associated element $j'$ has been accepted before, and the last  
inequality holds because $J_i\subseteq \{j: i\in A_j\}$. 

The inequality above implies that
\begin{align*}
    &1-\sum_{j''\in N_\tau:A_{j''}\cap \left(A_{j}\cup A_{j'}\right)\neq \emptyset}\frac{\alpha x_{j''}}{\bP(F_{j''})}\\
    =&1-\sum_{i=1}^{\mathcal{L}}\sum_{j''\in N_{\tau}\cap J_i} \frac{\alpha x_{j''}}{\mathbb{P}(F_{j''})}\\
    \ge& 1-\sum_{i=1}^{\mathcal{L}}\sum_{j''\in N_\tau\cap J_i}\frac{\alpha x_{j''}}{1-\alpha L +\alpha\sum\limits_{j'\in N_\tau\cap J_i}x_{ j'}}\\
    =&1-\sum_{i=1}^{\mathcal{L}}\frac{\alpha \sum\limits_{j'\in N_\tau\cap J_i}x_{j'}}{1-\alpha 
    L+\alpha\sum\limits_{j'\in N_\tau\cap J_i}x_{j'}}.
\end{align*}
For simplicity, let $y_{\tau i}=\sum_{j'\in N_\tau \cap J_i}x_{j'}$, then the term above can be written as $1-\sum\limits_{i=1}^{\mathcal{L}}\frac{\alpha y_{\tau i}}{1-\alpha L +\alpha y_{\tau i}}$. Note that the function $g(x)=\frac{\alpha x}{1-\alpha L +\alpha x}=1-\frac{1-\alpha L}{1-\alpha L +\alpha x}$ is concave when $1-\alpha L >0$. Therefore, $\sum\limits_{i=1}^{\mathcal{L}}\frac{\alpha y_{\tau i}}{1-\alpha L +\alpha y_{\tau i}}$ is concave in $(y_{\tau i})_{i=1}^{\mathcal{L}}$. By letting $y_{\tau}=\sum_{i=1}^{\mathcal{L}}y_{\tau i}\le 1$, this implies that 
\[ \sum_{i=1}^{\mathcal{L}} \frac{\alpha y_{\tau i}}{1-\alpha L +\alpha y_{\tau i}} \le \frac{\alpha \sum\limits_{i=1}^{\mathcal{L}} y_{\tau i}}{1-\alpha L + \alpha \sum\limits_{i=1}^{\mathcal{L}}y_{\tau i}/\mathcal{L}}=\frac{\alpha y_{\tau}}{1-\alpha L+ \alpha y_{\tau}/\mathcal{L}} \le \frac{\alpha }{1-\alpha L + \alpha /\mathcal{L}}\le \frac{\alpha }{1-\alpha L + \alpha/ (2L) }, \]
where the last inequality follows from the fact that $\mathcal{L}\le 2L$. By the assumption that $\alpha \le 1-\alpha L +\alpha/(2L)$, we have $1-\frac{\alpha y_{\tau}}{1-\alpha L +\alpha y_{\tau}/\mathcal{L}}\ge 0$. 

Combining discussions above, it holds that 
\[\prod_{\tau=1}^T\left(1-\sum_{j''\in N_\tau:A_{j''}\cap \left(A_{j}\cup A_{j'}\right)\neq \emptyset}\frac{\alpha x_{j''}}{\bP(F_{j''})}\right)=\prod_{\tau=1}^T\left(1-\sum_{i=1}^{\mathcal{L}}\sum_{j''\in N_\tau\cap J_i}\frac{\alpha x_{j''}}{\bP(F_{j''})}\right)\ge\prod_{\tau=1}^T\left(1-\frac{\alpha y_{\tau}}{1-\alpha L + \alpha y_{\tau}/ \mathcal{L}}\right).\]
To provide a lower bound on the term above, we consider an optimization problem. Note that 
\[\sum_{\tau=1}^T y_{\tau}=\sum_{\tau=1}^T \sum_{i=1}^{\mathcal{L}} y_{\tau i}=\sum_{\tau=1}^T \sum_{i=1}^{\mathcal{L}} \sum_{j'\in N_{\tau}\cap J_i} x_{j'}\le \sum_{i=1}^{\mathcal{L}} \sum_{\tau=1}^T \sum_{j'\in N_{\tau}: i\in A_{j'}} x_{j'}=\sum_{i=1}^{\mathcal{L}}\sum_{j': i\in A_{j'} } x_{j'}\le \mathcal{L}. \]
It is sufficient to consider the optimization problem as follows:
\begin{align*}
    &\min_{y_{\tau}\ge 0}~\prod_{\tau=1}^{T}\left(1-\frac{\alpha y_{\tau}}{1-\alpha L+\alpha y_{\tau}/\mathcal{L}}\right) ~\text{s.t.~} y_{\tau}\le 1, \forall \tau, ~ \sum_{\tau=1}^T y_{\tau}\le \mathcal{L},\\
    =&\exp\left(\min_{y_{\tau}\ge 0}~\sum_{\tau=1}^T\log \left(1-\frac{\alpha y_{\tau}}{1-\alpha L+\alpha y_{\tau}/ \mathcal{L}}\right)\right)~\text{s.t.~} y_{\tau}\le 1, \forall \tau, ~ \sum_{\tau=1}^{T}y_{\tau}\le \mathcal{L},\\
    \ge&\exp\left(\min_{y_{\tau}\ge 0}~\sum_{\tau=1}^T\log \left(1-\frac{\alpha y_{\tau}}{1-\alpha L+\alpha y_{\tau}/\mathcal{L}}\right)\right)~\text{s.t.~}y_{\tau}\le 1, \forall \tau, ~ \sum_{\tau=1}^{T} y_{\tau}\le \mathcal{L}.
\end{align*}

Therefore, we focus on
\begin{equation}\exp\left(\min_{y_\tau \ge 0}~\sum_{\tau =1}^T\log \left(1-\frac{\alpha y_{\tau}}{1-\alpha L+\alpha y_{\tau }/\mathcal{L}}\right)\right)~\text{s.t.~}y_\tau\le 1,~ \forall \tau,~\sum_{\tau=1}^{T} y_{\tau}\le \mathcal{L}.\label{eq:reduction_opt}\end{equation}
We claim that $f(x)=\log\left(1-\frac{\alpha x}{1-\alpha L+\alpha x/\mathcal{L}}\right)$ is also a concave function when $0\le \alpha \le 1/L$. Note that
\begin{align*}
    &f'(x)=\frac{1}{1-\frac{\alpha x}{1-\alpha L+\alpha x/\mathcal{L}}}\frac{-\alpha(1-\alpha L+\alpha x/\mathcal{L})+\alpha^2x/\mathcal{L}}{(1-\alpha L+\alpha x/\mathcal{L})^2}=\frac{-\alpha(1-\alpha L)}{(1-\alpha L+\alpha x/\mathcal{L})(1-\alpha L-(\mathcal{L}-1)\alpha x/\mathcal{L})},\\
    &f''(x)=\frac{\alpha^2(1-\alpha L)}{\mathcal{L}(1-\alpha L+\alpha x/\mathcal{L})^2(1-\alpha L-(\mathcal{L}-1)\alpha x/\mathcal{L})^2}\left(\left(\mathcal{L}-2\right)(\alpha L-1)-\frac{2(\mathcal{L}-1)\alpha x}{\mathcal{L}}\right)\le 0.
\end{align*}
Therefore, the optimization problem \eqref{eq:reduction_opt} is to minimize a concave function with linear constraints, thus the optimal solution is obtained at an extreme point of the feasible region. Moreover, the coefficient matrix is totally unimodular, therefore, all vertices are integral. If $T\le \mathcal{L}$, the optimal value is $\left(1-\frac{\alpha}{1-\alpha L+\alpha /\mathcal{L}}\right)^T\ge \left(1-\frac{\alpha}{1-\alpha L+\alpha/\mathcal{L}}\right)^{\mathcal{L}}$, otherwise, the optimal value is $\left(1-\frac{\alpha}{1-\alpha L+\alpha/\mathcal{L}}\right)^{\mathcal{L}}$. Therefore, we can conclude that a lower bound to Problem \eqref{eq:reduction_opt} is $\left(1-\frac{\alpha}{1-\alpha L+\alpha/\mathcal{L}}\right)^{\mathcal{L}}$. Moreover, this bound is decreasing in $\mathcal{L}$ and we have $\mathcal{L}\le 2L$, thus the result follows.

\subsection{Proof of Lemma \ref{lem:prophet_disjoint_mass}} \label{pf:lem:prophet_disjoint_mass}

Recall that in the worst case, all elements intersect with the set of items $\{1,\dots,L\}$ at most once, then we have
\begin{align*}
&\sum_{i=1}^{L}\sum_{i'\neq i}\sum_{t=1}^T\sum_{t'\neq t}\sum_{\substack{j\in N_t,j'\in N_{t'}:\\A_j\cap A_{j'}=\emptyset\\i\in A_j, i'\in A_{j'}}} x_{j}x_{j'}
=\sum_{i=1}^{L}\sum_{i'\neq i}\sum_{\substack{(j,j'):\\A_j\cap A_{j'}=\emptyset\\i\in A_j, i'\in A_{j'}}}x_{j}x_{j'}\\
=&\sum_{i=1}^L\sum_{i'\neq i}\sum_{j:i\in A_j}\sum_{j':i'\in A_{j'}}x_{j}x_{j'}\left(1-\mathds{1}\left\{A_j\cap A_{j'}\neq \emptyset\right\}\right)\\
=&\sum_{i=1}^L\sum_{i'\neq i}\left(\sum_{j:i\in A_j}x_{j}\right)\left(\sum_{j':i'\in A_{j'}}x_{j'}\right)-\sum_{i=1}^L\sum_{i'\neq i}\sum_{j:i\in A_j}\sum_{j':i'\in A_{j'}}x_{j}x_{j'}\mathds{1}\left\{A_j\cap A_{j'}\neq\emptyset\right\}\\
=&L(L-1)-\sum_{i=1}^L\sum_{i'\neq i}\sum_{j:i\in A_j}\sum_{j':i'\in A_{j'}}x_{j}x_{j'}\mathds{1}\left\{A_j\cap A_{j'}\neq\emptyset\right\}\\
\ge&L(L-1)-\sum_{i=1}^L\sum_{i'\neq i} \sum_{j:i\in A_j}\sum_{i''\in A_j\backslash \{i\}}\sum_{j':i', i''\in A_{j'}}x_{j}x_{j'}\\
=&L(L-1)-\underbrace{\sum_{i=1}^L\sum_{i''\in M\backslash\{1,\dots,L\}}\left(\sum_{j:i,i''\in A_j}x_{j}\right)\left(\sum_{i'\neq i}\sum_{j':i',i''\in A_{j'}}x_{j'}\right)}_{(a)}.
\end{align*}
where the first equality holds because $|N_t|=1$ for any $t$ in the standard OCRS model and the last equality holds because if $i\in A_j$ for $i\in \{1,\dots,L\}$, then $A_j \backslash \{i\}\cap\{1,\dots,L\}=\emptyset$. In order to upper bound term $(a)$, for simplicity, let 
\[\alpha_{ii''}=\sum_{j:i,i''\in A_j}x_{j}, \ \beta_{ii''}=\sum_{i'\neq i}\sum_{j':i',i''\in A_{j'}}x_{j'}.\]
Note that for any fixed $i\in \{1,\dots,L\}$, it holds that
\[\alpha_{ii''}+\beta_{ii''}=\sum_{j:i,i''\in A_j}x_j+\sum_{i'\neq i}\sum_{j:i',i''\in A_{j'}}x_{j'}=\sum_{i=1}^L \sum_{j:i,i''\in A_{j}}x_j\le \sum_{j:i''\in A_{j}}x_j\le 1, \forall i''\in M\backslash \{1,\dots, L\},\]
\[\sum_{i''\in M\backslash [L]}\alpha_{ii''}=\sum_{i''\in M\backslash [L]}\sum_{j:i,i''\in A_j}x_j=\sum_{j:i\in A_j}\sum_{i''\in M\backslash [L]}x_j\mathds{1}\{i''\in A_j\}\le (L-1)\sum_{j:i\in A_j}x_j\le (L-1),\]
\[\sum_{i''\in M\backslash [L]}\beta_{ii''}=\sum_{i''\in M\backslash [L]}\sum_{i'\neq i}\sum_{j:i',i''\in A_{j}}x_{j}\le \sum_{i'\neq i}(L-1)\sum_{j:i'\in A_j}x_j\le (L-1)^2.\]
Therefore, the optimization problem below provides an upper bound to term $(a)$:
\begin{equation}
\max_{K,\alpha_k,\beta_k} \sum_{k=1}^K \alpha_k\beta_k, \ \text{s.t. } \alpha_k+\beta_k\le 1, \forall k, \sum_{k=1}^K \alpha_k\le L-1, \sum_{k=1}^K \beta_k\le (L-1)^2.\label{eq:stan_ocrs_opt}\end{equation}
We claim the optimal value to Problem \eqref{eq:stan_ocrs_opt} is $(L-1)^2/L$, which is achieved at $\alpha_k=1/L$, $\beta_k=1-1/L$ for all $k$ and $K=L(L-1)$.
We first show it is sufficient to consider $K^*=L(L-1)$. Suppose $K>L(L-1)$, let $(\alpha^*_k, \beta^*_k)_k$ denote an optimal solution. Without loss of generality, assume $\alpha_1^*\ge \alpha_2^*\ge \dots \alpha_{K}^*$, then $(\beta^*_k)_k$ is optimal when $\beta^*_k$ is set as large as possible following the index order until the sum reaches $(L-1)^2$. That is, there exists an index $k^*$ where $k^*$ is the smallest number such that for any $k\ge k^*+1$, $\beta^*_k=0$ and 
\[\beta_1^*=1-\alpha^*_1, \ \beta_2^*=\min\left\{1-\alpha_2^*, (L-1)^2-\beta_1^*\right\}, \dots, \beta^*_{k^*}=\min\left\{1-\alpha_{k^*}^*, (L-1)^2-\sum_{k=1}^{k^*-1}\beta_k^*\right\}. \]
By the definition of $k^*$, it holds that
\[\beta_k^*=1-\alpha_k^*, \forall k<k^*, \ (L-1)^2-\sum_{i=1}^{k^*-1}\beta_i^*\le 1-\alpha_{k^*}^*, \]
otherwise, if $\beta_k^*\neq 1-\alpha^*_k$ for some $k<k^*$, then $\beta^*_{k+1}=0$, contradicting to the fact that $k^*$ is the smallest index. Now suppose $k^*\ge L(L-1)+1$ and $\alpha^*_{L(L-1)+1}>0$, then we have
\[\beta^*_{L(L-1)}=1-\alpha_{L(L-1)}^*\le (L-1)^2-\sum_{k=1}^{L(L-1)-1}\beta^*_k=(L-1)^2-\sum_{k=1}^{L(L-1)-1}(1-\alpha_k^*),\]
which implies that
\[(L-1)^2\ge \sum_{k=1}^{L(L-1)}(1-\alpha_k^*)=L(L-1)-\sum_{k=1}^{L(L-1)}\alpha_k^*>L(L-1)-(L-1)=(L-1)^2, \]
which leads to a contradiction. Thus, we either have $k^*\le L(L-1)$ or $\alpha^*_{L(L-1)+1}=0$. In both cases, since $\alpha_k^*\beta_k^*=0$ for any $k>L(L-1)$, it is sufficient to consider $K^*=L(L-1)$. Therefore, the optimization problem \eqref{eq:stan_ocrs_opt} can be reduced to 
\[\max_{\alpha_k,\beta_k} \sum_{k=1}^{L(L-1)} \alpha_k\beta_k,\ \text{s.t.} \ \alpha_k+\beta_k\le 1, \forall k, \sum_{k=1}^{L(L-1)}\alpha_k\le L-1, \sum_{k=1}^{L(L-1)}\beta_k\le (L-1)^2. \]
Note that it is sufficient to consider the case where all constraints are tight. Therefore, the problem is equivalent to 
\[\min_{\alpha_k}\sum_{k=1}^{L(L-1)}\alpha_k^2, \ \text{s.t. }\sum_{k=1}^{L(L-1)}\alpha_k=L-1.    \]
Since the problem is to minimize a convex function, we have $\alpha^*_k=1/L$ for any $k$. Thus, we can conclude that term $(a)$ is upper bounded by $(L-1)^2$.

\section{Additons to \Cref{sec:rcrs}}
\subsection{Proof of \Cref{lem:batch_upper_bound}}\label{ssec:pf_batch_upper_bound}
We first define a function $f$ from $\partial_{N_t}(\jc)$ to $\{1, \ldots ,L\}$.
Here, $f$ assigns each $j \in \partial_{N_t}(\jc)$ to an arbitrary $i \in \{1, \ldots , L\}$
for which $i \in A_{j}$. Observe that this is well-defined, as each $j \in \partial_{N_t}(\jc)$ satisfies $A_j \cap A_{\jc} \neq \emptyset$ by definition.
On the other hand, observe that for each $j \in \partial_{N_t}(\jc)$,
if $f(j) = i$, then $j \in f^{-1}(i) \subseteq \partial_{N_t}(j)$,
and so $x(f^{-1}(i)) \le x(\partial_{N_t}(j)) =: x_{j,t}$.
Thus, since $b$ is non-increasing, we get that
\begin{equation}
    \sum_{j \in \partial_{N_t}(\jc)} x_j b(x_{j,t}) = \sum_{i=1}^{L} \sum_{j \in N_t \cap f^{-1}(i)} x_j  b(x_{j,t}) \le \sum_{i=1}^{L} x(f^{-1}(i)) b(x(f^{-1}(i))).
\end{equation}
Now, if we focus on the term $\sum_{i=1}^{L} x(f^{-1}(i)) b(x(f^{-1}(i)))$,
then we know that $\sum_{i=1}^{L} x(f^{-1}(i)) = x_{t,\jc}$, as 
$(f^{-1}(i))_{i=1}^{L}$ partition $\partial_{N_t}(\jc)$. Since $b$ is non-increasing, one can argue via induction over $L$ that this term is maximized when $x(f^{-1}(i)) = x_{t,\jc}/L$ for each $i=1,\ldots, L$.
Thus,
$$
\sum_{j \in \partial_{N_t}(\jc)} x_j b(x_{j,t}) \le x_{t,\jc} b(x_{t,\jc}/L),
$$
and so the proof is complete.

\subsection{Proof of \eqref{eqn:schur_concave}} \label{pf:eqn:schur_concave}
Given $\bm{z} =(z_1, \ldots ,z_T) \in [0,1]^T$ with $\sum_{t=1}^T z_t = L -z_0$, let us assume that $z_2 > z_1$ w.l.o.g. We then must show that $\partial_2 \psi(\bm{z}) - \partial_1 \psi(\bm{z}) \le 0$. First observe that we may exchange the order
of partial differentiation and integration, such that
\begin{equation*}
    \partial_2 \psi(\bm{z}) = \int_{0}^{1} -y \left( b(z_{2}/L) + \frac{z_{2} b'(z_2/L)}{L} \right) \left(1-  y z_{1} b(z_{1}/L)\right) \prod_{t = 3}^{T}\left(1-  y z_{t} b(z_{t}/L)\right).
\end{equation*}
A similar expression holds for $\partial_1 \psi(\bm{z})$, and so after some algebraic simplifications,
we can write
$\partial_2 \psi(\bm{z}) - \partial_1 \psi(\bm{z})$ as:
\small
\begin{align*}
    \int_{0}^{1} \frac{y}{L} \left(\left( b(z_{1}/L) L + z_{1} b(z_1/L) \right) \left(1-  y z_{2} b'(z_{2}/L)\right)  - \left( b(z_{2}/L) L + z_{2} b(z_2/L) \right) \left(1-  y z_{1} b'(z_{1}/L)\right) \right)
    \prod_{t = 3}^{T}\left(1-  y z_{t} b(z_{t}/L)\right) dy.
\end{align*}
\normalsize
Now, the function $y \rightarrow \prod_{t = 3}^{T}\left(1-  y z_{t} b(z_{t}/L)\right) dy$ is non-negative for $y \in [0,1]$,
so we'll first focus on upper bounding the function 
\begin{equation} \label{eqn:critical_function}
f(y):= \frac{y}{L} \left(\left( b(z_{1}/L) L + z_{1} b'(z_1/L) \right) \left(1-  y z_{2} b(z_{2}/L)\right)  - \left( b(z_{2}/L) L + z_{2} b'(z_2/L) \right) \left(1-  y z_{1} b(z_{1}/L)\right) \right).
\end{equation}
Unfortunately, $f$ is positive for certain values of $y$. However, since $z_2 > z_1$, it has a single non-zero root
on $[0,1]$ at
\begin{equation*}
    y_c :=  \frac{-L b(z_1/L) + Lb(z_2/L) - z_1 b'(z_1/L) + z_2 b'(z_2/L)}{L z_1 b(z_1/L) b(z_2/L) - L z_2 b(z_1/L)b(z_2/L)-z_1 z_2 b(z_2/L) b'(z_1/L) + z_1 z_2 b(z_1/L) b'(z_2/L)}.
\end{equation*}
Specifically, $f(y) \ge 0$ for $y \in [0,y_c]$ and $f(y) \le 0$ for $y \in [y_c, 1]$. On the other hand, since
$b$ is decreasing,  we can apply elementary inequalites to get that
\begin{equation} \label{eqn:upper_bound_integral}
          \prod_{t = 3}^{T}\left(1-  y z_{t} b(z_{t}/L)\right) \le e^{- y (L - z_0 - z_1 - z_2) b(1)}.   
\end{equation}
Similarly, since $b(z) \le 1$, $\sum_{t=3}^{T} z_t = L - z_1 - z_2 - z_0$ and $z_t \le 1$, we 
have that
\begin{equation} \label{eqn:lower_bound_integral}
     \prod_{t = 3}^{T}\left(1-  y z_{t} b(z_{t}/L)\right) \ge \left(1-  y\right)^{L - z_1 - z_2 - z_0}.
\end{equation}
Thus, by applying \eqref{eqn:upper_bound_integral} on $[0,y_c]$ and \eqref{eqn:lower_bound_integral} on $(y_c,1]$,
we can lower bound $\partial_2 \psi(\bm{z}) - \partial_1 \psi(\bm{z})$ by
\begin{equation} \label{eqn:reduced_integral}
    \int_{0}^{y_c} f(y) e^{- y (L - z_0 - z_1 - z_2) b(1)} dy + \int_{y_c}^1 f(y) \left(1-  y\right)^{L - z_1 - z_2 - z_0} dy.
\end{equation}
Now, \eqref{eqn:reduced_integral} has a closed-form expression, which is non-positive for any $z_0 \in [0,1]$
and $z_2 > z_1$. The proof is thus complete.

\subsection{Proof of \Cref{thm:random_element_positive}}\label{ssec:proof_random_element_pos}
Fix $\jc \in N$, where we again assume w.l.o.g. that $\jc \in N_0$. By applying \Cref{lem:minimum_attained} to \eqref{eqn:accepted_lower_bound}, we know that
\begin{equation} \label{eqn:final_accepted_lower_bound}
\mb{P}(Z_{\jc}(1) \mid X_{\jc} = 1) \ge b(x_{0,\jc})\int_{0}^{1}\left(1- y(1- x_{0,\jc})b\left(\frac{1- x_{0,\jc}}{L} \right) \right) (1 -yb(1/L))^{L-1}) dy. 
\end{equation}
Now, the left-hand side of \eqref{eqn:final_accepted_lower_bound} has a closed-form expression which
is minimized when $x_{0,\jc} = 0$ if $L=2$ and $x_{0,\jc} =1$ if $L \ge 3$. In the former case,
we get that
$$
\mb{P}(Z_{\jc}(1) \mid X_{\jc} = 1) \ge \frac{3 - 6 e^{1/2} + e + 8 e^{3/2} + 21 e^2 + 14 e^{5/2} + 
 7 e^3}{16 e (1 + e^{1/2} + e)^2} \ge 0.397,
$$
and in the latter case, we get that
$$
\mb{P}(Z_{\jc}(1) \mid X_{\jc} = 1) \ge \frac{(e^L - e^{1/L}) \left(1- \left(1 + \frac{(e^L -1) (L^2 -1)}{(e^{1/L} - e^L) L^2} \right)^L  \right)}{(e^L - e)(1+L)}.
$$
Since the function $L \rightarrow \frac{(e^L - e^{1/L}) \left(1- \left(1 + \frac{(e^L -1) (L^2 -1)}{(e^{1/L} - e^L) L^2} \right)^L \right)}{(e^L - e)}$ is decreasing for $L \ge 3$, in both cases we beat $1/(1+L)$.

\subsection{Proof of \Cref{thm:rcrs_standard}}\label{ssec:pf_rcrs_standard}
Our proof relies on the two results below. We prove Lemma \ref{lem:base_case} in Appendix \ref{pf:lem:base_case} and Lemma \ref{lem:inductive_step} in Appendix \ref{pf:lem:inductive_step}.
\begin{lemma}[Base case] \label{lem:base_case}
For $q=0$, \eqref{eqn:rcrs_induction_hyp} holds.
\end{lemma}
\begin{lemma}[Inductive Step] \label{lem:inductive_step}
Fix $1 \le k \le K-1$. Suppose that \eqref{eqn:rcrs_induction_hyp} holds for all $0 \le q \le k -1$.
Then \eqref{eqn:rcrs_induction_hyp} holds for $k$.
\end{lemma}

Fix $j \in N$. Due Lemmas \ref{lem:base_case} and \ref{lem:inductive_step}, we have that for each $y \in (0,1]$,
\begin{equation*}
   \mb{P}(Z_{j}(y) =1 \mid Y_j = y, X_j =1) \ge  c(y)\left(1 - \frac{L}{K c(1)} \right).
\end{equation*}
Since $Y_j$ is distributed u.a.r., the result follows after integrating over $y \in [0,1]$.

\subsection{Proof of \Cref{lem:base_case}} \label{pf:lem:base_case}

Fix $y \in (0,1/K]$. Observe that $\hF_{j}(0)=1$, so since $c(y) \le 1$, we have that $\mb{P}(B_j \mid Y_j =1) = c(y)$, which implies
that
\begin{equation}
    \mb{P}(Z_{j}(y) \mid Y_j = y, X_j =1) = c(y) \mb{P}(\cap_{i \in A_j} F_{i}(y) \mid Y_j = y, X_j =1) \le c(y).
\end{equation}
The upper bound of the induction hypothesis \eqref{eqn:rcrs_induction_hyp} thus holds.
In order to verify the lower bound and complete the proof, it suffices to argue that 
\begin{equation} \label{eqn:lower_bound_induction}
\mb{P}(\cap_{i \in A_j} F_{i}(y) \mid Y_j = y, X_j =1) \ge \left(1 - \frac{L}{K c(1)} \right).
\end{equation}
We instead upper bound $\mb{P}(\cup_{i \in A_j} \neg F_{i}(y) \mid Y_j = y, X_j =1)$, with explanations following afterwards:
\begin{align*}
    \mb{P}(\cup_{i \in A_j} \neg F_{i}(y) \mid Y_j = y, X_j =1) &\le \sum_{i \in A_j} \mb{P}(\neg F_{i}(y) \mid Y_j = y, X_j =1) \\
        &= \sum_{i \in A_j} \sum_{\substack{j' \neq j: \\ i \in A_{j'}}} \mb{P}(Z_{j'}(y) \mid Y_j =y, X_j =1) \\
        &\le \sum_{i \in A_j} \sum_{\substack{j' \neq j: \\ i \in A_{j'}}} y x_{j'} \le \frac{L}{K} \le \frac{L}{K c(1)}.
\end{align*}
Here the first inequality applies the union bound, the second equality uses that $\neg F_{i}(y)$ occurs  if and only if there exists some $j' \in N \setminus \{j\}$ with $i \in A_{j'}$ for which $Z_{j'}(y)$ occurs. The third inequality uses that $Z_{j'}(y)$ occurs only if $Y_{j'} < y$ and $X_{j'}=1$, and the final inequalities follow via from the constraints \eqref{eqn:feasInExpn} and that $c(1) \le 1$. Thus, \eqref{eqn:lower_bound_induction} holds, and so the proof is complete.

\subsection{Proof of \Cref{lem:continuous_induction}}\label{ssec:pf_continuous_induction}
Let us fix an element $j_0 \in N$, whose items we denote by $A_{j_0} = \{1, \ldots ,L\}$ for simplicity. Our goal is then to lower bound $\mb{P}(\cap_{i=1}^{L} F_{i}(y_k) \mid Y_{j_0} > y_k)$. We first argue that we can remove the conditioning on $Y_{j_0} > y_k$,
and instead focus on lower bounding $\mb{P}(\cap_{i =1}^{L} F_{i}(y_k))$. As discussed in
Section \ref{ssec:rcrs_standard}, this step of the proof only works because we are in the standard RCRS setting. We provide a proof in \Cref{pf:prop:rcrs_coupling}.
\begin{proposition} \label{prop:rcrs_coupling}
$\mb{P}(\cap_{i=1}^L F_{i}(y_k) \mid Y_{j_0} > y_k) \ge \mb{P}(\cap_{i=1}^L F_{i}(y_k))$.
\end{proposition}

We next apply the same simplifying assumption that we used when we lower bounded \eqref{eqn:de_morgan_law} of Section \ref{ssec:beating_standard}.
Specifically, for the analogous term $\mb{P}(\cap_{i =1}^L F_{i}(y_k))$, the worst-case input with respect to minimization occurs when the constraints \eqref{eqn:feasInExpn} on the items $\{1, \ldots , L\}$ of $\jc$ are tight; that is,
\begin{align} 
    \sum_{j: i\in A_{j}} x_{j}&=1, &\forall i\in \{1, \ldots , L\}.\label{eq:worst_capacity_bound_random_order}
\end{align}

We are now ready to lower bound $\mb{P}(\cap_{i =1}^{L} F_{i}(y_k))$, where we derive a similar
sequence of inequalities as done for \eqref{eqn:de_morgan_law} of Section \ref{ssec:beating_standard}, yet adjusted to random-order arrivals. Specifically, we lower bound $\bP\left(\cap_{i=1}^L F_{i}(y_k)\right)$ in the following way, with explanations following afterwards:
\begin{align*}
    \ge& 1-\sum_{i=1}^L \bP\left(\neg F_{i}(y_k)\right)+\max_{i}\sum_{i'\neq i}\bP\left(\neg F_{i}(y_k)\cap \neg F_{i'}(y_k)\right)\\
    \ge&1-\sum_{i=1}^L\bP\left(\neg F_{i}(y_k)\right)+\frac{1}{L}\sum_{i=1}^L\sum_{i'\neq i}\bP\left(\neg F_{i}(y_k)\cap \neg F_{i'}(y_k)\right)\\
    \ge&1- L \int_{0}^{y_k} c(z) dz +\frac{1}{L}\sum_{\substack{i,i' \in [L]: \\ i \neq i'}}\bP\left(\neg F_{i}(y_k)\cap \neg F_{i'}(y_k)\right)\\    
    \ge&1- L \int_{0}^{y_k} c(z) dz +\frac{1}{L}\sum_{\substack{i,i': \\ i \neq i'}}\left(\sum_{j:\{i,i' \}\subseteq A_j}\bP(\Z_j(y_k))+ \sum_{\substack{j, j' \in N:\\A_j\cap A_{j'}=\emptyset\\ A_j\cap[L]=i,  A_{j'}\cap [L]=i'}}\bP\left(\Z_j(y_k) \cap \Z_{j'}(y_k)\right)\right)\\
    \ge&1-L \int_{0}^{y_k} c(z) dz +\frac{1}{L} \sum_{\substack{i,i': \\ i \neq i'}}\left(\sum_{j:\{i,i'\}\subseteq A_j}x_j \int_{0}^{y_k} \left(1- \frac{L}{Kc(1)} \right) c(z) dz +\sum_{\substack{j, j' \in N:\\A_j\cap A_{j'}=\emptyset\\ A_j\cap[L]=i,  A_{j'}\cap [L]=i'}}\bP\left(\Z_j(y_k) \cap \Z_{j'}(y_k)\right)\right)
\end{align*}
The first inequality follows by inclusion-exclusion, the second by an averaging argument,
and the third by the upper bound of the induction hypothesis \eqref{eqn:rcrs_induction_hyp}. The fourth inequality holds by considering a subset of the events in which $\neg F_{i}(y_k) \cap\neg F_{i'}(y_k)$ holds,
and the final inequality applies the lower bound of the induction hypothesis \eqref{eqn:rcrs_induction_hyp}.

We next state the analog of \Cref{lem:pair_prob} for random-order arrivals. The proof of
\Cref{lem:pair_prob_random} appears in \Cref{pf:lem:pair_prob_random}.
\begin{lemma} \label{lem:pair_prob_random}
    For any $j,j' \in N$ with $A_{j} \cap A_{j'} = \emptyset$,
    \begin{equation*}
        \mb{P}(Z_{j}(y_k) \cap Z_{j'}(y_k)) \ge \left(\int_{0}^{y_k} c(z) (1-z)^L dz \right)^2 x_{j} x_{j'}.
    \end{equation*}
\end{lemma}
Observe now that after applying \Cref{lem:pair_prob_random}, we get that 
$$
\sum_{\substack{j, j' \in N:\\A_j\cap A_{j'}=\emptyset\\ A_j\cap[L]=i,  A_{j'}\cap [L]=i'}}\bP\left(\Z_j(y_k) \cap \Z_{j'}(y_k)\right) \ge \left(\int_{0}^{y_k} c(z) (1-z)^L dz \right)^2 \sum_{\substack{j, j' \in N:\\A_j\cap A_{j'}=\emptyset\\ A_j\cap[L]=i,  A_{j'}\cap [L]=i'}} x_{j} x_{j'}.
$$
Thus, we are left with analyzing
\begin{equation} \label{eqn:minimize_pair_rcrs}
\sum_{j:\{i,i'\}\subseteq A_j}x_j \int_{0}^{y_k} \left(1-  \frac{L}{c(1)K} \right)c(z) dz + \left(\int_{0}^{y_k} c(z) (1-z)^L dz \right)^2 \sum_{\substack{j, j' \in N:\\A_j\cap A_{j'}=\emptyset\\ A_j\cap[L]=i,  A_{j'}\cap [L]=i'}} x_{j} x_{j'} 
\end{equation}
Now, since $K \ge 2L/c(1)$, $c(z) \le 1$ and $y_k \le 1$, 
$$
1-  \frac{L}{c(1)K}\ge \frac{1}{1+L} = \int_{0}^{1}(1-z)^L dz  \ge \int_{0}^{y_k} c(z) (1-z)^L dz,
$$
and so,
$$
\int_{0}^{y_k} \left(1-  \frac{L}{c(1)K} \right)c(z) dz \ge \left(\int_{0}^{y_k} c(z) (1-z)^L dz \right)^2.
$$
Thus, in order to minimize \eqref{eqn:minimize_pair_rcrs},  one should set $x_{j} = 0$ for each $j \in N$
with $|A_{j} \cap \{1, \ldots ,L\}| \ge 2$. Combined with the simplifying assumption we made in \eqref{eq:worst_capacity_bound_random_order}, this leaves us with a lower bound on
$\mb{P}(\cap_{i=1}^L F_{i}(y_k))$
of
$$
1-L \int_{0}^{y_k} c(z) dz +\frac{1}{L} \left( \int_{0}^{y_k} c(z) (1-z)^L dz \right)^2 \sum_{\substack{i,i': \\ i \neq i'}}\sum_{\substack{j, j' \in N:\\A_j\cap A_{j'}=\emptyset\\ A_j\cap[L]=i,  A_{j'}\cap [L]=i'}} x_j x_{j'},
$$
subject to the constraints $\sum_{j:i\in A_j}x_j=1$ for any $i\in [L]$ and $|A_j\cap [L]|\le 1$ for any element $j$.
Now, the right-most term is precisely the term of \eqref{eqn:club_suit_proof}. Thus, we can lower bound this by $L-1$ via \Cref{lem:prophet_disjoint_mass}. By applying \Cref{prop:rcrs_coupling}, we
get that
\begin{align*}
    \mb{P}(\cap_{i=1}^L F_{i}(y_k) \mid Y_{j_0} > y_k) \ge 1-L \int_{0}^{y_k} c(z) dz +\frac{(L-1)}{L} \left( \int_{0}^{y_k} c(z) (1-z)^L dz \right)^2 \ge c(y_k),
\end{align*}
where the second inequality follows since $c$ is a selection function (see \eqref{eqn:selection_function}). The proof of \Cref{lem:continuous_induction} is therefore complete.

\subsection{Proof of \Cref{lem:inductive_step}} \label{pf:lem:inductive_step}

Fix $y \in (y_{k}, y_{k+1}]$ and $j \in N$. Our goal is to show that
\begin{equation} \label{eqn:goal_inductive_step}
    c(y)\left(1 - \frac{L}{K c(1)} \right) \le \mb{P}(Z_{j}(y) =1 \mid Y_j = y, X_j =1) \le c(y). 
\end{equation}
First observe that conditional on $Y_j =y$ and $X_{j} =1$, $Z_{j}(y)$ occurs if and only if
$B_j =1$ and $\cap_{i \in A_j} F_{i}(y)$ both occur. Thus,
\begin{equation}
    \mb{P}(Z_{j}(y) \mid Y_j = y, X_j =1) = \mb{P}(B_{j}=1 \mid Y_{j} = y) \cdot \mb{P}(\cap_{i \in A_{j}} F_{i}(y) \mid Y_j > y).
\end{equation}
On the other hand, by \Cref{lem:continuous_induction}, we know that
$\mb{P}(\cap_{i \in A_{j}} F_{i}(y_k) \mid Y_{j} > y_k ) \ge c(y_k)$.
Thus, since $c$ is decreasing, $c(y) \le c(y_k)$, and so
$
\frac{c(y)}{\hF_{j}(q)} = \frac{c(y)}{\mb{P}(\cap_{i \in A_{j}} F_{i}(y_k) \mid Y_{j} > y_k )} \le 1.
$
Recalling the definition of $B_j$ in \Cref{alg:recursive_rcrs}, we get that
\begin{equation}
    \mb{P}(B_{j}=1 \mid Y_{j} = y) = \frac{c(y)}{\mb{P}(\cap_{i \in A_{j}} F_{i}(y_k) \mid Y_{j} > y_k )}.
\end{equation}
Thus, in order to prove \eqref{eqn:goal_inductive_step}, it suffices to show that    
\begin{equation} \label{eqn:small_change_in_arrival}
       \left(1 - \frac{L}{K c(1)} \right) \le \frac{\mb{P}(\cap_{i \in A_{j}} F_{i}(y) \mid Y_j > y)}{\mb{P}(\cap_{i \in A_{j}} F_{i}(y_k) \mid Y_{j} > y_k )} \le 1.
\end{equation}
First observe that since $y > y_k$,
\begin{align}
\mb{P}(\cup_{i \in A_{j}} \neg F_{i}(y) \mid Y_j > y) =& \mb{P}(\cup_{i \in A_{j}} \neg F_{i}(y_k) \mid Y_j > y) + \mb{P}(\cup_{i \in A_{j}} \neg F_{i}(y) \setminus \cap_{i \in A_{j}} F_{i}(y_k) \mid Y_j > y) \nonumber \\
=& \mb{P}(\cup_{i \in A_{j}} \neg F_{i}(y_k) \mid Y_j > y_k) + \mb{P}(\cup_{i \in A_{j}} \neg F_{i}(y) \setminus \cap_{i \in A_{j}} F_{i}(y_k) \mid Y_j > y), \label{align:small_change_in_arrival} 
\end{align}
where the second step changes the conditioning on the left-most term of \eqref{align:small_change_in_arrival} from $Y_j > y$ to $Y_j > y_k$.
By applying the trivial lower bound of $0$ to the right-most term of \eqref{align:small_change_in_arrival}, we get that
$\mb{P}(\cup_{i \in A_{j}} \neg F_{i}(y) \mid Y_j > y) \ge \mb{P}(\cup_{i \in A_{j}} \neg F_{i}(y_k) \mid Y_j > y_k)$,
and so after taking complements and dividing by $\mb{P}(\cap_{i \in A_{j}} F_{i}(y_k) \mid Y_{j} > y_k )$, the upper bound of \eqref{eqn:goal_inductive_step} follows.

In order to prove the lower bound of \eqref{eqn:goal_inductive_step}, we first upper bound $\mb{P}(\cup_{i \in A_{j}} \neg F_{i}(y) \setminus \cap_{i \in A_{j}} F_{i}(y_k) \mid Y_j > y)$ by $L(y_{k+1} - y_k) = L/k$. This follows in
the same way as \eqref{eqn:lower_bound_induction} in the proof of \Cref{lem:base_case}, so we omit the details. Observe then
that 
$$
\mb{P}(\cup_{i \in A_{j}} \neg F_{i}(y) \mid Y_j > y) \le \mb{P}(\cup_{i \in A_{j}} \neg F_{i}(y_k) \mid Y_j > y_k) + \frac{L}{K},
$$
and so after taking complements and dividing by $\mb{P}(\cap_{i \in A_{j}} F_{i}(y_k) \mid Y_{j} > y_k )$,
$$
\frac{\mb{P}(\cap_{i \in A_{j}} F_{i}(y) \mid Y_j > y)}{\mb{P}(\cap_{i \in A_{j}} F_{i}(y_k) \mid Y_{j} > y_k )} \ge 1 - \frac{L}{K \cdot \mb{P}(\cap_{i \in A_{j}} F_{i}(y_k) \mid Y_{j} > y_k )}.
$$
Now, by \Cref{lem:continuous_induction}, we know that $\mb{P}(\cap_{i \in A_{j}} F_{i}(y_k) \mid Y_{j} > y_k ) \ge c(y_k) \ge c(1)$, where the second inequality uses that $c$ is decreasing. Thus, the lower bound of \eqref{eqn:goal_inductive_step} also holds, and so the proof is complete.

\subsection{Proof of \Cref{prop:rcrs_coupling}} \label{pf:prop:rcrs_coupling}
   It will be easier to instead prove that
\begin{equation}\label{eqn:increasing_matching_prob}
 \mb{P}(\cup_{i=1}^L \neg F_{i}(y_k)) \ge \mb{P}(\cup_{i=1}^L \neg F_{i}(y_k) \mid Y_{j_0} > y_k).
\end{equation}
Consider two executions of \Cref{alg:recursive_rcrs} for elements which arrive before time $y_k$. The first is the \textit{regular} execution on $N$ with random variables $(B_j, Y_j, X_j)_{ j \in N}$. Observe then that the left-hand side of \eqref{eqn:increasing_matching_prob}
is the probability an item of $\{1, \ldots ,L\}$ is sold before time $y_k$.
The second is the \textit{parallel} execution of the RCRS on $N \setminus \{\jc\}$ with the random variables $(B_j, Y_j, X_j)_{j \in N \setminus \{\jc\}}$. Observe that the probability an item of $\{1, \ldots , L\}$ is sold before time $y_k$ in the parallel execution is precisely the right-hand side of \eqref{eqn:increasing_matching_prob}. More, if an item of $\{1, \ldots ,L\}$ is sold in the parallel execution, then it is also sold in the regular execution. Thus, \eqref{eqn:increasing_matching_prob} follows after taking expectations over the random variables.

\subsection{Proof of \Cref{lem:pair_prob_random}} \label{pf:lem:pair_prob_random}
First recall for an arbitrary element $j \in N$ the definition of random bit $B_{j}$ of \Cref{alg:recursive_rcrs}. Specifically, conditional on $Y_{j} = y_j$, $B_j$ is distributed as
a Bernoulli of parameter $\min \left( \frac{c(y_j)}{\hF_{j}(k)}, 1\right)$,
where $\hF_{j}(k):=\mb{P}(\cap_{i \in A_{j}} F_{i}(y_k) \mid Y_j > y_k )$ since $k \ge 1$ by assumption.
Thus, since $c(y_j) \le 1$, we have that
\begin{equation} \label{eqn:bounds_on_bits}
    c(y_j) x_j \le \mb{P}(B_j X_j =1 \mid Y_j = y_j) \le x_j
\end{equation}
Let us say that $j$ \textit{survives}, provided $B_j X_j =1$. Otherwise, we say that $j$ \textit{dies}. Consider now the following three events: 
\begin{enumerate}
    \item[\mylabel{item:event_1}{(i)}] Both $j$ and $j'$ survive, and $\max\{Y_j, Y_{j'}\} < y_k$.
    \item[\mylabel{item:event_2}{(ii)}] Each $j'' \in N \setminus \{j\}$ with $A_{j''} \cap A_{j} \neq \emptyset$ either dies, or has $Y_{j''} > Y_j$.
    \item[\mylabel{item:event_3}{(iii)}] Each $j'' \in N \setminus \{j'\}$ with $A_{j''} \cap A_{j'} \neq \emptyset$ either dies, or has $Y_{j''} > Y_{j'}$.
\end{enumerate}
Observe that if all three events occur, then  $Z_{j}(y_k) \cap Z_{j'}(y_k)$ will occur. Thus, we focus on lower bounding their joint probability, conditional on $Y_{j} = y_j$ and $Y_{j'} = y_{j'}$
for $y_{j}, y_{j'} < y_k$. By starting with \ref{item:event_1} and applying \eqref{eqn:bounds_on_bits}, we first get a lower bound of 
\begin{equation} \label{eqn:first_event_lower_bound}
    c(y_j) c(y_{j'})  x_{j} x_{j'}. 
\end{equation}
Now, to handle \ref{item:event_2}, observe that by applying \eqref{eqn:bounds_on_bits}, we get a lower bound
of $\prod_{\substack{j'' \in N: \\ A_{j''} \cap A_{j} \neq \emptyset}} \left(1- x_{j''} y_{j} \right)$,
subject to $\sum_{\substack{j'' \in N \setminus \{j\}: \\ A_{j''} \cap A_j \neq \emptyset}} x_{j''} \le L(1-x_{j})$.  Due to this constraint, it is easy to see that this element is lower bounded by $(1- y_j)^{L}$. The same argument applies to \ref{item:event_3}, leading to an analogous lower bound of $(1- y_{j'})^{L}$. Finally, 
it is easy to see that conditional on $Y_{j} = y_j$ and $Y_{j'} = y_{j'}$, these events are positively correlated. Thus, combined with \eqref{eqn:first_event_lower_bound}, we get that 
\begin{equation}
    \mb{P}(Z_{j}(y_k) \cap Z_{j'}(y_k) \mid Y_{j} = y_j, Y_{j'} = y_{j'}) \ge  c(y_j) c(y_{j'}) (1- y_j)^L (1- y_{j'})^L x_{j} x_{j'}.
\end{equation}
Since $Y_j$ and $Y_{j'}$ are independent, we can integrate over each separately to complete the proof.

\section{Random-element OCRS with an $L$-partite Graph}\label{sec:Lpartite}

Theorem \ref{thrm:prophet_guarantee} shows that $1/(1+L)$ is beatable for any value of $L$ under the standard OCRS,
and so combined with \Cref{thrm:tightness}, we have proven a separation between standard OCRS and random-element OCRS when $L$ is a prime power. We now show that if the underlying graph has some structural properties, then $1/(1+L)$ is beatable even for random-element OCRS. We focus on the case where the elements and items form an $L$-partite hypergraph. Specifically, the set of items can be partitioned into $L$ disjoint subsets, such that every element contains at most one item from each subset. 

\begin{definition}[$L$-partite Hypergraph] \label{def:LpartiteHypergraph}
We say that the feasibility structure forms an \textit{$L$-partite hypergraph} if the item set $M$ can be partitioned into $M_1\cup\cdots\cup M_L$ such that $|A_j\cap M_{\ell}|\le 1$ for all elements $j\in N$ and $\ell=1,\ldots,L$.
Put in words, the items can be divided into $L$ groups such that each element contains at most one item from each group.
\end{definition}
Hypergraphs of this form have been widely studied in NRM. For example, in the assemble-to-order system, all products are assembled from a set of components so that different combinations of items for each component lead to different products. Without loss of generality, we assume each element $j$ is consists of $L$ items with exactly one item from each $M_i$. If there exists a element which contains less than $L$ items, we can add a dummy item to the group which is consumed by this element. 
We now argue that $1/(1+L)$ is beatable in this setting. As in the case of standard OCRS inputs, 
it suffices to lower bound \eqref{eqn:club_suit_proof}.
\begin{lemma}\label{lem:partite_disjoint_mass}
For an $L$-partite hypergraph, it holds that $\eqref{eqn:club_suit_proof}\ge 1$.
\end{lemma}
\begin{proof}[Proof of Lemma \ref{lem:partite_disjoint_mass}]
    Analogous to the proof of Lemma \ref{lem:prophet_disjoint_mass}, we have
\begin{align*}
&\sum_{i=1}^{L}\sum_{i'\neq i}\sum_{t=1}^T\sum_{t'\neq t}\sum_{\substack{j\in N_t,j'\in N_{t'}\\A_j\cap A_{j'}=\emptyset\\i\in A_j, i'\in A_{j'}}} x_{j}x_{j'}\\
=&\sum_{i=1}^{L}\sum_{i'\neq i}\sum_{t=1}^T\sum_{t'=1}^T\sum_{\substack{j\in N_t,j'\in N_{t'}\\A_j\cap A_{j'}=\emptyset\\i\in A_j, i'\in A_{j'}}} x_{j}x_{j'}-\sum_{i=1}^{L}\sum_{i'\neq i}\sum_{t=1}^T\sum_{\substack{j\in N_t,j'\in N_{t}\\A_j\cap A_{j'}=\emptyset\\i\in A_j, i'\in A_{j'}}} x_{j}x_{j'}\\
=&\sum_{i=1}^{L}\sum_{i'\neq i}\sum_{\substack{(j,j'):\\A_j\cap A_{j'}=\emptyset\\i\in A_j, i'\in A_{j'}}}\left(\left(\sum_{t=1}^Tx_{j}\mathds{1}\{j\in N_t\}\right)\left(\sum_{t=1}^Tx_{j'}\mathds{1}\{j'\in N_t\}\right)-\sum_{t=1}^Tx_{j}x_{j'}\mathds{1}\{j,j'\in N_t\}\right)\\
=&\sum_{i=1}^L\sum_{i'\neq i}\sum_{\substack{(j,j'):\\ A_j\cap A_{j'}=\emptyset\\i\in A_j, i'\in A_{j'}}}\left(x_j x_{j'}-\sum_{t=1}^Tx_jx_{j'}\mathds{1}\{j,j'\in N_t\}\right)
\end{align*}
where the final equality holds because the $N_t$'s are disjoint across time $t$.  Continuing this derivation, the final expression equals
\begin{align*}
&\sum_{i=1}^L\sum_{i'\neq i}\sum_{j:i\in A_j}\sum_{j':i'\in A_{j'}}\left(x_j x_{j'}-\sum_{t=1}^Tx_jx_{j'}\mathds{1}\{j,j'\in N_t\}\right)\left(1-\mathds{1}\left\{A_j\cap A_{j'}\neq \emptyset\right\}\right)\\
\ge&\sum_{i=1}^L\sum_{i'\neq i}\sum_{j:i\in A_j}\sum_{j':i'\in A_{j'}}\left(x_j x_{j'}-\sum_{t=1}^Tx_jx_{j'}\mathds{1}\{j,j'\in N_t\}-x_jx_{j'}\mathds{1}\left\{A_j\cap A_{j'}\neq\emptyset\right\}\right)\\
=&\sum_{i=1}^L\sum_{i'\neq i}\left(\sum_{j:i\in A_j}x_{j}\right)\left(\sum_{j':i'\in A_{j'}}x_{j'}\right)-\sum_{i=1}^L\sum_{i'\neq i}\sum_{t=1}^T\left(\sum_{j\in N_t:i\in A_j}x_{j}\right)\left(\sum_{j'\in N_t:i'\in A_{j'}}x_{j'}\right)\\
&-\sum_{i=1}^L\sum_{i'\neq i}\sum_{j:i\in A_j}\sum_{j':i'\in A_{j'}}x_{j}x_{j'}\mathds{1}\left\{A_j\cap A_{j'}\neq \emptyset\right\}\\
=&L(L-1)-\sum_{i=1}^L\sum_{i'\neq i}\sum_{t=1}^T\left(\sum_{j\in N_t:i\in A_j}x_{j}\right)\left(\sum_{j'\in N_t:i'\in A_{j'}}x_{j'}\right)-\sum_{i=1}^L\sum_{i'\neq i}\sum_{j:i\in A_j}\sum_{j':i'\in A_{j'}}x_{j} x_{j'}\mathds{1}\left\{A_j\cap A_{j'}\neq\emptyset\right\} \\
\ge&L(L-1)-\sum_{i=1}^L\sum_{i'\neq i}\sum_{t=1}^T\left(\sum_{j\in N_t:i\in A_j}x_{j}\right)\left(\sum_{j'\in N_t:i'\in A_{j'}}x_{j'}\right)-\sum_{i=1}^L\sum_{i'\neq i} \sum_{j:i\in A_j}\sum_{i''\in A_j\backslash \{i\}}\sum_{j':i', i''\in A_{j'}}x_{j}x_{j'}
\end{align*}
\begin{align*}
=&L(L-1)-\underbrace{\sum_{i=1}^L\sum_{i'\neq i}\sum_{t=1}^T\left(\sum_{j\in N_t:i\in A_j}x_{j}\right)\left(\sum_{j'\in N_t:i'\in A_{j'}}x_{j'}\right)}_{(a)}\\
&-\underbrace{\sum_{i=1}^L\sum_{i''\in M\backslash [L]\cup M_i}\left(\sum_{j:i,i''\in A_j}x_{j}\right)\left(\sum_{\substack{i'\neq i:\\i'\notin M_{\ell} \text{ if } i''\in M_{\ell}}}\sum_{j':i',i''\in A_{j'}}x_{j'}\right)}_{(b)},
\end{align*}
where the final equality holds because all elements intersect with items $\{1,\dots,L\}$ at most once and every element has exactly one item from the set $M_{\ell}$. We now analyze the two terms $(a)$ and $(b)$ separately. 

    For term $(a)$, let $y_{ti}=\sum_{j\in N_t:i\in A_j}x_{j}$ for simplicity, then for any $i\in \{1,\dots,L\}$, it holds that
    \[\sum_{t=1}^Ty_{ti}=\sum_{t=1}^T\sum_{j\in N_t: i\in A_j}x_j=\sum_{j:i\in A_j}x_j= 1, \]
    \[\sum_{t=1}^T\sum_{i'\neq i}y_{ti'}=\sum_{t=1}^T\sum_{i'\neq i}\sum_{j\in N_t:i'\in A_j}x_{j}=\sum_{i'\neq i}\sum_{j: i'\in A_j}x_j=L-1, \]
    \[\sum_{i=1}^L y_{ti}=\sum_{i=1}^L \sum_{j\in N_t:i\in A_j}x_j\le \sum_{j\in N_t}x_j\le 1, \forall t.\]
    To upper bound term $(a)$, for any fixed $i$, we consider the optimization problem:
    \[\max \sum_{t=1}^Ty_{ti}\left(\sum_{i'\neq i}y_{ti'}\right), \ \text{s.t.} \sum_{t=1}^Ty_{ti}\le 1, \sum_{t=1}^T\sum_{i'\neq i}y_{ti'}\le (L-1), \sum_{i=1}^L y_{ti}\le 1, \forall t. \]
    Similar to the proof of Lemma \ref{lem:prophet_disjoint_mass}, it is sufficient to consider $T=L$, the optimal value is $(L-1)/L$ achieved at $y_{ti}=1/L$. Thus, it follows that term $(a)$ is upper bounded by $L-1$.
    
    For term $(b)$, note that for any fixed $i$, since every element $j$ has exactly one item from the set $M_{\ell}$, it holds that
    \[\sum_{i''\in M\backslash [L]\cup M_i}\sum_{j:i,i''\in A_j} x_{j} =\sum_{\ell\neq i}\sum_{i''\in M_{\ell}}\sum_{j:i,i''\in A_j}x_j=\sum_{\ell\neq i}\sum_{j:i\in A_j}x_j=L-1, \]
    and
    \begin{align*}
        &\sum_{i''\in M \backslash [L]\cup M_i}\sum_{\substack{i'\neq i:\\ i'\notin M_{\ell} \text{ if }i''\in M_{\ell}}}\sum_{j':i',i''\in A_{j'}}x_{j'}\\
        =&\sum_{\ell\neq i}\sum_{i''\in M_{\ell}}\sum_{i'\notin \{i,\ell\}}\sum_{j':i',i''\in A_{j'}}x_{j'}=\sum_{\ell\neq i}\sum_{i'\notin \{i,\ell\}}\sum_{j':i'\in A_{j'}}x_{j'}=(L-2)(L-1).
    \end{align*}
    Moreover, for any fixed $i''$, we have
    \[\sum_{j:i,i''\in A_j}x_{j}+\sum_{\substack{i'\neq i:\\i'\notin M_{\ell} \text{ if }i''\in M_{\ell}}}\sum_{j':i',i''\in A_{j'}}x_{j'}=\sum_{j:i''\in A_j}x_{j}\le 1.  \]
    For simplicity, let 
    \[\alpha_{ii''}=\sum_{j:i,i''\in A_j}x_j, \ \beta_{ii''}=\sum_{\substack{i'\neq i:\\i'\notin M_{\ell} \text{ if }i''\in M_{\ell}}}\sum_{j':i',i''\in A_{j'}}x_{j'},\]
    for any fixed $i$, we consider the optimization problem:
    \begin{align*}
        \max &\sum_{i''\in M\backslash [L]}\alpha_{ii''}\beta_{ii''}, \\ \text{s.t.} &\sum_{i''\in M\backslash [L]} \alpha_{ii''}=L-1, \forall i, \sum_{i''\in M\backslash [L]}\beta_{ii''}=(L-2)(L-1), \forall i, \alpha_{ii''}+\beta_{ii''}\le 1, \forall i, i''. 
    \end{align*}
    Again, similar to the proof of Lemma \ref{lem:prophet_disjoint_mass}, the optimal value is $L-2$ which is obtained when $\alpha_{ii''}=1/(L-1)$ and $\beta_{ii''}=(L-2)/(L-1)$, and it then follows that term $(b)$ is upper bounded by $L(L-2)$.

    In conclusion, we have that
    \[\sum_{i=1}^{L}\sum_{i'\neq i}\sum_{t=1}^T\sum_{t'\neq t}\sum_{\substack{j\in N_t,j'\in N_{t'}\\A_j\cap A_{j'}=\emptyset\\i\in A_j, i'\in A_{j'}}} x_{j}x_{j'}\ge L(L-1)-(L-1)-L(L-2)=1. \]
\end{proof}
Combined with the previous discussion, Theorem \ref{thrm:L_partite_guarantee} then follows.
\begin{theorem}\label{thrm:L_partite_guarantee}
Given $L \ge 2$, suppose that $\pi$ of \Cref{def:ocrs} is passed $\alpha$  which satisfies
    \[1-\alpha(1+L)+\frac{\alpha^2}{L}\left(\frac{1-\alpha (1+L)+\alpha/(2L)}{1-\alpha L+\alpha/(2L)}\right)^{2L}\ge 0. \]
Then, $\pi$ is $\alpha$-selectable on $L$-partite hypergraphs.
\end{theorem}
The left-hand side function of \Cref{thrm:L_partite_guarantee} is decreasing in $\alpha$ and greater than $0$ at $\alpha = 1/(1+L)$.
Thus, $1/(1+L)$ is beatable for $L$-partite hypergraphs.

\section{Upper Bound of $(1-\frac1{(1+L)^{1+L}})/L$ for Random-element Offline CRS} \label{sec:negOffline}

Our negative result in Section \ref{ssec:tightness} exploited not only the negative correlation between realizations of elements, but also the fact that the online algorithm did not know future realizations in advance.
We now tweak the construction to provide a upper bound on \textit{offline} algorithms, which know future realizations in advance.
In particular, we can consider an offline algorithm for the accept-reject NRM problem (\Cref{def:discrete_time_nrm}) that knows in advance the active product in each batch $N_t$. 

\begin{theorem}\label{thrm:negRandomOrder}
No offline algorithm for the accept-reject NRM problem is better than $(1-\frac1{(1+L)^{1+L}})/L$-competitive against \eqref{eq:fluid_lp}, when $L$ is a prime power. 
\end{theorem}

\begin{proof}[Proof of Theorem \ref{thrm:negRandomOrder}]
Take the configuration of batches from \Cref{def:worst_case_problem} based on a finite affine plane, which exists assuming $L$ is a prime power.
We set the arrival probability to be $\lambda_j=1/(1+L)$ for every product $j$ (in any batch).
Because there are $L$ products in each batch, this means that the probability of no arrival in a batch is $1-L/(1+L)=1/(1+L)$.
We set the reward to be $r_j=1$ for all products $j$.

It is easy to check that defining $x_j=\lambda_j=1/(1+L)$ for all $j$ forms a feasible solution to the fluid LP~\eqref{eq:fluid_lp}.
Indeed, for any item $i$,
$$
\sum_{j:i\in A_j} x_j
=\sum_{t=1}^{1+L}\sum_{j\in N_t:i\in A_j}\frac{1}{1+L}
=\sum_{t=1}^{1+L}\frac{1}{1+L}
=1,
$$
where again the penultimate equality holds by property~(i) from \Cref{def:worst_case_problem}.
This shows that the optimal value of the fluid LP is at least $$\sum_j x_j=L(1+L)\frac1{1+L}=L.$$

Meanwhile, any offline algorithm can accept at most one product by property~(ii) from \Cref{def:worst_case_problem}, and the probability of no product arriving in a given batch is $1/(1+L)$.
Therefore, the expected reward of any online algorithm is at most $1-(1/(1+L))^{1+L}$, whose ratio relative to the optimal value of the fluid LP can be at most
$
(1-\frac1{(1+L)^{1+L}})/L.
$

\end{proof}

An $\alpha$-selectable random-element offline contention resolution scheme would imply an offline algorithm for the accept-reject NRM problem that is an $\alpha$-approximation against~\eqref{eq:fluid_lp}.  Therefore, a corollary of \Cref{thrm:negRandomOrder} is that no offline contention resolution scheme can be better than $(1-\frac1{(1+L)^{1+L}})/L$-selectable when there are random elements, and $L$ is a prime power.

\end{document}